\documentclass[journal]{IEEEtran}
\usepackage{cite}
\usepackage{amsmath,amssymb,amsfonts}
\usepackage{algorithmic}
\usepackage{graphicx,color}
\usepackage{subcaption}  
\usepackage{booktabs} 
\usepackage{textcomp}
\usepackage{multirow}
\usepackage{caption} 

\usepackage{geometry}
\usepackage[T1]{fontenc}
\usepackage{aecompl}
\def\BibTeX{{\rm B\kern-.05em{\sc i\kern-.025em b}\kern-.08em
		T\kern-.1667em\lower.7ex\hbox{E}\kern-.125emX}}
\begin{document}

\title{Cross Far- and Near-field Beam Management Technologies in Millimeter-Wave and Terahertz MIMO Systems}

\author{
    Yuhang~Chen,~Heyin Shen and Chong Han, \IEEEmembership{Senior~Member,~IEEE}
    \thanks{
    \textit{(Corresponding author: Chong Han.)}
    Y. Chen and H. Shen are with the Terahertz Wireless Communications (TWC) Laboratory, Shanghai Jiao Tong University, Shanghai 200240,
    China.

    C. Han is with Terahertz Wireless Communications (TWC) Laboratory,
    and with the Department of Electronic Engineering and Cooperative Medianet
    Innovation Center (CMIC), Shanghai Jiao Tong University, Shanghai 200240,
    China (e-mail: chong.han@sjtu.edu.cn).

		}
	}
	\maketitle
	\thispagestyle{empty}

\begin{abstract} 
The evolution of wireless communication toward next-generation networks introduces unprecedented demands on data rates, latency, and connectivity.
To meet these requirements, two key trends have emerged: the use of higher communication frequencies to provide broader bandwidth, and the deployment of massive multiple-input multiple-output systems with large antenna arrays to compensate for propagation losses and enhance spatial multiplexing.
These advancements significantly extend the Rayleigh distance, enabling near-field (NF) propagation alongside the traditional far-field (FF) regime.
As user communication distances dynamically span both FF and NF regions, cross-field (CF) communication has also emerged as a practical consideration.
Beam management (BM)—including beam scanning, channel state information estimation, beamforming, and beam tracking—plays a central role in maintaining reliable directional communications.
While most existing BM techniques are developed for FF channels, recent works begin to address the unique characteristics of NF and CF regimes.
This survey presents a comprehensive review of BM techniques from the perspective of propagation fields.
We begin by building the basic through analyzing the modeling of FF, NF, and CF channels, along with the associated beam patterns for alignment.
Then, we categorize BM techniques by methodologies, and discuss their operational differences across propagation regimes, highlighting how field-dependent channel characteristics influence design tradeoffs and implementation complexity.
In addition, for each BM method, we identify open challenges and future research directions, including extending FF methods to NF/CF scenarios, developing unified BM strategies for field-agnostic deployment, and designing low-overhead BM solutions for dynamic environments.
\end{abstract}


\IEEEspecialpapernotice{(Invited Paper)}

\maketitle

\section{INTRODUCTION}
\label{sec_Introduction}

With the commercialization of the fifth generation (5G), the development and expectations surrounding the sixth generation (6G) and beyond wireless networks have garnered increasing attention from both academia and industry~\cite{ref_6G1}. The anticipated features of 6G include ultrafast speeds reaching up to terabits per second and ultra-low latency as small as 1 ms, supporting emerging applications such as extended reality, holographic communications, and intelligent interaction~\cite{ref_6G2, ref_6GKPI}. To meet the extreme requirements of future communications, two fundamentally interconnected technological trends can be observed: the expansion of the frequency spectrum and the deployment of massive antenna arrays.
 
On one hand, the deployment of higher frequencies up to millimeter-wave (mmWave, 30-300 GHz) and terahertz (THz, 0.1-10 THz) bands, enables wider bandwidths~\cite{ref_minband,ref_6GmmW_THz,ref_THz}. This expansion is expected to overcome the limitations of bandwidth and meet the demand for terabit-per-second wireless links~\cite{ref_THz_Old_revisit,ref_rol_mmW}. On the other hand, multiple-input multiple-output (MIMO) technology has been widely adopted in wireless communications. One notable trend in the evolution of MIMO is the integration of more antennas, leading to the development of massive MIMO (16-256 elements) and ultra-massive MIMO (UM-MIMO, 256+ elements)~\cite{ref_6G_ELAA,ref_bjornson2024towards}. The incorporation of more antenna elements to compose large MIMO systems is advantageous for compensating propagation losses at higher frequency bands. Additionally, increasing the number of elements in MIMO systems supports higher network densification by providing greater spatial degrees of freedom (SDoF), positioning large MIMO systems as a key technology for 6G and beyond~\cite{ref_6G_network,ref_6G_frontiers}.

\subsection{Far-field, Near-field and Cross-field Communications}
In communication systems prior to 5G, the communication distance typically falls within the far-field (FF) region of the antenna array, where the wavefront can be approximated as planar.  
The boundary between the far-field and near-field (NF) regions is defined by the Rayleigh distance, which is proportional to the square of the array aperture $S^2$ divided by the wavelength $\lambda$, i.e., $S^2/\lambda$.  
In conventional MIMO systems, the array aperture is relatively small and the operating wavelengths are large~\cite{ref_rayleigh_Mag}.  
For example, an antenna array with a 0.3-meter aperture operating at 2.1~GHz yields a Rayleigh distance of approximately 2.1 meters, which is negligible for typical outdoor communication scenarios.

As antenna arrays become larger and carrier frequencies increase, the Rayleigh distance extends significantly, reaching practical communication distances.  
Recently, there has been growing interest in MIMO NF communications~\cite{ref_Near_Cui, ref_nearfield_tut, ref_NF_OJCOMS}.  
For instance, an antenna array with a 0.5-meter aperture operating at 28~GHz yields a Rayleigh distance of 46.7 meters, sufficient to cover many outdoor users.  
At even higher frequencies, the Rayleigh distance can extend to hundreds or even thousands of meters, motivating the study of NF communications.

\begin{figure}
\centerline{\includegraphics[width=0.45\textwidth]{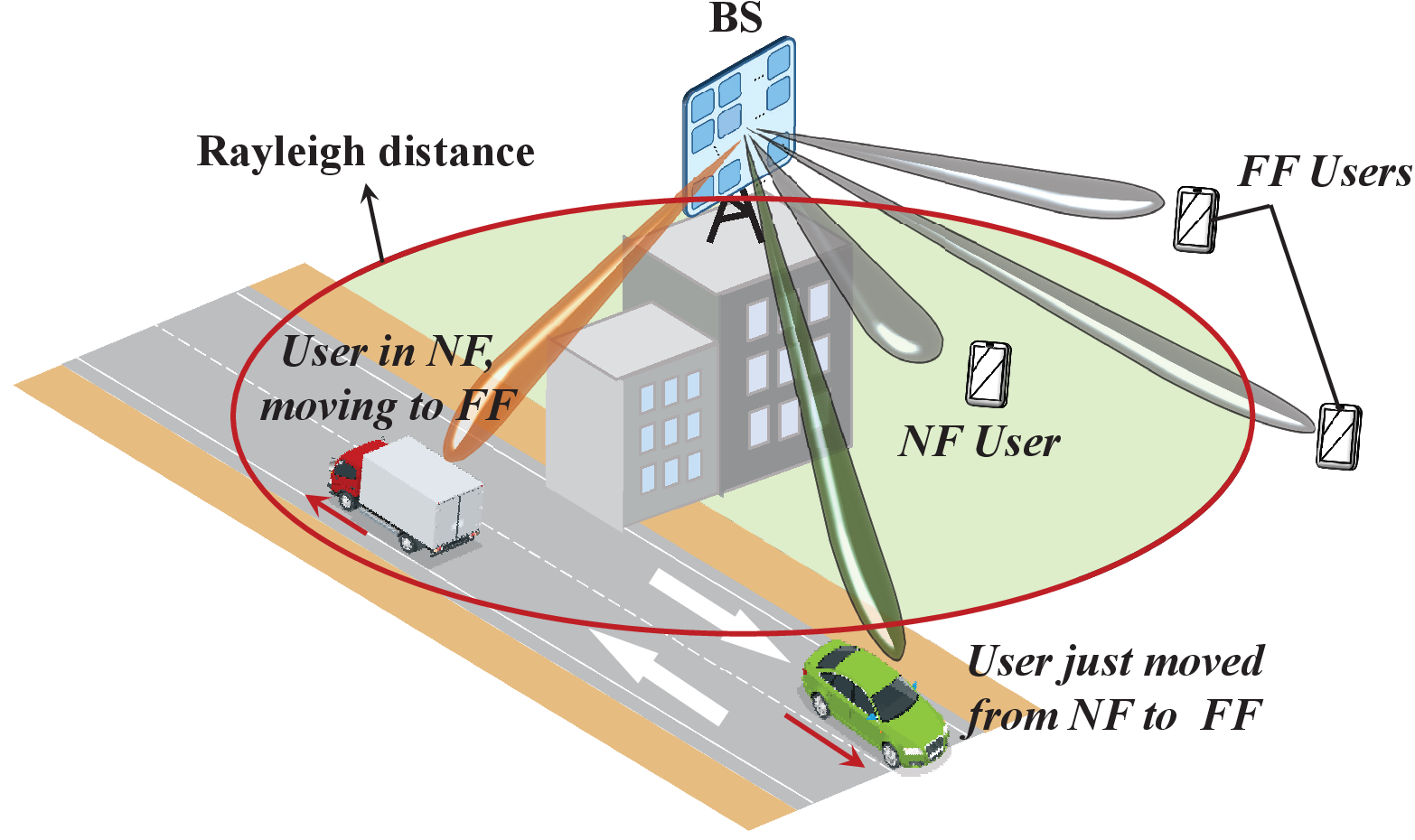}}
    \caption{Illustration of users in different propagation regions.}
   \label{fig_FF_NF_CF}
\end{figure}

However, treating the NF and FF regions as strictly separate is impractical, since real-world communication distances often span both regions.  
Moreover, user mobility can cause transitions between the NF and FF, leading to the notion of {cross-field (CF)} communication~\cite{ref_Cross, ref_hybrid_field}.  
This concept captures the coexistence and dynamic transition between NF and FF regions, and highlights the need for unified communication techniques that can operate effectively across both.  
A related concept, referred to as {hybrid-field communication}, has also been explored in recent studies~\cite{ref_hybrid_field}.
Fig.~\ref{fig_FF_NF_CF} illustrates users located in different propagation regions.

\subsection{Beam Management}

At high frequencies, large antenna arrays are typically deployed to generate narrow beams with high beamforming gain, thereby compensating for severe propagation losses. To establish a reliable communication link, the beams at the transmitter (Tx) and receiver (Rx) must be properly aligned. This process involves a set of procedures collectively referred to as beam management (BM)~\cite{ref_BM_3GPP}, which includes beam scanning, channel state information (CSI) estimation, beamforming, and beam tracking.

\subsubsection{Beam Scanning}
\label{subsec_beam_scanning}

At high frequencies, omnidirectional pilot transmission during initial access suffers from significant propagation loss, resulting in an extremely low signal-to-noise ratio (SNR) at Rx, which may prevent successful signal detection. To overcome this, directional beams are used for scanning even before the link is established. This process, known as beam scanning, employs high-gain beams to sweep the spatial domain, enhancing the received SNR.
Beam scanning typically uses a predefined codebook, where each codeword corresponds to a specific beamformer. The beams in the codebook are designed to cover different regions of the angular and distance domains to ensure full spatial coverage. Hence, the design of the codebook is critical for efficient beam scanning.

A typical scanning procedure involves the transmitter first sweeping its beam across all directions, transmitting pilot signals with each beam while the receiver remains fixed. Once the Tx sweep is complete, the receiver performs a similar sweep while the transmitter uses a fixed beam. This process continues until all combinations of Tx and Rx beams are covered. The number of pilot signals transmitted determines the scanning overhead. 


\begin{table}[t]
    \centering
    \caption{Comparison of beam estimation and channel estimation-based methods for beamforming.}
    \label{Tab_Comp_BA_CE}
    \renewcommand{\arraystretch}{1.3}  
    \setlength{\tabcolsep}{7pt}
    \begin{tabular}{|c|c|c|}
        \hline
        \textbf{Metric} & \textbf{Beam Estimation} & \textbf{Channel Estimation} \\
        \hline
        Scanning overhead & Low & High \\
        \hline
        Fast variation support & High & Low \\
        \hline
        Spectral efficiency & Low & High \\
        \hline
    \end{tabular}
\end{table}

\begin{figure*}[t]
    \centering
    \begin{subfigure}[t]{1\textwidth}
        \centering
        \includegraphics[width=\textwidth]{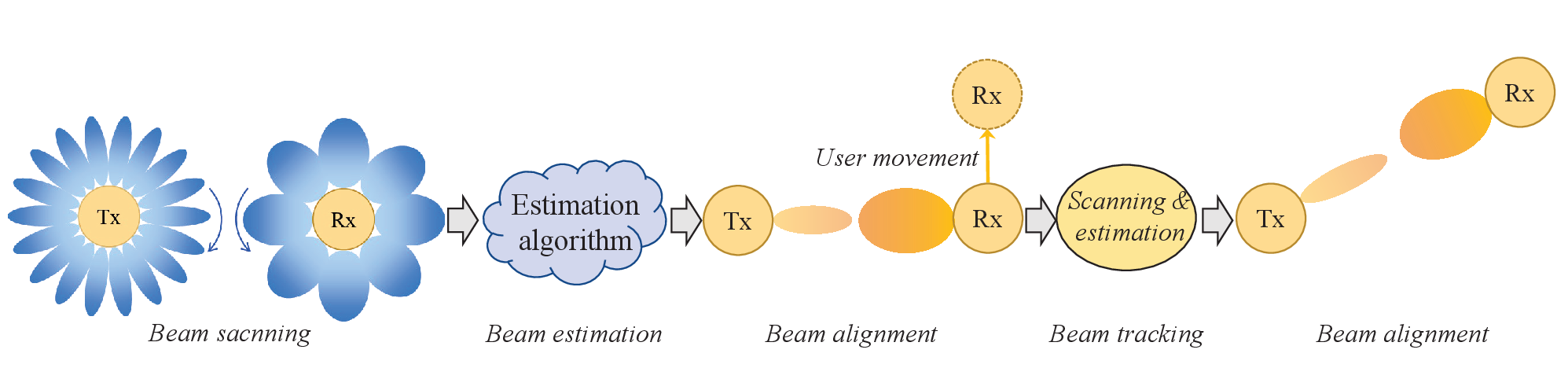}
        \caption{Beam estimation and alignment.}
        \label{fig_overall_procedure_a}
    \end{subfigure}
    \hspace{0.04\textwidth}
    \begin{subfigure}[t]{0.7\textwidth}
        \centering
        \includegraphics[width=\textwidth]{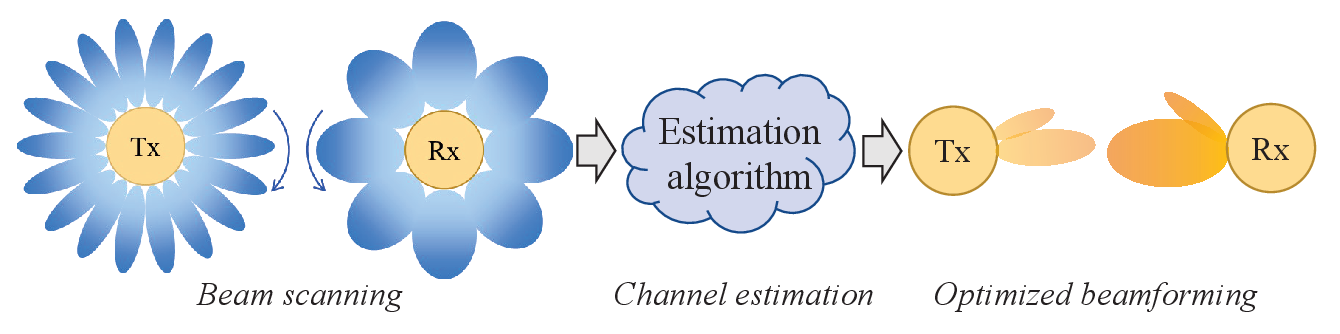}
        \caption{Channel estimation and beamforming.}
        \label{fig_overall_procedure_b}
    \end{subfigure}
    \caption{Illustration of two representative beam management procedures: (a) beam estimation followed by beam alignment, and (b) channel estimation followed by beamforming.}
    \label{fig_overall_procedure_combined}
\end{figure*}

\subsubsection{CSI Estimation}

Following beam scanning, Rx performs CSI estimation using the received pilot signals. Depending on the target application, CSI estimation can be divided into two types: beam estimation and channel estimation.

Beam estimation aims to identify the optimal beam direction, often aligning with the strongest propagation path. Since beam direction is itself a form of CSI, this method is sufficient for beam alignment. In some literature, the combination of beam scanning and beam estimation is referred to as beam training. 
Channel estimation, in contrast, aims to recover the full channel response, providing more detailed information than just the beam direction. This enables the design of beamforming weights that optimize system performance beyond simple alignment, improving spectral efficiency and reducing interference.


\subsubsection{Beamforming}
Beamforming is performed based on the estimated CSI. Two main strategies are commonly used, i.e., beam estimation-based beam alignment and channel-based beamforming.
Particularly, when only beam direction is estimated, beamforming is performed by aligning the beams of Tx and Rx. 
By contrast, when full CSI is available, beamforming weights can be optimized according to system objectives, such as maximizing spectral efficiency or minimizing interference in multi-user scenarios. 

The trade-offs between beam and channel estimation are summarized in Table~\ref{Tab_Comp_BA_CE} and illustrated as follows: 
\begin{itemize}
  \item {Overhead:} Beam estimation requires fewer pilot signals, resulting in lower scanning overhead. Channel estimation incurs higher overhead due to the need for more measurements.
  \item {Channel dynamics:} Beam estimation is more suitable for fast-varying channels, where frequent full CSI estimation is impractical. Channel estimation is better suited for slowly-varying scenarios, such as fixed wireless backhaul.
  \item {Performance:} Channel-based beamforming typically achieves higher spectral efficiency, since it enables globally optimized weight design. Beam alignment may suffer from suboptimal performance due to its directional nature.
\end{itemize}

\subsubsection{Beam Tracking}

The aforementioned BM techniques apply to both static and dynamic scenarios. In the presence of channel variations, such as user mobility, beam tracking becomes essential. It is considered a specialized BM process focused on maintaining beam alignment over time. 
However, to reduce the overhead, tracking schemes often exploit the temporal correlation of the channel to predict future beam directions efficiently.

To summarize the aforementioned beam management techniques, Fig.~\ref{fig_overall_procedure_combined} presents two representative procedures. 
Fig.~\ref{fig_overall_procedure_combined}(a) shows the process based on beam estimation and alignment, while Fig.~\ref{fig_overall_procedure_combined}(b) illustrates the procedure involving channel estimation followed by beamforming.
 
\subsection{Motivations and Contributions}

Inspired by the growing importance of FF, NF, and CF communications, as well as the central role of BM, this survey provides a comprehensive review of BM techniques from the perspective of propagation fields.  
Rather than treating BM techniques in each field separately, we examine how various methods are influenced by and can be adapted to different field regimes.  
Furthermore, we identify open challenges and research opportunities to guide future work in this area.  
The main contributions of this survey are summarized as follows:

\begin{itemize}
    \item We begin by reviewing the fundamentals of field-aware BM, including channel models and beam patterns.  
    Specifically, we introduce the NF, FF, and CF channel models and analyze the approximation errors of the FF and CF models relative to the ground-truth NF channel.  
    In addition, we present beam patterns designed for beam alignment under each model, showing that FF beams provide resolution only in the angular domain, while NF and CF beams offer resolution in both angle and distance domains.
    
    \item We present a systematic and comprehensive review of BM techniques across FF, NF, and CF communications, covering beam scanning, CSI estimation, beamforming, and beam tracking.  
    Instead of introducing techniques separately by the propagation field, we adopt a unified methodology-based classification and compare how each technique operates under different propagation conditions.  
    Based on the key difference that FF channels offer only angular resolution, while NF and CF channels provide both angular and distance resolution, we highlight how the distinct characteristics and constraints of each regime affect design choices, performance tradeoffs, and implementation complexity.

    \item Building on the field-aware analysis of BM techniques, we outline potential research directions and technical challenges for each BM category.  
    These mainly include: extending existing FF techniques to NF/CF communications under constraints of low overhead and complexity,  
    conducting comprehensive comparisons of BM techniques to support methodology selection across different scenarios, 
    developing unified BM technologies applicable across propagation fields for practical deployment, 
    and designing BM strategies tailored to emerging communication systems.
\end{itemize}

\begin{figure}[t]
    \centering
    {\includegraphics[width= 0.4\textwidth]{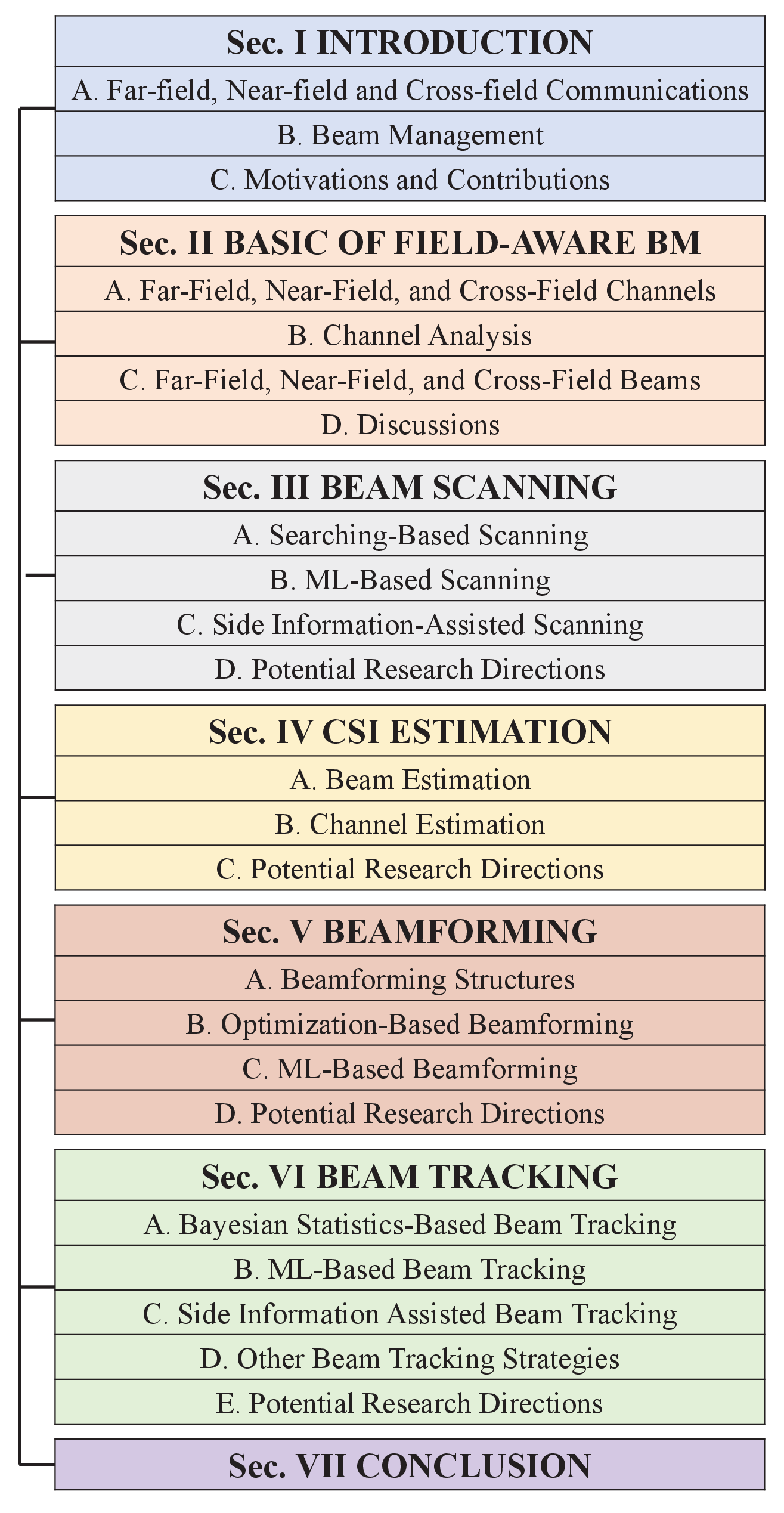}}
    \caption{{Organization of this paper} }
    \label{fig_organization}
\end{figure}

The organization of this paper is as follows. 
Sec.~\ref{sec_basic} presents the fundamentals of field-aware BM. It covers the characteristics of FF, NF, and CF channels, provides channel analysis, discusses beam properties in different propagation fields, and highlights key observations.
Sec.~\ref{sec_beam_scanning} reviews beam scanning techniques. These are categorized into searching-based, machine learning (ML)-based, and side information-assisted methods. Potential research directions for beam scanning are also discussed.
Sec.~\ref{sec_Csi_estimation} focuses on CSI estimation, dividing the discussion into beam estimation and channel estimation approaches. Their respective advantages and research opportunities in different field regimes are also summarized.
Sec.~\ref{sec_beamforming} discusses beamforming design from a field-aware perspective. It introduces beamforming structures and analyzes beamforming algorithms. Potential directions for improving field-adaptive beamforming are also provided.
Sec.~\ref{sec_Beam_tracking} covers beam tracking strategies under dynamic channel conditions. 
It includes Bayesian statistics-based methods, ML-based approaches, and techniques leveraging side information, along with a discussion of other strategies and open problems.
Potential research directions on beam tracking are provided. 
Sec.~\ref{sec_conclusion} concludes the entire survey.
The organization of this survey is illustrated in Fig.~\ref{fig_organization}. 

\section{BASIC OF FIELD-AWARE BM}
\label{sec_basic}
Technologies involved in the BM procedure are inherently field-specific, as the beam pattern strongly depends on the underlying channel model.  
In the literature, two primary MIMO channel models are commonly considered: the spherical-wave model (SWM), typically used in the NF, and the planar-wave model (PWM), generally adopted in the FF~\cite{ref_SW_PW_Modeling,ref_spherical_fronts}.  
More recently, a hybrid spherical- and planar-wave model (HSPM) has been proposed in~\cite{ref_HSPM}, which captures spherical-wave propagation across subarrays and planar-wave propagation within each subarray to realize the combination of NF and FF. 

In this section, we first introduce the field-specific channel models and analyze their applicability across different propagation conditions.  
We then discuss the corresponding beam patterns in the FF, NF, and CF regimes.

\subsection{Far-Field, Near-Field and Cross-Field Channels}
\begin{figure*}[t]
    \centering
    {\includegraphics[width= 0.95\textwidth]{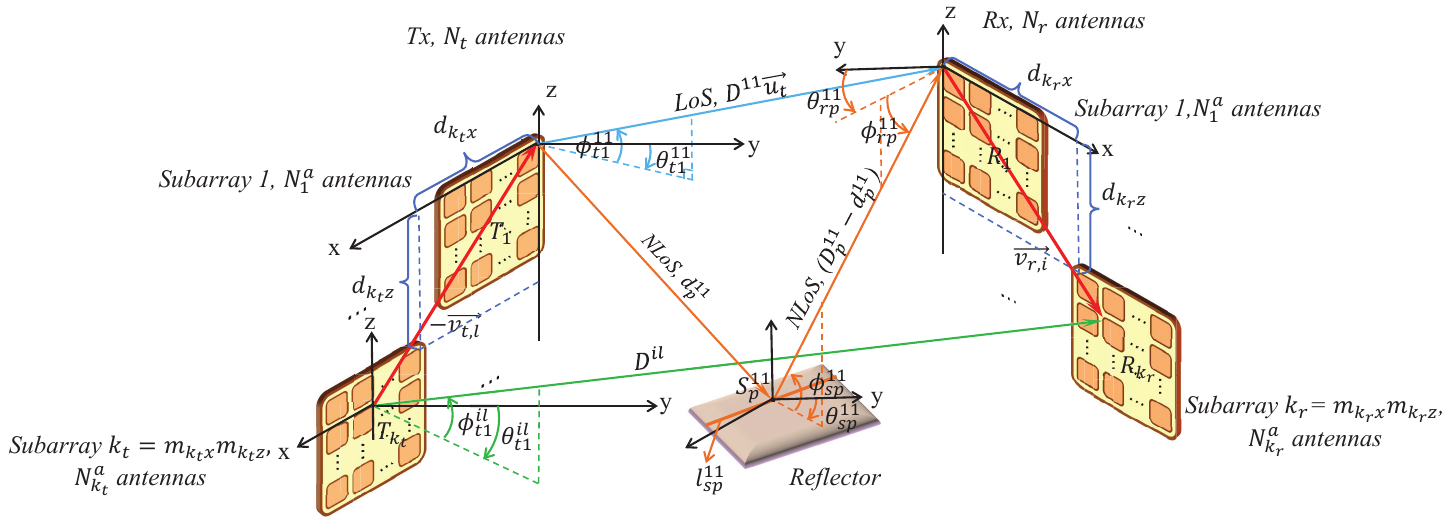}}
    \caption{{Channel model in 2D planar array systems.} }
    \label{fig_gen_ULAA_system_model}
\end{figure*} 

\subsubsection{Spherical-Wave Channel Model}
\label{subsec_CHAN_SWM}
The SWM accounts for the spherical radio-wave front and is considered the most accurate model for characterizing radio-wave propagation~\cite{ref_SW_PW_Modeling}.  
As illustrated in Fig.~\ref{fig_gen_ULAA_system_model}, we consider a two-dimensional (2D) planar antenna array deployed at both the Tx and Rx. Let $D^{il}$ denote the distance between the $l^{\rm th}$ transmit antenna and the $i^{\rm th}$ receive antenna, where $i = 1,\ldots, N_r$ and $l = 1,\ldots, N_t$. Here, $N_t$ and $N_r$ represent the number of antennas at the Tx and Rx, respectively.  
In the SWM, the complex path gain between the $l^{\rm th}$ transmit antenna and the $i^{\rm th}$ receive antenna, denoted by $\alpha^{il}$, is modeled as
\begin{equation}
\label{equ_CHAN_gain}
\alpha^{il} = \lvert\alpha^{il}\rvert e^{-j\frac{2\pi}{\lambda}D^{il}}.
\end{equation}

The overall channel response between the $l^{\rm th}$ transmit and $i^{\rm th}$ receive antennas is expressed as
\begin{equation}
\label{equ_CHAN_SW_channel}
\mathbf{H}_{\rm S}(i,l)=\sum_{p=1}^{N_p}\lvert\alpha^{il}_p\rvert e^{-j\frac{2\pi}{\lambda}D^{il}_p},
\end{equation}
where $\mathbf{H}_{\rm S}$ is the $N_r\times N_t$ spherical-wave channel matrix, and $p = 1,\ldots, N_p$ indexes the propagation paths. Specifically, $p = 1$ corresponds to the line-of-sight (LoS) path, while $p > 1$ represents non-line-of-sight (NLoS) paths.  

The channel matrix $\mathbf{H}_{\rm S}$ in~\eqref{equ_CHAN_SW_channel} depends on the parameter set $\mathcal{P}_{S} = \{|\alpha^{il}_p|, D^{il}_p\}$, which contains $2N_p(N_t \times N_r)$ elements. The number of parameters grows quadratically with the number of antennas and linearly with the number of paths, leading to increased modeling and computational complexity in large-scale systems.

\subsubsection{Planar-Wave Channel Model}\label{subsec_PWM Method}

The PWM is an approximation of the SWM when the array size is far less than the communication distance. 
In particular, the signal transmission is approximated as parallel and the wavefront is analyzed as a plane. 
We denote $(\theta_{tp}^{il}, \phi_{tp}^{il})$, $(\theta_{rp}^{il}, \phi_{rp}^{il})$ as the direction-of-departure (DoD) and direction-of-arrival (DoA) pairs for the $p^{\rm th}$ path between the $l^{\rm th}$ transmitted and the $i^{\rm th}$ received antennas, in which $\theta$ and $\phi$ denote the azimuth and elevation angles, respectively. 
As shown in Fig.~\ref{fig_gen_ULAA_system_model}, we consider the $l^{\rm th}$ transmitted antenna is located in the $k_t^{{\rm th}}$ subarray, and use the index pair $(n_{k_tx}, n_{k_tz})$ to denote the position of this antenna in the subarray, where $n_{k_tx} = 0,1,..., N_{k_tx}^{a}-1$, $n_{k_tz} = 0,1,..., N_{k_tz}^{a}-1$, and $N_{k_tx}^{a}$ and $N_{k_tz}^{a}$ represent the number of antennas of the $k_t^{{\rm th}}$ subarray along x-axis and z-axis, respectively. 
Similarly, the index pair $(n_{k_rx}, n_{k_rz})$ determines the position of the $i^{\rm th}$ received antenna in the $k_r^{{\rm th}}$ subarray, where $n_{k_rx} = 0,1,...,N_{k_rx}^{a}-1$, $n_{k_rz} = 0,1,...,N_{k_rz}^{a}-1$, and $N_{k_rx}^{a}$ and $N_{k_rz}^{a}$ represent the number of antennas of the $k_r^{{\rm th}}$ subarray along x-axis and z-axis, respectively.

By considering the plane-wave transmission, $D^{il}$ is approximated as~\cite{ref_two_level,ref_SW_PW_Modeling}
\begin{equation}
D^{il}= D^{11}+\Delta D^{il}, 
\end{equation}
where $\Delta D^{il} =d( \psi_t-\psi_r)$, and 
$\psi_t = (m_{tx}+n_{tx}){\sin}\theta_{tp}^{11}{\cos}\phi_{tp}^{11}-(m_{tz}+n_{tz}){\sin}\phi_{tp}^{11}$,
$\psi_r = (m_{rx}+n_{rx}){\sin}\theta_{rp}^{11}{\cos}\phi_{rp}^{11}-(m_{rz}+n_{rz}){\sin}\phi_{rp}^{11}$ according to the geometric relationships in Fig.~\ref{fig_gen_ULAA_system_model}.
Since $\Delta D^{il}$ is far less than $D^{11}$, $|\alpha^{il}|$ is considered approximately the same. The complex path gain between the $l^{\rm th}$ transmitted antenna and the $i^{\rm th}$ received antenna is approximated as
\begin{equation}
    \label{equ_CHAN_approx_gain}
\alpha^{il}\approx|\alpha^{11}|e^{-j\frac{2\pi}{\lambda} D^{il}} = \alpha^{11} e^{-j\frac{2\pi}{\lambda}\Delta D^{il}}.
\end{equation}
Therefore, the planar-wave channel response between the $l^{\rm th}$ transmitted and $i^{\rm th}$ received antennas is represented as
\begin{equation}
\mathbf{H}_{\rm P}(i,l)=\Sigma_{p=1}^{N_p}\alpha^{11}_pe^{-j\frac{2\pi d}{\lambda}
    (\psi_t-\psi_r)}.
\label{channel_planar}
\end{equation}

The planar-wave channel matrix in~\eqref{channel_planar} can be further arranged in a compact form as
\begin{equation}\label{equ_CHAN_P_model}
\mathbf{H}_{\rm P}=\Sigma_{p=1}^{N_p}\alpha^{11}_p
\mathbf{a}_{rp}(\theta_{rp}^{11},\phi_{rp}^{11})\mathbf{a}_{tp}^{{\rm H}}(\theta_{tp}^{11},\phi_{tp}^{11}),
\end{equation}
where $\mathbf{a}_{rp}(\theta_{rp}^{11},\phi_{rp}^{11})$ and $\mathbf{a}_{tp}(\theta_{tp}^{11},\phi_{tp}^{11})$ stand for the array response vectors of the $p^{\rm th}$ path at Tx and Rx, respectively. 
Specifically, $\mathbf{a}_{tp}(\theta_{tp}^{11},\phi_{tp}^{11}) $ is expressed as
\begin{equation}\label{equ_CHAN_array_steering_vector}
\mathbf{a}_{tp}(\theta_{tp}^{11},\phi_{tp}^{11})=\left[1 \dots  \mathrm{e}^{-j\frac{2\pi d}{\lambda}\psi_t}\dots \mathrm{e}^{-j\frac{2\pi d}{\lambda}\Psi_t}\right]^{\mathrm{T}},
\end{equation}
where $\Psi_t = (M_{K_tx}+N^a_{K_tx}-2){\sin}\theta_{tp}^{11}{\cos}\phi_{tp}^{11}-(M_{K_tz}+N^a_{K_tz}-2){\sin}\phi_{tp}^{11}$, $N^a_{K_tx}$ and $N^a_{K_tz}$ denote the number of antennas of the $K_t^{\rm th}$ subarray along x-axis and z-axis, respectively.
Moreover, $\Psi_r = (M_{K_rx}+N^a_{K_rx}-2){\sin}\theta_{rp}^{11}{\cos}\phi_{rp}^{11}-(M_{K_rz}+N^a_{K_rz}-2){\sin}\phi_{rp}^{11}$, where $N^a_{K_rx}$ and $N^a_{K_rz}$ denote the number of antennas of the $K_r^{\rm th}$ subarray along x-axis and z-axis, respectively. 
Similarly, $\mathbf{a}_{rp}(\theta_{rp}^{11},\phi_{rp}^{11})$ is constructed by replacing the superscript $t$ as $r$ in~\eqref{equ_CHAN_array_steering_vector}.

The parameter set $\mathcal{P}_{P} = \{|\alpha^{11}_p|, D^{11}_p, \theta^{11}_{tp},\phi^{11}_{tp}, \theta^{11}_{rp}, \phi^{11}_{rp} \}$ with $6N_p$ elements uniquely determines the planar-wave channel matrix in~\eqref{equ_CHAN_P_model}. 
Given the sparsity of high-frequency channels in the millimeter-wave and THz band, that $N_p\leq 10$~\cite{ref_DAoSA}, the required number of parameters to determine the PWM is much smaller than the SWM in~\eqref{equ_CHAN_SW_channel}.

\subsubsection{Hybrid-Spherical and Planar-Wave Channel Model}
\label{sec_CHAN_HSPM_UMMIMO}

In the HSPM, the PWM is employed within one subarray, which remains precise due to the relatively small array size. 
Among the subarrays, the SWM is utilized to improve the modeling accuracy. As shown in Fig.~\ref{fig_gen_ULAA_system_model}, we consider that the sub-channel between the $k_{t}^{\rm th}$ transmitted subarray and the $k_{r}^{\rm th}$ received subarray is $\Sigma_{p=1}^{N_p}\alpha^{k_{r}k_{t}}_p\mathbf{a}_{rp}^{k_{r}k_{t}}(\mathbf{a}_{tp}^{k_{r}k_{t}})^{\rm H}$, where $\alpha^{k_{r}k_{t}}_p$ denotes the path gain between the reference antennas of the $k_t^{\rm th}$ transmitted and $k_r^{\rm th}$ received subarrays.
Moreover, $\mathbf{a}_{tp}^{k_{r}k_{t}}=\mathbf{a}_{tp}^{k_rk_t}(\theta_{tp}^{k_rk_t},\phi_{tp}^{k_rk_t})$ and $\mathbf{a}_{rp}^{k_{r}k_{t}}=\mathbf{a}_{rp}^{k_rk_t}(\theta_{rp}^{k_rk_t},\phi_{rp}^{k_rk_t})$ represent the array response vector of the transmitted and received subarrays as~\eqref{equ_CHAN_array_steering_vector}, respectively.
Therefore, the HSPM for MIMO systems can be represented by~\eqref{equ_CHAN_HSPM}.
\begin{figure*}[ht]
\centering
\begin{equation}
\label{equ_CHAN_HSPM}
    \mathbf{H}_{\rm HSPM} = \sum_{p=1}^{N_p} \vert\alpha^{11}_p\vert \left[\begin{array}{ccc}
     e^{-j\frac{2\pi}{\lambda}D^{11}_p}\mathbf{a}_{rp}^{11} (\mathbf{a}_{tp}^{11})^{\rm H} & \ldots & e^{-j\frac{2\pi}{\lambda}D^{1K_t}_p}\mathbf{a}_{rp}^{1K_t}(\mathbf{a}_{tp}^{1K_t})^{\rm H} \\
    \vdots & \ddots & \vdots \\
     e^{-j\frac{2\pi}{\lambda}D^{K_r1}_p}\mathbf{a}_{rp}^{K_r1}(\mathbf{a}_{tp}^{K_r1})^{\rm H} & \ldots & e^{-j\frac{2\pi}{\lambda}D^{K_rK_t}_p}\mathbf{a}_{rp}^{K_rK_t}(\mathbf{a}_{tp}^{ K_rK_t})^{\rm H} \\
    \end{array}\right].
\end{equation}
\end{figure*}

    In \eqref{equ_CHAN_HSPM}, different subarrays share common reflectors, leading to the number of multi-path being the same for different subarray pairs. 
    Moreover, the amplitude of path gain is the same for different subarrays, while the DoD, DoA and phase of path gain are different among different subarrays.

The HSPM in~\eqref{equ_CHAN_HSPM} is characterized by the parameter set $\mathcal{P}_{HSPM} =\left\{|\alpha_{p}^{11}|,D^{k_rk_t}_p, \theta_{rp}^{k_rk_t},\phi_{rp}^{k_rk_t},\right.$ $\left.\theta_{tp}^{k_rk_t},\phi_{tp}^{k_rk_t} \right\}$, which contains $N_p(1+5K_rK_t)$ elements.
The spherical-wave MIMO channel model~\eqref{equ_CHAN_SW_channel} and the planar-wave MIMO channel model~\eqref{equ_CHAN_P_model} are special cases of the HSPM when $K_t=N_t, K_r=N_r$ and $K_t=K_r=1$, respectively.
With various $K_t$ and $K_r$, the HSPM achieves high accuracy with a relatively small number of channel parameters, compared to the PWM and SWM, respectively, which will be shown in Sec.~\ref{subsec_Analysis}.
Notably, HSPM is a framework for modeling the cross-field channels. Detailed parameters in the HSPM need to be further characterized by channel measurement and analysis as studied in~\cite{ref_cross_measure}.



\subsection{Channel Analysis}
\label{subsec_Analysis}

\subsubsection{Accuracy and Applicability of the PWM}
\label{subsec_CHAN_Accuracy_Applicability_PWM}

To evaluate the accuracy of the planar-wave model (PWM), we consider the approximation error of the path gain for the line-of-sight (LoS) path. The extension to non-line-of-sight (NLoS) paths follows a similar procedure.  
Specifically, the normalized approximation error between the $l^{\rm th}$ transmit antenna and the $i^{\rm th}$ receive antenna is defined as
\begin{equation} \label{equ_error_planar_1}
\epsilon^{il}_{\rm PWM} = \frac{
\left\lvert 
|\alpha^{il}| e^{-j\frac{2\pi}{\lambda}D^{il}} 
- 
|\alpha^{11}| e^{-j\frac{2\pi}{\lambda} D^{11}} e^{-j\frac{2\pi}{\lambda}\Delta D^{il}} 
\right\rvert
}{|\alpha^{il}|},
\end{equation}
where the first and second terms correspond to the path gain computed by the SWM and PWM (from~\eqref{equ_CHAN_gain} and~\eqref{equ_CHAN_approx_gain}), respectively.

Since the amplitude $|\alpha^{il}|$ is approximately proportional to the distance $D^{il}$, and the array size is assumed much smaller than the communication distance, we have $|\alpha^{11}| \approx \dots \approx |\alpha^{N_rN_t}|$.  
Thus, \eqref{equ_error_planar_1} simplifies to:
\begin{subequations}
\label{equ_error_planar}
\begin{align}
\epsilon^{il} &\approx \left\lvert 
e^{-j\frac{2\pi}{\lambda}D^{il}} - 
e^{-j\frac{2\pi}{\lambda} D^{11}} e^{-j\frac{2\pi}{\lambda}\Delta D^{il}} 
\right\rvert, \\
&= \left\lvert 
2\sin\left( \frac{\pi}{\lambda} (D^{il} - D^{11} - \Delta D^{il}) \right)
\right\rvert.
\end{align}
\end{subequations}

Following the derivations in~\cite{ref_HSPM}, the approximation error in~\eqref{equ_error_planar} can be expressed as
\begin{equation} \label{equ_CHAN_approximation_error}
\begin{split}
\epsilon^{il}_{\rm PWM} &\approx 
\Bigg| 2\sin\Bigg( \frac{\pi d^2}{2 D^{11} \lambda}
\Big[
(m_{k_rx} - m_{k_tx} + n_{k_rx} - n_{k_tx})^2 \\
& \quad \cdot (\sin^2 \theta^{11}_{t1} \cos^2 \phi^{11}_{t1} + \cos^2 \theta^{11}_{t1}) \\
& \quad + (m_{k_rz} - m_{k_tz} + n_{k_rz} - n_{k_tz})^2 \cos^2 \phi^{11}_{t1}
\Big] + \mathcal{P} \Bigg) \Bigg|,
\end{split}
\end{equation}
where $\mathcal{P}$ includes higher-order residual terms.

From~\eqref{equ_CHAN_approximation_error}, we observe that the PWM approximation error $\epsilon^{il}_{\rm PWM}$ increases with the antenna index differences in both horizontal and vertical directions, and decreases with the communication distance $D^{11}$ and wavelength $\lambda$.  
We define the maximum values of the antenna index terms as 
${\rm max} \{|M_{K_{tx}} + N_{K_{tx}}|^2, |M_{K_{rx}} + N_{K_{rx}}|^2\}$ and 
${\rm max} \{|M_{K_{tz}} + N_{K_{tz}}|^2, |M_{K_{rz}} + N_{K_{rz}}|^2\}$, respectively.  
Let the normalized array sizes be defined as
\begin{equation}
\begin{split}
\label{equ_ltlr}
\mathcal{L}_t &= \sqrt{(M_{K_{tx}} + N_{K_{tx}})^2 + (M_{K_{tz}} + N_{K_{tz}})^2},\\
\mathcal{L}_r & = \sqrt{(M_{K_{rx}} + N_{K_{rx}})^2 + (M_{K_{rz}} + N_{K_{rz}})^2}.
\end{split}
\end{equation}
The corresponding physical array apertures are $d \mathcal{L}_t$ and $d \mathcal{L}_r$ at Tx and Rx, respectively.  
The quantity $\frac{\pi d^2 \mathcal{L}_t \mathcal{L}_r}{\lambda D^{11}}$ characterizes the transition between near-field and far-field, which aligns with the Rayleigh distance expression $D_{\rm ray} = \frac{2 S^2}{\lambda}$~\cite{ref_rayleigh_Mag}, where $S$ denotes the array aperture.  

Rewriting the Rayleigh condition as $\frac{2 S^2}{\lambda D_{\rm ray}} = 1$, we note that the expressions are similar up to a constant.  
Thus, when $\frac{\pi d^2 \mathcal{L}_t \mathcal{L}_r}{\lambda D^{11}} \gtrsim 1$, the far-field assumption is violated and the PWM becomes inaccurate.  
In contrast, when this quantity is much less than 1, the PWM provides a sufficiently accurate approximation.

For example, consider a system with carrier frequency $f = 3$~GHz and $D^{11} = 100$~m, with both Tx and Rx equipped with $4 \times 4$ uniform planar arrays (UPAs) and half-wavelength spacing. Then, $\frac{\pi d^2 \mathcal{L}_t \mathcal{L}_r}{\lambda D^{11}} \approx 1.6 \times 10^{-3} \ll 1$, and the PWM is accurate.  
By contrast, at $f = 0.3$~THz and $D^{11} = 10$~m under the same array size, the expression yields $\frac{\pi d^2 \mathcal{L}_t \mathcal{L}_r}{\lambda D^{11}} \approx 1.57$, and the PWM approximation becomes inaccurate.

\subsubsection{Accuracy and Applicability of the HSPM}
\label{subsec_CHAN_Accuracy_Applicability_HSPM}

The analysis of the HSPM follows a similar approach.  
Since the HSPM considers spherical-wave propagation between subarrays and planar-wave propagation within each subarray, the approximation error is effectively bounded by the error at the subarray level.  
Assuming negligible error at the subarray reference element, the overall approximation error becomes
\begin{equation}
\label{equ_CHAN_approximation_error_HSPM}
\begin{split}
\epsilon^{il}_{\rm HSPM} &\approx 
\Bigg| 2\sin\Bigg( \frac{\pi d^2}{2 D^{k_rk_t} \lambda}
\Big[
(n_{k_{rx}} - n_{k_{tx}})^2 \\
& \quad \cdot (\sin^2 \theta^{k_rk_t}_{t1} \cos^2 \phi^{k_rk_t}_{t1} + \cos^2 \theta^{k_rk_t}_{t1}) \\
& \quad + (n_{k_{rz}} - n_{k_{tz}})^2 \cos^2 \phi^{k_rk_t}_{t1}
\Big] \Bigg) \Bigg|,
\end{split}
\end{equation}
where the index differences are now within each subarray.

\begin{figure}[t]
    \centering
    {\includegraphics[width= 0.45\textwidth]{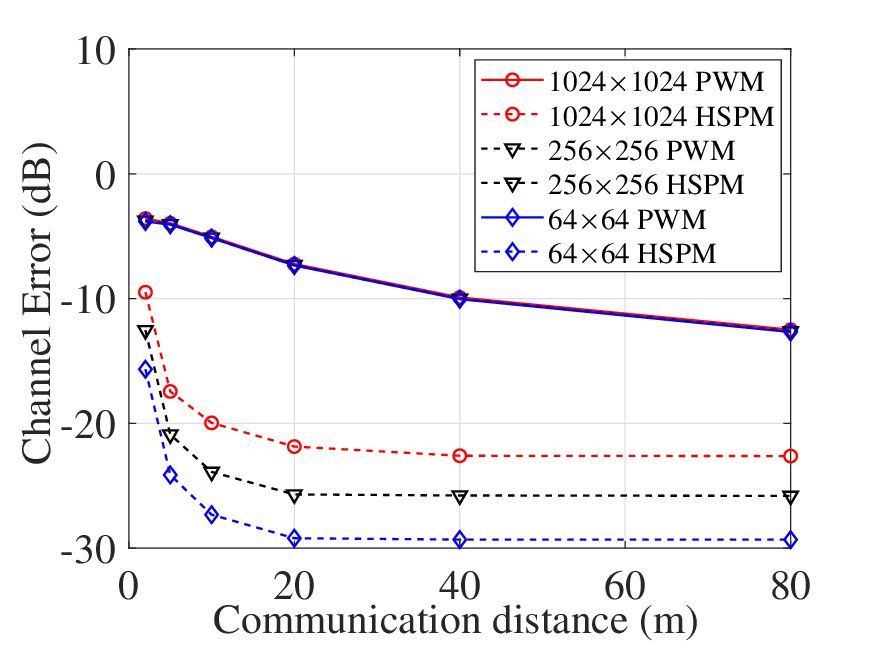}}
    \caption{Errors of PWM and HSPM versus communication distance. }
    \label{fig_channelerror_distance}
\end{figure} 

The approximation error $\epsilon^{il}_{\rm HSPM}$ increases with the subarray size and decreases with communication distance and wavelength, similar to the PWM case.  
Let the subarray dimensions be defined as
\[
\mathcal{L}_{ts} = \sqrt{N_{K_{tx}}^2 + N_{K_{tz}}^2}, \quad 
\mathcal{L}_{rs} = \sqrt{N_{K_{rx}}^2 + N_{K_{rz}}^2},
\]
and the physical apertures be $d\mathcal{L}_{ts}$ and $d\mathcal{L}_{rs}$.  
Then, $\frac{\pi d^2 \mathcal{L}_{ts} \mathcal{L}_{rs}}{\lambda D^{11}}$ defines the far-field condition for subarrays.

Returning to the earlier example with $f = 0.3$~THz and $D^{11} = 10$~m, if the entire array is divided into 16 subarrays, we obtain $\frac{\pi d^2 \mathcal{L}_{ts} \mathcal{L}_{rs}}{\lambda D^{11}} \approx 6.1 \times 10^{-3} \ll 1$.  
In this case, the approximation error is negligible, and the HSPM provides an accurate model.  
Therefore, we conclude that the HSPM is applicable when the communication distance is larger than the Rayleigh distance of each subarray.

As shown in Fig.~\ref{fig_channelerror_distance}, we evaluate the errors of the PWM and HSPM by computing $\Vert\mathbf{H}_{\rm{P}} - \mathbf{H}_{\rm{S}}\Vert_F^2 / \Vert\mathbf{H}_{\rm{S}}\Vert_F^2$ and $\Vert\mathbf{H}_{\rm{HSPM}} - \mathbf{H}_{\rm{S}}\Vert_F^2 / \Vert\mathbf{H}_{\rm{S}}\Vert_F^2$, respectively.
The evaluation is conducted under a LoS scenario with a carrier frequency of $0.3$~THz. The antenna spacings are set as $d_{tx}^{k_t} = d_{tz}^{k_t} = d_{rx}^{k_r} = d_{rz}^{k_r} = 32\lambda$.
As observed in the figure, at a communication distance of 40~m, the HSPM achieves a 12~dB lower approximation error compared to the PWM.  
When $N_t = N_r = 1024$ at 40~m, the PWM becomes inaccurate, with an approximation error of $-10$~dB. In this case, $\frac{\pi d^2 \mathcal{L}_t \mathcal{L}_r}{\lambda D^{11}}$ is 0.707, which is comparable to 1. Here, $\mathcal{L}_t$ and $\mathcal{L}_r$ are defined as in~\eqref{equ_ltlr}.  
Furthermore, the errors of both PWM and HSPM decrease as the communication distance increases.  
When $N_t = N_r = 1024$, the PWM error decreases by approximately 6~dB, and the HSPM error decreases by approximately 14~dB, as the communication distance increases from 5~m to 80~m.
Overall, the HSPM achieves lower errors than the PWM at all evaluated communication distances.

\subsection{Far-Field, Near-Field, and Cross-Field Beams}
\label{Sec_Beampattern}
The beam pattern generated by a MIMO system can be expressed as the inner product between the array response vector $\mathbf{a} \in \mathbb{C}^{N_t}$ and the beamforming weight vector $\mathbf{w} \in \mathbb{C}^{N_t}$, given by
\begin{equation}
    \label{equ_B}
    B = \left\vert  \mathbf{w}^{\rm H}\mathbf{a} \right\vert^2.
\end{equation}
The array response vector $\mathbf{a}$ varies with the underlying channel model, i.e., FF, NF, or CF, while the beamforming weight vector $\mathbf{w}$ is typically designed based on the desired steering direction or beam pattern.
To achieve beam alignment, the beamforming weights are commonly constructed to match the array response vector in the intended direction, thereby maximizing the signal strength.
It is also noted that, in the NF, the beamforming weights can be designed to produce non-Gaussian wavefronts, such as non-diffractive Bessel beams and curved Airy beams~\cite{ref_Airy}.
However, BM for these types of beams remains underexplored in the literature.
In the following, we focus on beam patterns specifically designed for beam alignment, which are widely studied in existing works.

\begin{figure*}[t]
    \centering
    {\includegraphics[width= 0.9\textwidth]{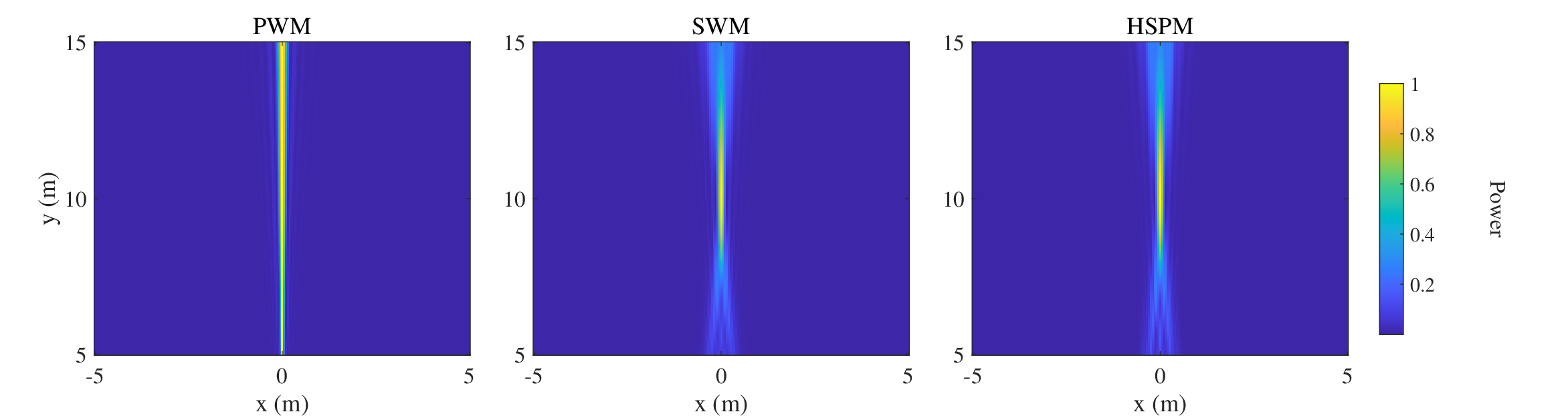}}
    \caption{Beamforming gain of the PWM, SWM, and HSPM-based beams.} 
    \label{fig_PWM_SWM_HSPMbeams}
\end{figure*} 

\subsubsection{SWM-Based Beam}
According to the SWM in~\eqref{equ_CHAN_SW_channel}, the array response vector for a NF channel is expressed as
\begin{equation}
\mathbf{a}_{\rm SWM}(\theta,\phi,D) = \left[1, \dots,  \mathrm{e}^{-j\frac{2\pi d}{\lambda} \Delta \Phi_i}, \dots, \mathrm{e}^{-j\frac{2\pi d}{\lambda} \Delta \Phi_{N_t}} \right]^{\mathrm{T}},
\end{equation}
where the phase shift $\Delta \Phi_i$ is given by
\begin{equation}
\label{equ_phase_difference_initial}
    \Delta \Phi_i \!=\! \frac{2\pi D}{\lambda} \left( \sqrt{1 + \frac{2(\Gamma_{ix} \varpi_x + \Gamma_{iz} \varpi_z)}{D} + \frac{\Gamma_{ix}^2 + \Gamma_{iz}^2}{D^2}} - 1 \right)\!\!,
\end{equation}
with $\varpi_x = \sin\theta \cos\phi$ and $\varpi_z = \sin\phi$. Here, $\Gamma_{ix}$ and $\Gamma_{iz}$ denote the position offsets of the $i^{\rm th}$ antenna element along the $x$- and $z$-axes, respectively.

To steer the beam toward a specific NF direction $(\theta_0, \phi_0, D_0)$, the beamforming vector is constructed as
\begin{equation}
\begin{split}
&\mathbf{w}_{\rm SWM}(\theta_0, \phi_0, D_0) =
\\
&\frac{1}{\sqrt{N}} \left[1, \dots, \mathrm{e}^{-j\frac{2\pi d}{\lambda} \Delta \Phi_{0,i}}, \dots, \mathrm{e}^{-j\frac{2\pi d}{\lambda} \Delta \Phi_{0,N_t}} \right]^{\mathrm{T}},
\end{split}
\end{equation}
where $\Delta \Phi_{0,i}$ is calculated using~\eqref{equ_phase_difference_initial} with the parameters $(\theta_0, \phi_0, D_0)$.  
The resulting NF beam pattern is
\begin{equation}
\label{equ_NF_beam}
B_{\rm NF}(\theta, \phi, D) = \left\vert \mathbf{w}_{\rm SWM}^{\rm H}(\theta_0, \phi_0, D_0) \, \mathbf{a}_{\rm SWM}(\theta, \phi, D) \right\vert^2,
\end{equation}
which depends on both the angular and distance parameters.

\subsubsection{PWM-Based Beam}
The array response vector for the PWM channel follows~\eqref{equ_CHAN_array_steering_vector}.  
To steer the beam toward direction $(\theta_0, \phi_0)$, the beamforming weight is constructed as
\begin{equation}
\mathbf{w}_{\rm PWM}(\theta_0,\phi_0) = \frac{1}{\sqrt{N}} \left[1, \dots, \mathrm{e}^{-j\frac{2\pi d}{\lambda} \Psi_i}, \dots, \mathrm{e}^{-j\frac{2\pi d}{\lambda} \Psi_{N_t}} \right]^{\mathrm{T}},
\end{equation}
where $\Psi_i$ represents the phase steering term corresponding to the desired angles. Specifically, 
\[
\Psi = (M_{K_{tx}} + N^a_{K_{tx}} - 2) \sin\theta_0 \cos\phi_0 - (M_{K_{tz}} + N^a_{K_{tz}} - 2) \sin\phi_0,
\]
and $N^a_{K_{tx}}$, $N^a_{K_{tz}}$ denote the number of antennas in the $K_t^{\rm th}$ subarray along the $x$- and $z$-axes, respectively.  
The resulting FF beam pattern is given by
\begin{equation}
\label{equ_PWM_pattern}
B_{\rm FF}(\theta, \phi) = \left\vert \mathbf{w}_{\rm PWM}^{\rm H}(\theta_0,\phi_0) \, \mathbf{a}_{\rm PWM}(\theta, \phi) \right\vert^2,
\end{equation}
where $\mathbf{a}_{\rm PWM}(\theta, \phi)$ follows the FF array response model.  
It is evident from~\eqref{equ_PWM_pattern} that the FF beam pattern is solely a function of angle.

\subsubsection{HSPM-Based Beam}
Based on the HSPM in~\eqref{equ_CHAN_HSPM}, the array response vector is constructed by combining per-subarray responses as
\begin{equation}
\mathbf{a}_{\rm HSPM}(\theta, \phi, D) = \left[\mathbf{a}^{1}, \dots, \mathbf{a}^{K_b} e^{-j \Delta \Phi_K} \right]^{\rm T},
\end{equation}
where $\mathbf{a}^k = \mathbf{a}_{\rm PWM}(\theta_k, \phi_k)$ is the FF response vector for the $k^{\rm th}$ subarray, and the angles $(\theta_k, \phi_k)$ are calculated as:
\begin{subequations}
\label{equ_angles}
\begin{align}
\theta_k &= \arccos \left( \frac{D \cos\theta \cos\phi}{\sqrt{(D \cos\phi)^2 + \Gamma_{kx}^2 + 2D \Gamma_{kx} \sin\theta \cos\phi}} \right), \\
\phi_k &= \arccos \left( \frac{D \cos\theta \cos\phi}{\sqrt{(D \cos\theta)^2 + \Gamma_{kz}^2 + 2D \Gamma_{kz} \cos\theta \sin\phi}} \right),
\end{align}
\end{subequations}
where $\Gamma_{kx}$ and $\Gamma_{kz}$ denote the position offsets of the $k^{\rm th}$ subarray relative to the reference point along the $x$- and $z$-axes.

The phase shift $\Delta \Phi_K$ is computed in the same form as~\eqref{equ_phase_difference_initial}, with $\Gamma_{ix}$ and $\Gamma_{iz}$ replaced by subarray-level distances $\Gamma_{kx}$ and $\Gamma_{kz}$.

To steer the beam toward $(\theta_0, \phi_0, D_0)$, the CF beamforming weight is constructed as
\begin{equation}
\mathbf{w}_{\rm HSPM}(\theta_0, \phi_0, D_0) = \frac{1}{\sqrt{N}} \left[\mathbf{a}_0^{1}, \dots, \mathbf{a}_0^{K_b} e^{-j\Delta \Phi_{K0}} \right]^{\rm T},
\end{equation}
where $\mathbf{a}_0^k$ and $\Delta \Phi_{K0}$ are computed using the target parameters $(\theta_0, \phi_0, D_0)$.  
The resulting beam pattern is given by
\begin{equation}
\label{equ_CF_beam}
B_{\rm HSPM}(\theta, \phi, D) = \left\vert \mathbf{w}_{\rm HSPM}^{\rm H}(\theta_0, \phi_0, D_0) \, \mathbf{a}_{\rm HSPM}(\theta, \phi, D) \right\vert^2,
\end{equation}
which, like the NF case, depends on both angular and distance information.

To compare the differences among the PWM- SWM- and HSPM-based beam patterns, we consider a uniform linear array with $N = 128$ antenna elements spaced at half-wavelength intervals, operating at a carrier frequency of 9~GHz. The base station (BS) is placed at the origin $(0, 0)$, and the user equipment (UE) is located at $(0, 10)$~m. In the HSPM, the array is partitioned into $4$ subarrays. The beam is steered toward the UE, and the beam pattern is evaluated over a 2D Cartesian plane ($x \in [-5, 5]$\,m, $y \in [5, 15]$\,m).

As shown in Fig.~\ref{fig_PWM_SWM_HSPMbeams}, the PWM-based beam depends solely on the angle, with the maximum gain observed along the direction of the UE. In contrast, the SWM and HSPM beam patterns incorporate both angle and distance information, resulting in the highest gain occurring precisely at the UE location. 
Notably, the HSPM achieves similar spatial focusing performance compared to the SWM and offers a favorable trade-off between modeling accuracy and complexity.

It is worth noting that the actual beam pattern depends on the beamforming architecture. In particular, in hybrid beamforming structures, the beamforming weights are subject to hardware constraints. For example, in array-of-subarrays (AoSA) systems, the analog beamforming weights typically follow a block-diagonal structure~\cite{ref_AoSA_training}, which may lead to beam patterns that differ from the ideal patterns discussed above. Hence, beamforming design must be conducted with strict consideration of hardware constraints.

\subsection{Discussions}

Most existing research efforts have focused on the PWM and SWM. The PWM has been extensively studied in earlier MIMO systems, while the SWM is regarded as the most accurate model for capturing spatial and field-specific characteristics. However, as discussed above, the SWM requires a large number of channel parameters to construct the channel matrix, resulting in significant modeling complexity~\cite{ref_3D_channel, ref_Channel_smart_radio}. On the other hand, the PWM offers low complexity but suffers from limited accuracy when applied to high-frequency MIMO channels with large-scale antenna arrays.

To address this trade-off, several low-complexity channel models have been proposed in addition to the HSPM. For instance, authors in~\cite{ref_two_level, ref_WSMS} introduce channel models based on subarrays and units, assuming that common reflectors are shared among subarrays. Unlike the HSPM, which builds subarray-wise responses from physical propagation geometry, these works use shared channel parameters—including DoD, DoA, and complex gain amplitude—augmented with phase-shift coefficients to account for spherical-wave propagation. While this approach reduces complexity, it compromises modeling accuracy compared to the HSPM.
Another line of work explores hybrid-field channel models~\cite{ref_hybrid_Channel_Estimation, ref_HuHybrid_Field_Channel_Estimation}, where scatterers in the FF and NF are modeled separately using PWM and SWM, respectively. Due to the use of PWM in the FF component, this class of models is also effective for capturing cross-field characteristics with reduced complexity.

In summary, the evolution of channel modeling for high-frequency MIMO systems has progressed from PWM to SWM, and more recently toward hybrid approaches that aim to balance accuracy and tractability.

\section{BEAM SCANNING}
\label{sec_beam_scanning}
As introduced in Sec.~\ref{sec_Introduction}, during the beam scanning process, both the Tx and Rx employ directional beams selected from predefined codebooks with high beamforming gain to scan the entire spatial domain. Each codeword corresponds to a specific realization of the precoder and combiner.
To enable effective beam scanning, both the codebook and the scanning strategy must be carefully designed. Since the beam pattern is determined by the production of the codeword and the array response vector in the channel, codebook design is inherently dependent on the underlying channel model and the adopted beamforming structure.

Beam scanning methods can be broadly categorized into three types, the searching-based methods, methods leveraging ML tools, and approaches that exploit side information.  
In this section, we first provide a comprehensive review of these beam scanning methods in the context of FF, NF, and CF communications. Then, we discuss potential research directions for future beam scanning design.

\subsection{Searching-Based Scanning}

Searching-based beam scanning can be categorized into exhaustive search, hierarchical search, and concurrent search methods.
{Exhaustive search} involves scanning all possible beam combinations, ensuring comprehensive coverage at the cost of high training overhead. {Hierarchical search} uses a multi-layer codebook to progressively narrow the beam width, significantly reducing overhead while maintaining accuracy. Beam scanning techniques have evolved from exhaustive to more efficient hierarchical methods over time. However, in both exhaustive and hierarchical methods, only one beam is generated per time slot, leading to low search efficiency. {Concurrent search} addresses this limitation by generating multiple beams simultaneously, reducing the overall scanning overhead.

\subsubsection{Exhaustive Search}

Before link establishment, the exhaustive search explores all possible beam combinations and selects the beam pair with the highest received power.
To design the exhaustive search codebook, the channel model and optimization-based methods are studied in the literature. 

Particularly, exhaustive search codebooks can be directly \emph{designed based on the channel model}. In this case, beamforming weights $\mathbf{w}$ in~\eqref{equ_B} follows the structure of the array response vector, utilizing channel phase properties to maximize gain at the desired position.
In FF PWM, where the phase depends only on angles (see~\eqref{equ_PWM_pattern}), codebooks are formed by combining beams from different directions. For example, IEEE 802.15.3c~\cite{IEEE_Std_802_15_3c} uses exhaustive search in its beamforming protocol. Similarly, authors in~\cite{ref_QUPA} propose an angle-domain exhaustive codebook for 3D coverage in a quadric quadruple-planar-array (QUPA) architecture.

In NF SWM and CF HSPM, phases depend on both angle and distance~\cite{ref_Steering_focusing}. Thus, codebooks must span both domains~\cite{ref_codebookdesign_Dai}. Uniformly dividing the spatial and distance domains is one approach. 
However, due to non-uniform phase variation in the distance domain~\cite{ref_NearorFar}, this leads to inefficient, mismatched beams. 
To address this, a polar-domain codebook~\cite{ref_NearorFar} is proposed, using second-order Taylor expansions to match the distance-phase relationship. A similar non-uniform design is proposed for the uniform-circular array (UCA) in~\cite{ref_near_training_UCA}.

However, hardware constraints in hybrid beamforming (e.g., limited-resolution phase shifters~\cite{ref_hybrid_beamforming}) may prevent the direct use of analytically derived beams. Instead, \emph{optimization-based codebook design} is used. An ideal beam is defined over a target angular (or angular-distance) region with high flat gain and zero elsewhere. The actual codeword is optimized to approximate this ideal pattern within hardware constraints. Examples include restricting ripple levels~\cite{ref_zhang2017codebook} and minimizing beam pattern mean-squared error (MSE)~\cite{ref_Song2017Codebook}.

Although exhaustive search guarantees optimal beam alignment, its application in massive MIMO with hybrid beamforming is limited by high training overhead, due to the narrow beam widths and limited number of simultaneously generated beams.

\subsubsection{Hierarchical Search}

Hierarchical search has emerged as an efficient alternative to exhaustive search. It reduces overhead by iteratively refining beam widths.
Based on this principle, the hierarchical search beams have varying coverage across different search layers, requiring careful design.

A widely used approach for designing hierarchical beams is tree-based division~\cite{ref_xiao2016hierarchical}. Particularly, a tree-based codebook is constructed by partitioning the spatial domain in a way that resembles a tree structure. Initially, a coarse set of wide beams is used to cover the entire domain. In subsequent layers, each beam is further split into smaller sub-regions, allowing for progressively finer division of the beam direction. 
For FF communications, tree-based hierarchical codebooks typically divide the angular domain, with each layer refining the beam directions based on prior scanning results~\cite{ref_xiao2016hierarchical}.
In NF and CF communications, additional distance-based partitions are required since the phase of the array response vector depends on both angle and distance~\cite{ref_NF_Hie_Training,ref_Codebook_NearorFar}.

Similar to the exhaustive search codebook design, hierarchical codebooks can also be designed using channel models or optimization. 
One common approach to designing hierarchical search codebooks using the channel structure is to divide the entire antenna array into multiple subarrays. Since beams generated by a subarray have a wider width than those generated by the entire array, this property can be leveraged for efficient scanning. Based on this principle, authors in~\cite{ref_QUPA} propose hierarchical codebooks for QUPA to achieve three-dimensional coverage in FF communications.
Coordination among subarrays can further improve beam coverage. For example, an AoSA structure with time-delay phase shifters is proposed in~\cite{ref_AoSA_training} to mitigate wideband beam squint. The use of AoSA introduces a block-diagonal structure in the analog beamformer, necessitating careful coordination among subarrays during codebook design to achieve the desired beam coverage. 
A similar approach is explored in~\cite{ref_xiao2016hierarchical} by incorporating subarray partitioning and selective antenna deactivation into the hierarchical codebook design. 

Similar principles are also applied in NF communications.
However, it is worth noticing that when only part of the antenna array is activated, the generated beams even in NF communications provide only angular resolution, since the communication distance is usually larger than the Rayleigh distance for the activated subarray. 
For instance, authors in~\cite{ref_Near_TWO_Training} propose a two-stage hierarchical NF beam scanning method. In the first stage, only a subset of the antenna array is activated, ensuring that all distances remain in the far-field of the array, allowing traditional FF scanning methods to determine angular information. In the second stage, the full array is used for NF scanning to refine distance estimation, thereby reducing the overall scanning overhead.
The successive deployment of FF and NF beams for NF scanning has also been explored. Since FF beams can effectively acquire angular information, NF beams can then be deployed within a fixed angular range to further reduce scanning overhead. In~\cite{ref_fast_near_training}, a two-phase scanning method is proposed, where FF beams first estimate the angles of users, followed by NF beams refining the distance information. In~\cite{ref_NF_Training_Sparse_DFT}, a three-phase NF scanning method is introduced. First, a sparse DFT codebook is employed for angular scanning based on angular periodicity to identify candidate angles.
Then, a subarray is activated for further angular domain scanning to resolve ambiguities caused by angular periodicity.
Finally, an NF polar-domain codebook is applied at the estimated angle to determine the user’s range.
Similarly, authors in~\cite{ref_NF_Training_2learning} design a hierarchical NF codebook that first applies FF scanning followed by NF scanning, mitigating the power leakage effect of FF beams in NF scenarios.

Apart from designing the codebook based on subarrays, the additional distance-dependent phase term in NF and CF channels provides an opportunity to enhance hierarchical codebook design and improve scanning efficiency. By leveraging this property, scanning strategies can be optimized to achieve better coverage while reducing training overhead.
For example, authors in~\cite{ref_eltraining_near} project the channel into a slope-intercept domain, where each spatial-chirp beam functions as a narrow beam at a specific distance ring while simultaneously covering a broader area at other distances. This dynamic beam coverage significantly reduces training overhead compared to the exhaustive search. To further refine the scanning performance, the final codebook is optimized using manifold optimization and alternative minimization, ensuring that the realized beam pattern closely approximates the ideal design.

For CF communications, hierarchical scanning has also been explored to efficiently handle the joint angle and distance domain variations. In \cite{ref_efficient_Hybrid}, a two-stage scanning strategy is introduced, where only a subset of the antennas is activated to conduct far-field beam scanning, which helps obtain a coarse estimate of the angular range. Following this, a refined hierarchical scanning process is performed, jointly optimizing the angle and distance domains for precise beam alignment. Building on this, authors in~\cite{ref_triple_training} propose a three-stage CF beam training method, where each stage’s codebook is systematically designed based on the CF beamforming gain with angle and distance-dependent terms derived from the channel structure. 
The codebook is designed to ensure that the beam coverage of each codeword does not overlap with others, preventing redundant scanning, while the entire potential region is fully covered by the union of all beam codewords, maximizing scanning efficiency.

Hierarchical search codebooks can also be designed using optimization algorithms. With ideal codebooks determined, algorithms that fix the beam range and optimize the beam codebook, such as those proposed in~\cite{ref_zhang2017codebook} and~\cite{ref_Song2017Codebook}, can be employed to enhance beam selection efficiency. In NF communications, authors in~\cite{ref_NF_Hie_Training} utilize a Gerchberg-Saxton-based algorithm to derive a theoretical codeword under a fully digital architecture. This is followed by an alternating optimization algorithm to transform the theoretical codeword into a practical hybrid beamforming structure, ensuring compatibility with hardware constraints while maintaining scanning efficiency.

While hierarchical search significantly reduces scanning overhead compared to exhaustive search, it relies on continuous feedback between the Rx and Tx, introducing additional complexity. Developing effective and scalable scanning methods remains an open challenge, particularly in the context of large-scale MIMO systems and hybrid beamforming architectures.

\subsubsection{Concurrent Search}

In both exhaustive and hierarchical search methods, only one beam can be generated at a time, leading to low scanning efficiency. To address this limitation, concurrent search methods have been proposed, enabling the simultaneous generation of multiple beams in different directions to accelerate the scanning process.

One approach to concurrent search is to exploit the structural characteristics of the MIMO system. In wideband beamforming, true time-delay (TTD) devices leverage frequency-dependent phase variations to steer beams toward distinct spatial locations. 
By carefully designing the time delays of the TTD devices, beams at different frequencies can be flexibly directed to specific locations, enabling simultaneous scanning of multiple spatial positions in both FF~\cite{ref_wideband_tracking_dai} and NF~\cite{ref_NF_rainbow}, thereby reducing scanning overhead. In addition, a subarray-based approach is proposed in~\cite{ref_simu_training_Zhang} where each subarray functions as an independent unit, generating multiple beams simultaneously to serve different user groups. This method enhances concurrent search efficiency by leveraging subarray-specific beamforming characteristics.

Beyond structural-based approaches, codebook design can also be utilized for concurrent search by generating codewords with multiple main lobes. In~\cite{ref_hie_Qi}, a simultaneous multi-user hierarchical beam training method is introduced for FF communications. A codeword design algorithm is first used to generate beam weights based on an ideal codeword, followed by an adaptive hierarchical codebook, where codewords at each layer are determined dynamically based on previous beam training results. This approach enables the simultaneous scanning of multiple UEs, effectively reducing overhead. Similarly, authors in~\cite{ref_wang2024fast} propose a CF beam scanning strategy using a multi-directional beam sequence, where multiple directional NF beams are generated and steered toward different directions. This approach efficiently covers the entire angular space, achieving fast scanning with minimal overhead.
In addition, in~\cite{ref_coded_training}, a channel coding theory inspired concurrent search method is proposed with both codebook design and scanning strategy. 
Particularly, the duality between hierarchical search and channel coding is first established. Based on that, the ideal codewords for scanning are designed based on channel coding. Then, the direction for alignment can be determined based on the proposed channel coding beams with low overhead and high accuracy. 

The concurrent search method is useful in generating multiple beam at one time to reduce the searching overhead. However, the complexity of the codebook design, especially when the codeword needs to be adjusted frequently based on the result of the previous search, is too high. Moreover, since multiple beams are generated, the beamforming gain is reduced thus may be harmful to the scanning performance.

\subsection{ML-Based Scanning}

Recently, a variety of ML-inspired beam scanning methods have been proposed, offering superior performance compared to traditional search-based scanning. ML techniques can assist both codebook design and the beam scanning process, addressing key limitations in conventional approaches.

In codebook design, ML is typically used to tackle challenges such as hardware impairments and interference, which are difficult to mitigate with traditional codebooks. For example, deep reinforcement learning (DRL) is employed in~\cite{ref_Reinforcement_Learning_codebook} to design beam scanning codebooks for FF mmWave and THz MIMO systems. This approach enables the system to adapt beam patterns dynamically based on environmental conditions, user distribution, hardware impairments, and array geometry. The learned codebook significantly improves SNR performance compared to exhaustive search-based methods.
Similarly, authors in~\cite{ref_neural_codebook} introduce an end-to-end ML-based codebook design algorithm, optimizing beam patterns to mitigate both inter- and intra-cell interference while maximizing the achievable rate without limitation to the communication scenario (FF, NF or CF). 
This approach enables more efficient spectrum utilization and enhances overall system performance.

For beam scanning, ML models can be deployed to optimize the selection of scanning beams, reducing search overhead while maintaining high accuracy. In~\cite{ref_DRL_sweeping}, a DRL-based method is proposed to intelligently select scanning beams in pre-fixed codebook. The network is trained to learn environmental and user distribution information and makes reward-driven decisions about the optimal scanning beams, significantly lowering the scanning overhead.
Beyond simple selection, ML models can also enhance scanning adaptivity across different scenarios. In~\cite{ref_meta_alignment}, a meta-learning-based framework is introduced for adaptive beam alignment in FF communications. During scanning, the system successively generates probing and refining beams based on UE-reported beam gain information. This data is then processed by a trained network to determine the best beam direction. The fine-tuned meta-learning framework enables the system to adapt to diverse deployment environments, overcoming the limited generalization ability of conventional beam alignment techniques.
For NF communications, authors in~\cite{ref_Near_DL_training} propose a deep-learning (DL)-based NF beam training method that reduces scanning complexity while maintaining high accuracy. Instead of performing an exhaustive NF beam search, the first stage selects a subset of codewords for initial scanning. Based on the results, a second-stage refinement is conducted using additional codewords to improve accuracy. A trained network processes the received pilot signals and determines the optimal angle and distance directions for beam alignment, ensuring efficient NF beam scanning.

ML-driven beam scanning approaches significantly enhance scanning efficiency and enable more adaptive, scalable, and efficient BM in next-generation MIMO systems. However, since these methods are often designed to address specific challenges, comprehensive performance comparisons across different approaches remain difficult. Additionally, computational complexity is a critical factor for practical deployment, as the overhead of training and inference in ML-based models must be carefully balanced against the available hardware resources and real-time processing constraints in future communication systems. Developing standardized evaluation frameworks and optimizing ML architectures for low-complexity deployment will be key to fully leveraging ML-driven beam scanning in practical scenarios.

\begin{table*}[t]
\centering
\caption{Comparison of Beam Scanning Methods}
\renewcommand{\arraystretch}{1.2}
\begin{tabular}{|p{2.3cm}|p{1.7cm}|p{1.7cm}|p{1.3cm}|p{1.3cm}|p{1.3cm}|p{1.3cm}|p{1.8cm}|}
\hline
\textbf{Method Type} & \multicolumn{2}{c|}{\textbf{Example Studies}} & \textbf{Scanning Overhead} & \textbf{Beams per Time Slot} & \textbf{Codebook Complexity} & \textbf{Alignment Accuracy} & \textbf{Applicability to All Beamforming Structures} \\
\cline{2-3}
 & \textbf{FF} & \textbf{NF/CF} & & & & & \\
\hline
\textbf{Exhaustive search with channel structure} & 
IEEE Std. 802.15 \cite{IEEE_Std_802_15_3c} \newline B. Ning~\cite{ref_QUPA} &
X. Wei~\cite{ref_codebookdesign_Dai}\newline M. Cui~\cite{ref_NearorFar}\newline Y. Xie~\cite{ref_near_training_UCA} &
High & One & Low & High & No \\
\hline 

\textbf{Exhaustive search with optimization} & 
J. Zhang~\cite{ref_zhang2017codebook}\newline J. Song~\cite{ ref_Song2017Codebook} &
J. Song~\cite{ref_Song2017Codebook} &
High & One & High & High & Yes \\
\hline

\textbf{Hierarchical search with channel structure} & 
C. Lin~\cite{ref_AoSA_training}\newline B. Ning~\cite{ref_QUPA} \newline Z.Xiao~\cite{ref_xiao2016hierarchical} &
C. Wu~\cite{ref_Near_TWO_Training} \newline C. Zhou~\cite{ref_NF_Training_Sparse_DFT} \newline C. Weng~\cite{ ref_NF_Training_2learning} \newline X. Shi~\cite{ref_eltraining_near} \newline J. Luo~\cite{ref_efficient_Hybrid} \newline K. Chen~\cite{ref_triple_training} &
Medium & One & Medium & Medium & No \\
\hline

\textbf{Hierarchical search with optimization} & 
J. Zhang~\cite{ref_zhang2017codebook} \newline J. Song~\cite{ ref_Song2017Codebook} &
Y. Lu~\cite{ref_NF_Hie_Training} \newline K. Chen~\cite{ref_triple_training} &
Medium & One & High & Medium & Yes \\
\hline

\textbf{Concurrent search with channel structure} & 
J. Tan~\cite{ref_wideband_tracking_dai} \newline R. Zhang~\cite{ref_simu_training_Zhang} &
M. Cui~\cite{ref_NF_rainbow} \newline R. Zhang~\cite{ref_simu_training_Zhang} &
Low & Multiple & Low & Medium & No \\
\hline

\textbf{Concurrent search with optimization} & 
C. Qi~\cite{ref_hie_Qi} &
H. Wang~\cite{ref_wang2024fast} \newline T. Zheng~\cite{ref_coded_training} &
Low & Multiple & High & Medium & Yes \\
\hline

\textbf{ML-based: Codebook Design} & 
Y. Zhang \cite{ref_Reinforcement_Learning_codebook} \newline R. M. Dreifuerst~\cite{ref_neural_codebook} &
R. M. Dreifuerst~\cite{ref_neural_codebook} &
Low--Medium & One & Medium--High & High & Yes \\
\hline

\textbf{ML-based: Scanning Method} & 
J. Che~\cite{ref_DRL_sweeping} \newline Z. Xu~\cite{ref_meta_alignment} &
G. Jiang~\cite{ref_Near_DL_training} &
Low--Medium & Multiple & Medium & High & Yes \\
\hline

\textbf{Side info-assisted (sensor/anchor)} & 
\multicolumn{1}{c|}{--} &
L. Liu~\cite{ref_NF_Side_Training} &
-- & -- & -- & Medium & Yes \\
\hline

\textbf{Side info-assisted (vision)} & 
\multicolumn{1}{c|}{--} &
Y. Ahn~\cite{ref_CV_alignment} &
-- & -- & -- & Medium & Yes \\
\hline
\end{tabular}
\label{tab_beam_scanning}
\end{table*}

\subsection{Side Information-Assisted Scanning}
Side information can also be leveraged to assist beam scanning in FF, NF, and CF communications, providing additional contextual data to enhance user localization and reduce scanning overhead. The side information refers to data not directly obtained from the communication frequency band but still useful for beam alignment. This may include signals from other frequency bands or even image-based information that aids in identifying the location of UEs.
For instance, in~\cite{ref_NF_Side_Training}, side-angle information acquired from anchor nodes or surrounding sensors is used to estimate the user’s range. By leveraging this information, beam scanning can be conducted over a narrowed search space, significantly reducing scanning overhead. Additionally, authors in~\cite{ref_CV_alignment} introduce a computer vision-aided beam alignment approach, where a camera attached to the BS captures image data. A deep-learning-based detector then processes the images to determine the mobile device’s position, facilitating more efficient beam alignment.

These methods demonstrate the potential of multi-modal data fusion in beam scanning, integrating side information to enhance scanning efficiency and localization accuracy. Furthermore, such information can be utilized beyond FF, NF, and CF communication scenarios, offering broader applications in BM.
However, the availability of side information is not always guaranteed, as factors such as sensor placement, environmental conditions, and hardware limitations may impact data reliability. Moreover, effectively integrating this auxiliary information into the communication process while maintaining robustness and minimizing additional complexity remains an open challenge. 

\subsection{Potential Research Directions}

Although FF, NF, and CF channels exhibit distinct propagation characteristics, beam scanning methods across these regimes generally fall into three categories: searching-based, ML-assisted, and side information-assisted techniques, as previously introduced. A comparison of these methods is summarized in Table~\ref{tab_beam_scanning}. These techniques have mostly been studied in isolation, with limited extensions from FF to NF domains.
A key difference between FF and NF/CF scanning lies in the incorporation of distance-domain beam resolution. Accordingly, both the codebook design criteria and scanning strategy must be revised to consider beam division and scanning order in the distance domain.

In addition to the existing research on beam scanning, several potential directions remain open for exploration:
\begin{itemize}
    \item From a physical propagation perspective, the boundaries between FF, NF, and CF channels are continuous rather than discrete. The gradual transition of channel characteristics with respect to communication distance suggests that the strict separation of beam scanning methods across these regimes may not be necessary. Because beam patterns are inherently coupled with the underlying channel model, variations in channel conditions can influence both the codebook design and scanning procedure. Nonetheless, practical communication systems typically adopt a fixed codebook and scanning strategy due to standardization constraints, limiting the possibility of real-time adaptation to changing propagation conditions.
    These observations motivate the development of a unified beam scanning framework that remains robust across different channel conditions without explicitly relying on FF or NF classifications. Such an approach would enhance compatibility with diverse deployment scenarios while avoiding excessive complexity.
    
    \item In addition, reducing scanning overhead remains a key challenge in multi-user systems for FF, NF and CF communication. Most existing beam scanning techniques are user-specific. During the beam scanning process, separate orthogonal resources are demanded for each user. Consequently, the overall scanning overhead increases linearly with the number of users, posing a scalability issue. To address this, future research should explore user-agnostic beam scanning strategies that allow efficient multi-user alignment without proportional increases in resource consumption.
\end{itemize}

\section{CSI ESTIMATION}
\label{sec_Csi_estimation}

Based on the received pilot signal of beam scanning, 
the CSI estimation process further processes these signals to obtain the CSI information to assist beamforming design. 
With different aims of beamforming, the CSI estimation can be categorized into beam estimation and channel estimation methods, where the prior aims to obtain the beam information for alignment, while the latter one focuses on obtaining the full channel information. In the following, we separately introduce these techniques. 

\subsection{Beam Estimation}

Beam estimation can be categorized into two main approaches: power-based methods and estimation algorithm-based methods.
In power-based methods, beam directions are selected directly from the predefined codebooks based on the highest received power. As a result, for FF, NF, and CF communications, power-based beam estimation is commonly applied following exhaustive search, hierarchical search, concurrent search, and other beam-scanning techniques, utilizing well-designed codebooks as introduced in Sec.~\ref{sec_beam_scanning}.
However, as the beam direction in the codebook is usually fixed, these methods usually suffer from limited alignment accuracy. 

On the other hand, estimation algorithm-based methods involve additional processing of received pilot signals after beam scanning to determine the beam direction with higher accuracy.
In this context, beyond exhaustive and hierarchical search-based scanning, tailored scanning methods can be designed to enhance beam estimation.

For FF communications, authors in~\cite{ref_fast_training} introduce a downlink beam-scanning technique for THz-integrated MIMO and intelligent reflective surface (IRS) systems, where each node simultaneously generates multiple directional beams to explore its angular space. The scanning results are then processed using a compressive sensing-based estimation algorithm to identify the optimal beam alignment. Additionally, high-precision beam estimation algorithms are proposed in~\cite{ref_Root_MUSIC_HDAPA} by leveraging hybrid beamforming structures. These algorithms first adjust phase shifter weights for beam scanning and subsequently apply the multiple signal classification (MUSIC) algorithm to estimate the beam direction accurately. Moreover, a two-stage beam-scanning approach is presented in~\cite{ref_milli} for hybrid beamforming in THz frequencies. This method iteratively refines the SNR during scanning, and the collected pilot signals are further processed using a MUSIC-based algorithm for high-precision beam estimation.

For NF and CF communications, estimation algorithms have been shown to effectively extract beam information across both angular and distance domains by utilizing FF scanning methods. Specifically, authors in~\cite{ref_cross_training, ref_DL_NFTraining, ref_MU_NF_BT, ref_NF_Training_DFT} employ FF beam scanning followed by dedicated beam estimation techniques to retrieve beam information. The rationale, as demonstrated in~\cite{ref_cross_training}, is that the SNR in NF scenarios is typically higher than in FF due to shorter communication distances, ensuring sufficient SNR for pilot signal reception when using FF beams. Additionally, advanced beam estimation algorithms further refine beam alignment accuracy.
Moreover, in~\cite{ref_NF_Training_DFT}, it is shown that the received beam pattern at the UE, when using FF beams, contains both angular and range information. Specifically, authors in~\cite{ref_DL_NFTraining, ref_MU_NF_BT} employ machine learning techniques to determine NF beam directions, while authors in~\cite{ref_cross_training} introduce cross-field MUSIC and maximum likelihood estimation (MLE) for beam direction estimation. Furthermore, an on-grid angular support-based method is proposed in~\cite{ref_NF_Training_DFT} to extract angular information, followed by a minimum mean-squared error (MMSE)-based approach to estimate user distance using the approximated angular support width.
ML techniques have also proven effective in determining beam direction. 
For example, in~\cite{ref_RIS_NF_lowOverhead_Training}, deep residual networks are utilized to identify the optimal NF IRS codeword. Specifically, the neural network processes received signals from either all FF IRS codewords or a subset of NF IRS codewords to estimate the optimal NF codeword, thereby achieving low pilot overhead while maintaining accurate beam alignment.
 
In conclusion, the estimation algorithm-based methods significantly improve beam alignment accuracy, enabling high-precision alignment across various communication scenarios. However, compared to the power-based method, these techniques suffer from increased complexity. 

\subsection{Channel Estimation}

The channel estimation studies can be broadly classified into three main categories: traditional \textit{on-grid} and \textit{off-grid} algorithms, and ML-powered solutions.
The on-grid channel estimation solutions involve determining the directions of propagation paths from predefined fixed spatial grids, while the off-grid solutions eliminate the assumption of fixed grids.
Additionally, ML-powered approaches utilize neural networks and machine learning techniques to learn and estimate the channel characteristics for channel estimation.

\subsubsection{On-Grid Channel Estimation Technologies}
On-grid channel estimation solutions primarily rely on compressive sensing (CS)-based approaches. These methods exploit the sparsity of channels in high-frequency large MIMO systems and typically involve three key steps: on-grid sparse channel representation, signal observation, and sparse recovery.
Specifically, the sparse channel representation is formulated using an on-grid codebook, where the channel matrix is expressed as the product of the codebook and a sparse matrix. The signal observation step captures the received pilot signal after beam scanning. Based on these components, various sparse recovery algorithms are applied to estimate the channel.

The design of the sparse channel representation codebook follows a similar approach to beam-scanning codebooks and varies based on the channel model. By employing the FF PWM, the sparse channel representation codebook is constructed using fixed-grid angular samples corresponding to the propagation paths. Based on this representation, well-known sparse recovery algorithms such as orthogonal matching pursuit (OMP)~\cite{ref_OMP, CS_OMP, ref_OMP_THz} and approximate message passing (AMP)~\cite{AMP} are commonly used for channel estimation.
For wideband systems, a CS-based channel estimation scheme is introduced in~\cite{ref_wideband_CE}. To mitigate the wideband beam/split effect, which is caused by the phase variation at different frequencies due to the deployment phase shifter, a beam-split-aware OMP dictionary is proposed in~\cite{ref_BSA_OMP}. Additionally, for MIMO and IRS-integrated systems, CS-based channel estimation methods have been extensively studied~\cite{ref_IRS_CE_CE, ref_trice, ref_OMP_IRS, ref_IRS_CE_CS_THz, ref_IRSCE_sparsity, ref_triple}.

Beyond FF scenarios, on-grid channel estimation techniques have been extended to NF and CF communications. However, in NF and CF, conventional FF assumptions—such as angle-domain sparsity—no longer hold. Consequently, FF CS-based channel estimation techniques relying on the structural FF channel model, experience significant performance degradation. To address this, several NF and CF channel estimation techniques are developed with customized codebooks and sparse recovery methods that exploit the distinct characteristics of NF and CF channels.

For NF communications, authors in~\cite{ref_NearorFar} introduce a channel estimation approach leveraging a polar-domain codebook, where the NF channel is estimated using on-grid sparse recovery methods. Furthermore, NF channel estimation techniques are proposed in~\cite{ref_NFCE_JSAC} for wideband MIMO mmWave systems with extremely large-scale IRSs. This work addresses the effects of extremely large-scale IRS, spatial wideband effect, and NF propagation by designing a wideband spherical-domain codebook. The codebook is constructed by minimizing the coherence between arbitrary codewords, and a CS-based channel estimation framework is then applied to recover the channel parameters based on spherical-domain sparsity.

To enable seamless NF-to-FF channel estimation, several CF CE methods have been proposed. For instance, a CS-based CE framework is presented in~\cite{ref_DSE_SSE} for THz-integrated MIMO and IRS systems, incorporating both sparse channel representation and sparse recovery algorithms. This work introduces a subarray-based codebook to sparsely represent the HSPM and address the spherical-wave effect. Based on this, two low-complexity sparse recovery algorithms—separate side estimation (SSE) and dictionary-shrinkage estimation (DSE)—are developed. SSE estimates the positions of non-zero grids separately on each side of the channel.
DSE further reduces the computational complexity of SSE by leveraging the fact that angles for different subarray pairs are spatially correlated.
Furthermore, authors in~\cite{ref_Hybrid_CE} investigate holographic MIMO systems, revealing that power diffusion (PD) effects cause channel path energy to spread across FF and NF transform domains, leading to false paths in estimation. Conventional techniques fail to detect the actual paths due to this phenomenon. To address this, a PD-aware orthogonal matching pursuit (PD-OMP) algorithm is proposed, which identifies the PD range and eliminates spurious paths, thereby improving estimation accuracy in CF scenarios.
In~\cite{ref_Cross_CE}, a CS-based CF channel estimation method for THz MIMO systems with AoSA architecture is presented.
The proposed method first determines the propagation field of the signal by comparing the received signals across the arrays, to determine the channel estimation method.  
Based on that, the field-specific channel estimation is conducted to reduce computational complexity.

Despite their advantages, on-grid channel estimation suffers from grid mismatch errors, commonly referred to as the power leakage effect~\cite{ref_Super_ChenHu} since propagation angles and distances are continuous-valued in practice. This mismatch limits the accuracy of on-grid approaches, motivating further research into more adaptive and robust channel estimation techniques.

\subsubsection{Off-Grid Channel Estimation Technologies}

Off-grid channel estimation techniques eliminate the fixed grid assumption to achieve higher estimation accuracy and can be broadly classified into gridless CS solutions and subspace-based approaches.

Gridless CS methods refine the grid resolution iteratively, overcoming the limitations of fixed-grid approaches at the cost of increased computational complexity. Several techniques have been developed in this category, such as reweight-based methods~\cite{ref_Super_ChenHu, ref_NearorFar}, expectation-maximization (EM) sparse Bayesian learning~\cite{ref_EM_Bayesian}, and space-alternating generalized EM (SAGE)\cite{EM_SAGE3}. These methods eliminate the need for predefined grid points by iteratively refining the sparse recovery process. However, requires an exponential increase in computational complexity. Additionally, their performance heavily depends on the initialization of the algorithm, where a poor initialization can lead to suboptimal estimation results. In~\cite{ref_IRS_CE_ANM}, a two-stage atomic norm minimization problem is formulated to achieve super-resolution channel parameter estimation.

Subspace-based methods estimate channel angles by leveraging eigenvalue decomposition (EVD). The MUSIC algorithm~\cite{2D_B_MUSIC} determines the angles of the channel by separating signal and noise subspaces, while estimating signal parameters via rotational invariance techniques (ESPRIT)~\cite{2D_Unitary_ESPRIT, ref_ESPRIT_OFDM} exploits the rotational invariance property of the signal subspace for precise angle estimation. After estimating the angles, the channel matrix is reconstructed based on these results. However, subspace-based methods are typically designed for specific hybrid beamforming architectures, such as AoSA or fully-connected (FC) structures, limiting their applicability to other beamforming models.

Despite their ability to improve estimation accuracy, off-grid techniques come with significantly higher computational costs. Gridless CS solutions require iterative optimization procedures, while subspace-based methods rely on eigenvalue decomposition, both of which introduce considerable processing overhead. This trade-off between accuracy and complexity must be carefully considered when selecting an appropriate channel estimation method for practical implementations.

\begin{table*}[t]
\centering
\caption{Comparison of Channel Estimation Methods}
\renewcommand{\arraystretch}{1.2}
\begin{tabular}{|p{3.5cm}|p{2.5cm}|p{2cm}|p{2cm}|p{1.5cm}|p{1.5cm}|}
\hline
\textbf{Method Type} & \multicolumn{3}{c|}{\textbf{Example Studies}} & \textbf{Estimation Accuracy} & \textbf{Estimation Complexity} \\
\cline{2-4}
 & \textbf{FF} & \textbf{NF} & \textbf{CF} & & \\
\hline
\textbf{On-grid: CS-based} & 
A. Alkhateeb~\cite{ref_OMP}\newline
R. Méndez-Rial~\cite{CS_OMP}\newline
K. Dovelos~\cite{ref_OMP_THz}\newline
D. L. Donoho~\cite{AMP}\newline
K. Dovelos~\cite{ref_wideband_CE}\newline
A. M. Elbir~\cite{ref_BSA_OMP}\newline
Z. Wan~\cite{ref_IRS_CE_CE}\newline
K. Ardah~\cite{ref_trice}\newline
P. Wang~\cite{ref_OMP_IRS}\newline
X. Ma~\cite{ref_IRS_CE_CS_THz}\newline
T. Lin~\cite{ref_IRSCE_sparsity}\newline
X. Shi~\cite{ref_triple}&
M. Cui~\cite{ref_NearorFar}\newline
S. Yang~\cite{ref_NFCE_JSAC} &
Y. Chen~\cite{ref_DSE_SSE}\newline
S. Yue~\cite{ref_Hybrid_CE}\newline
S. Tarboush~\cite{ref_Cross_CE}\newline &
Low-Medium & Low \\
\hline

\textbf{Off-grid: CS-based} & 
C. Hu~\cite{ref_Super_ChenHu}\newline
W. Shao~\cite{ref_EM_Bayesian}\newline
K. Mawatwal~\cite{EM_SAGE3} &
M. Cui~\cite{ref_NearorFar} &
\multicolumn{1}{c|}{--} &
High & High \\
\hline

\textbf{Off-grid: Subspace-based} & 
Z. Guo~\cite{2D_B_MUSIC}\newline
A. Liao~\cite{2D_Unitary_ESPRIT}\newline
W. Ma~\cite{ref_ESPRIT_OFDM} &
\multicolumn{1}{c|}{--} &
\multicolumn{1}{c|}{--} &
High & Medium--High \\
\hline

\textbf{ML-based: Model-driven} & 
H. He~\cite{LDAMP}\newline
X. Wei~\cite{ref_GM_LAMP} &
W. Yu~\cite{ref_hybrid_THz_CE} &
Y. Chen~\cite{ref_HSPM} &
High & Medium--High \\
\hline

\textbf{ML-based: Data-driven} & 
P. Dong~\cite{mmWave_CE_CNN}\newline
E. Balevi~\cite{ref_CE_GAN} &
W. Ma~\cite{ref_sparse_CE_DL} &
S. Liu~\cite{ref_IRS_CE_DnDL}\newline
A. Taha~\cite{ref_IRS_CE_active_DL} &
High & Medium--High \\
\hline

\end{tabular}
\label{tab_channel_estimation_comparison}
\end{table*}

\subsubsection{ML-Powered Channel Estimation Technologies}
ML-based channel estimation solutions can be categorized into model-driven and data-driven approaches.

The model-driven method is designed based on traditional iterative algorithms, where each layer of the neural network corresponds to an iteration of the algorithm. While this approach achieves a faster convergence rate compared to conventional iterative methods, its performance remains heavily dependent on the original iterative algorithm. In~\cite{LDAMP}, a learned-denoising-based approximate message passing (LDAMP) network is integrated into an iterative sparse signal recovery algorithm to enhance channel estimation accuracy. Similarly, a prior-aided Gaussian mixture learned-AMP (GM-LAMP) approach is proposed in~\cite{ref_GM_LAMP}, where the beamspace channel elements are modeled using a Gaussian mixture distribution. This method replaces the original shrinkage function in the learned AMP (LAMP) network with a Gaussian mixture shrinkage function, improving estimation accuracy. To account for both FF and NF channel effects, authors in~\cite{ref_hybrid_THz_CE} transform each iteration of orthogonal approximate message passing (OAMP) into a contractive mapping by replacing the nonlinear estimator with a trained convolutional neural network (CNN). In the context of CF channel estimation, a two-phase CNN-based mechanism is proposed in~\cite{ref_HSPM} for THz MIMO systems. In the first phase, a CNN is developed to estimate the channel parameters of a reference subarray. The second phase then exploits the parameter relationships between the reference and remaining subarrays, significantly improving estimation accuracy.

In contrast, data-driven methods rely on large datasets to train neural networks, making them model-independent and adaptable to various scenarios. A deep-learning-based CS channel estimation scheme is introduced in~\cite{ref_sparse_CE_DL}. In~\cite{mmWave_CE_CNN}, a CNN is employed to tackle the channel estimation problem in hybrid beamforming systems. Generative adversarial networks (GANs) have also been applied for frequency-selective channel estimation in mmWave and THz communication systems with limited pilots and low SNR, as demonstrated in~\cite{ref_CE_GAN}. In integrated MIMO and IRS systems, authors in~\cite{ref_IRS_CE_DnDL} and~\cite{ref_IRS_CE_active_DL} propose activating only a subset of IRS elements during pilot reception, leveraging DL for channel estimation of segmented channels, achieving considerable estimation accuracy.

Despite their advantages, data-driven DL-based solutions face significant computational challenges due to the extremely large channel dimensions in large antenna array systems. The need for massive computing resources and the difficulty in achieving network convergence remain major obstacles to practical deployment.

\subsection{Potential Research Directions}
Despite differences in the propagation field, channel estimation techniques can generally be categorized into three types: on-grid, off-grid, and ML-based methods. A comparison of these approaches is provided in Table~\ref{tab_channel_estimation_comparison}.
These methods are initially developed for FF communications and have only recently been extended to NF and CF communications. A key distinction in FF and NF/CF channel estimation is the incorporation of distance-domain resolution, which is essential for accurately capturing propagation characteristics for spherical-wave assumption. 
For instance, in CS-based estimation methods, FF and NF channels are typically modeled using sparse representations, while FF deploys angular domain~\cite{ref_OMP}, NF considers polar domain~\cite{ref_NearorFar}.
In addition, to achieve a unified channel estimation approach for both FF and NF scenarios, CF channel estimation methods often adopt approximated channel representations. Examples include treating each subarray as a single unit~\cite{ref_DSE_SSE}, or reconfiguring the PD to apply FF-based techniques in NF settings~\cite{ref_Hybrid_CE}.

Although extensive research has been conducted on various channel estimation techniques, several important directions remain unexplored.
\begin{itemize}
    \item In the context of channel estimation, on-grid, off-grid, and ML-inspired methods are adaptable for FF, NF and CF communications. Given these advancements, a fundamental research question arises: which channel estimation method is most suitable for FF, NF, and CF communications? A systematic comparison of these techniques is essential to assess their relative effectiveness across different propagation conditions. Such a comparison should consider not only the beam scanning approach and channel estimation method but also the resulting system performance.
    In the evaluation, while channel estimation accuracy is the predominant metric for evaluating performance, it does not directly correlate with system-level metrics such as spectral efficiency and energy efficiency. Thus, exploring alternative evaluation metrics that better reflect the impact of channel estimation on overall system performance remains an important direction for future research. One promising approach is to investigate the correlation between the estimated and actual channel, providing a more direct link between estimation accuracy and communication performance. 
    \item Both channel estimation and beam estimation are well-established approaches for obtaining CSI, yet they represent two extremes in terms of the amount of CSI acquired. Beam estimation extracts only limited information related to beam direction, making it computationally efficient but providing minimal details about the channel, such as path gain and multipath components. In contrast, channel estimation recovers the full channel, improving performance by enhancing multiplexing gain and facilitating multi-user interference cancellation in the following transmission, at the cost of significantly higher computational complexity. Despite the benefits of acquiring full CSI, a fundamental question remains: How much CSI is actually required for optimal transmission in FF, NF, and CF communications? While beam direction information may be sufficient for certain applications, others may require additional multipath information or even complete channel knowledge to achieve optimal performance. The trade-off between CSI accuracy, computational complexity, and transmission performance remains an open research challenge. Further investigation into adaptive CSI acquisition strategies is needed to dynamically balance efficiency and performance based on the system’s operational requirements and constraints.
\end{itemize}

\section{BEAMFORMING}
\label{sec_beamforming}

Based on the estimated CSI, beamforming can be employed to generate the desired beam patterns for signal transmission and reception by controlling the phase and amplitude of the electromagnetic waves.
Beam alignment and beamforming are two primary techniques for beam generation. Beam alignment directly aligns the beams between the transmitter and receiver based on the beam estimation results, ensuring the beamforming gain is optimized for signal propagation. In contrast, beamforming is designed using full CSI, which enables more capability of performing spatial multiplexing and interference cancellation. 

Typical beamforming aims at designing the digital and analog beamformers to maximize the spectral efficiency (SE), subject to structural constraints imposed by beamforming architectures. For point-to-point (P2P) system, the SE can be generally formulated as
\begin{equation}
\begin{aligned}
    SE & = \log_2\left(|\mathbf{I}+\mathbf{C}^{-1}\mathbf{W}^H\mathbf{H}\mathbf{F}
    \mathbf{F}^H\mathbf{H}^H\mathbf{W}|\right)
    \label{eq_SU_SE}
\end{aligned}
\end{equation}
where $\mathbf{W}$ and $\mathbf{F}$ are the combiner and the precoder, respectively.
Depending on the beamforming architecture, $\mathbf{W}$ and $\mathbf{F}$ must satisfy different constraints, which will be detailed in Sec.~\ref{sec_beamforming}--\ref{subsec_bf_structures}.
$\mathbf{C} = \sigma_n^2\mathbf{W}^H\mathbf{W}$ denotes the noise covariance matrix after combining where $\sigma_n$ is the noise power. For multi-user (MU) systems, the SE can be represented as
\begin{equation}
    SE = \sum_{u=1}^U \log _{2} \left(|\mathbf{I} + \mathbf{C}_u^{-1} \mathbf{W}_u^H \mathbf{H}_u \mathbf{F}_u \mathbf{F}_u^H \mathbf{H}_u^{H} \mathbf{W}_{u}\right),
\label{eq_MU_SE}
\end{equation}
where $U$ denotes the number of users, and 
\begin{equation}
    \mathbf{C}_u=  \sum_{n \neq u}^{U} \mathbf{W}_{u}^{H} \mathbf{H}_{u} \mathbf{F}_{n}\mathbf{F}_{n}^{H}\mathbf{H}_{u}^{H} \mathbf{W}_{u}+\sigma_{\mathrm{n}}^{2} \mathbf{W}_{u}^{H} \mathbf{W}_{u},
\end{equation}
denotes the covariance matrix of the user interference and the noise. 
In addition, $\mathbf{W}_{u}$ and $\mathbf{F}_{u}$ denote the combiner and the precoder for each user.

As the design of hybrid beamformers is fundamentally constrained by their underlying architectures, the beamforming design varies significantly across different structures. In the following, we first introduce fundamental MIMO architectures, followed by an in-depth discussion of beamforming techniques, including the optimization-based and ML-based beamforming, for FF, NF, and CF communications.



\subsection{Beamforming Structures}
\label{subsec_bf_structures}
Beamforming structures are generally independent of the propagation field and are designed to balance the power consumption, hardware complexity, and system rate. 
However, certain architectures are particularly well-suited for specific propagation conditions due to their design characteristics.
In the following, we will introduce representative beamforming architectures that are widely studied for FF, NF, and CF communications, including the fully-digital, FC, AoSA, dynamic array-of-subarray (DAoSA), widely-spaced multi-subarray (WSMS), TTD-based, and low-
resolution analog-to-digital converters (ADCs)-based architectures. 

The most fundamental beamforming structure is the fully digital architecture, where each antenna element is connected to a dedicated RF chain. While this structure provides maximum flexibility and beamforming capability, its hardware cost and power consumption become prohibitively high in large MIMO systems due to the large number of RF chains required. To address this challenge, hybrid beamforming architectures have been introduced, where each RF chain is connected to multiple antennas, significantly reducing hardware complexity while maintaining beamforming capabilities. In this case, we can unify the representation of the beamformers in (\ref{eq_SU_SE}) and (\ref{eq_MU_SE}) as
\begin{subequations}
\begin{align}
    \mathbf{F} &= \mathbf{F}_{\mathrm{RF}}\mathbf{F}_{\mathrm{BB}},
    \\ 
    \mathbf{W} &= \mathbf{W}_{\mathrm{RF}}\mathbf{W}_{\mathrm{BB}},
\end{align}
\end{subequations}
where $\mathbf{F}_{\mathrm{RF}}$ and $\mathbf{W}_{\mathrm{RF}}$ denote the analog beamformer and combiner, respectively, while $\mathbf{F}_{\mathrm{BB}}$ and $\mathbf{W}_{\mathrm{BB}}$ represent the digital beamformer and combiner. 

\begin{figure}
    \centerline{\includegraphics[width=2in]{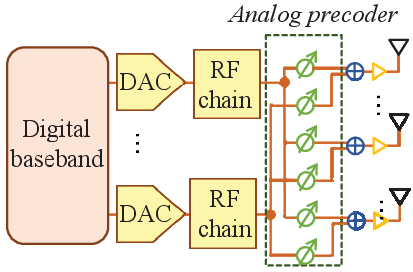}}
    \caption{FC hybrid beamforming architecture.}
   \label{fig_FC}
\end{figure}

In hybrid beamforming architectures, the FC and AoSA structures are two widely studied configurations~\cite{ref_hybrid_beamforming}. In the FC structure, as illustrated in Fig.~\ref{fig_FC}, each RF chain is connected to all antennas via phase shifters, such that the constant modulus (CM) constraint is imposed on the analog precoder and combiners, i.e.,
\begin{subequations}
\begin{align}
    |\mathbf{F}_{\rm RF}(i,j)| &= 1,\\
    |\mathbf{W}_{\rm RF}(i,j)| &= 1,
\end{align}
\end{subequations}
where $(i,j)$ denotes the element in the $i^{\rm th}$ line and $j^{\rm th}$ column of the matrix. 
This configuration ensures that the analog precoding and combining matrices are fully connected, meaning no zero elements exist. 

\begin{figure}
    \centerline{\includegraphics[width=2in]{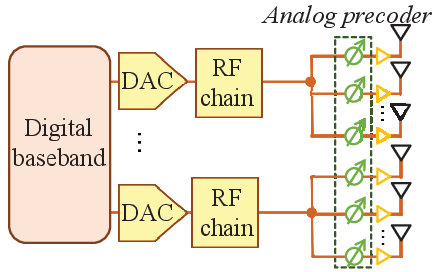}}
    \caption{AoSA hybrid beamforming architecture.}
   \label{fig_AoSA}
\end{figure}

In contrast, the AoSA structure, shown in Fig.~\ref{fig_AoSA}, partitions the antenna array into subarrays, with each RF chain connected to only one subset of antennas through phase shifters. Consequently, the analog precoding and combining matrices exhibit block-diagonal structures with many zero elements, i.e.,
\begin{subequations}
\begin{align}
    &\mathbf{F}_{\rm RF} = \operatorname{blkdiag}(\mathbf{f}_1,\dots,\mathbf{f}_K), \\&|\mathbf{f}_k(i)| = 1,
\end{align}
\end{subequations}
where $K$ is the number of subarrays and $\operatorname{blkdiag}(\cdot)$ denotes a block diagonal matrix constructed by arranging the input matrices/vectors along the main diagonal. 
Note that as digital signal processing generally provide unconstrained control over both amplitude and phase, the system constraints dominantly reside in the design of the analog beamforming. The presence of these constraints limits the flexibility of the hybrid beamforming design compared to the fully digital structure. 

Despite this limitation, hybrid beamforming algorithms can be optimized to improve system performance. The FC structure has demonstrated the ability to achieve near-optimal SE, making it a preferred choice for high-data-rate applications. However, this comes at the cost of increased power consumption due to the larger number of phase shifters and interconnections. In contrast, the AoSA structure is more power-efficient because it requires fewer hardware components, such as phase shifters and combiners. However, this reduction in hardware complexity comes at the expense of spectral efficiency, as the more constrained beamforming design limits the achievable performance~\cite{ref_hybrid_beamforming}.

\begin{figure}
    \centerline{\includegraphics[width=2.3in]{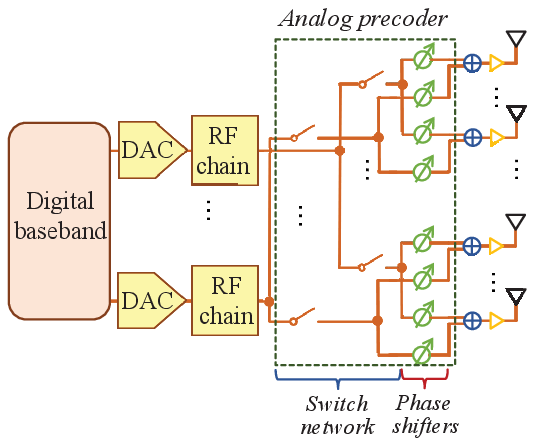}}
    \caption{DAoSA hybrid beamforming architecture.}
   \label{fig_DAoSA}
\end{figure}

In addition to the FC and AoSA structures, several hybrid beamforming architectures are proposed to cater to different requirements. 
First, in order to reduce power consumption while strike achieving high spectral efficiency, a DAoSA structure is proposed in~\cite{ref_DAoSA}. 
As depicted in Fig.~\ref{fig_DAoSA}, the DAoSA incorporates a switch network between the RF chains and subarrays. 
By dynamically controlling the state of the switches, specific phase shifters can be deactivated, resulting in reduced power consumption. This flexible control allows for achieving a trade-off between spectral efficiency and power consumption. In this case, the analog precoder should follow a special structure as
\begin{equation}
    \mathbf{F}_{\rm RF} = \begin{bmatrix}
        \mathbf{f}_{11} & \dots & \mathbf{f}_{1K}\\
        \vdots & \ddots & \vdots\\
        \mathbf{f}_{K1} &\dots&\mathbf{f}_{KK}
    \end{bmatrix},
\end{equation}
where $\mathbf{f}$ is either a zero vector or a vector satisfying the CM constraint.
The DAoSA serves as a generalization of the FC and AoSA architectures. 
In the DAoSA structure, the FC and AoSA architectures can be regarded as two extreme cases. 
When all switches are closed, the DAoSA reduces to the FC structure, while each RF chain having only one closed switch leads to the AoSA structure. With the DAoSA hybrid beamforming structure, the analog precoder and combiner contain zero vectors at random positions, reflecting the flexibility and adaptability of the DAoSA architecture.

\begin{figure}
    \centerline{\includegraphics[width=1.8in]{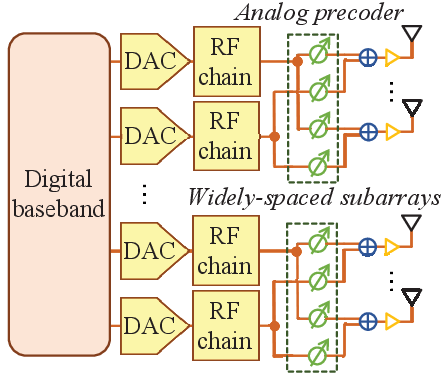}}
    \caption{WSMS hybrid beamforming architecture.}
   \label{fig_WSMS}
\end{figure}

Second, the FC, AoSA, and DAoSA structures have limited spatial multiplexing capabilities due to the sparse nature of high-frequency channels, especially in the THz band~\cite{ref_Multiray, ref_WSMS}. 
These architectures assume an antenna spacing of half the wavelength, consider communication distances that are much larger than the array aperture, and adopt the planar-wave assumption.
As a result, these systems face limitations in spatial multiplexing, which is constrained by the limited number of multipaths.
To address the challenge of limited spatial multiplexing in THz systems, a WSMS hybrid beamforming architecture is proposed~\cite{ref_WSMS}. 
As illustrated in Fig.~\ref{fig_WSMS}, the antenna array is divided into multiple widely-spaced subarrays.
Among the widely-spaced subarrays, the spherical-wave
propagation is considered, which enables the utilization of multiplexing gain provided by spherical-wave propagation even with only one LoS path.
In this case, the WSMS is useful to achieve the highest spatial multiplexing, thus the spectral efficiency among the introduced structures in this section. 
In this architecture, the analog precoder is represented in a block diagonal matrix as
\begin{subequations}
\begin{align}
    &\mathbf{F}_{\rm RF} = \operatorname{blkdiag}(\mathbf{F}_1,\dots,\mathbf{F}_K), \\
    &|\mathbf{F}_k(i,j)| = 1,
    \end{align}
\end{subequations}
where $\mathbf{F}_k$ represents the analog precoder for each subarray. 
In~\cite{ref_NF_dynamic1}, the authors investigate the trade-off between exploiting the NF SDoF and the hardware power consumption, and then provides a balanced solution where a dynamic beamforming architecture with an adaptive switch module is proposed.

Third, in wideband systems, the beam squint effect can lead to a significant loss in array gain in phase-shifter-based structures.
Specifically, the required optimal beamforming weights of the signal vary with the carrier frequency, while as a frequency-dependent device, all subcarriers are allocated with only the same weight. 
In this case, the beam directions deviate from the target direction as the frequency deviates from the central frequency, which thus reduces the array gain~\cite{ref_hybrid_beamforming}, i.e., the beam squint effect.
Inspired by this, some studies propose to insert the TTD devices into the hybrid beamforming architectures~\cite{ref_AoSA_training,ref_THzPrism,ref_wideband_tracking_dai,ref_DSFTTD}.
Compared to phase shifters, the beamforming weight generated by TTD is frequency-dependent, which can be used to solve the beam squint problem~\cite{ref_hybrid_beamforming}. The design of TTD-aided architectures for different communication fields is analyzed in~\cite{ref_cross_field1}. It first demonstrates that a threshold exists in FF at which increasing the number of TTDs no longer improves the performance, and a dense deployment of TTDs is required in the NF. Then, a cross near-and far-field beamforming (CNFB)
scheme is proposed which utilizes a limited number of TTDs.

In addition to the previously mentioned structures, there is growing research interest in hybrid beamforming architectures with low hardware costs, as the high hardware cost and energy consumption are major challenges in manufacturing high frequency MIMO systems.
To address these challenges, researchers have proposed the use of low-resolution ADCs and phase shifters. 
By employing low-resolution ADCs, the hardware cost and power consumption can be significantly reduced while still maintaining satisfactory system performance. 
Similarly, the use of low-resolution phase shifters allows for cost-effective beamforming implementation while minimizing the hardware complexity.
As an example, the energy-efficient dynamic-subarray with fixed true time delay (DS-FTTD) architecture proposed in~\cite{ref_DSFTTD} aims to further reduce the hardware cost and energy consumption associated with high-resolution TTD devices. By employing fixed true time delays in dynamic subarrays, the DS-FTTD architecture offers a cost-effective solution. 

In summary, various hybrid beamforming structures have been proposed in the literature to reduce the hardware complexity associated with fully digital architectures.
These architectures pose certain constraints on the beamformers. 
Among these structures, most are designed independently of the propagation field and can be applied across different communication distances. While others target at specific transmission field, such as the WSMS structure which enables the NF propagation and the TTD-aided CNFB to mitigate the beam squint effect in CF.

\subsection{Optimization-Based Beamforming}
In the cost-effective hybrid beamforming MIMO systems, the optimal beamformers to maximize the SE are generally hard to obtain due to the following reasons. Firstly, the design of the beamformers is a non-convex optimization problem due to the CM constraint. Secondly, the joint optimization of the digital and analog beamformers at both the transmitter and the receiver is required, leading to a high-dimensional design problem with multiple interdependent matrix variables. Thirdly, with diverse MIMO configurations, architecture-specific constraints can further complicate the beamforming design. 
Various optimization-based methodologies are developed to address these challenges, where solutions are derived via mathematical algorithms such as convex relaxation, alternating minimization, matrix approximation, etc. In the following, these methods are broadly categorized into two classes based on their treatment of channel properties, i.e., the field-independent beamforming and the field-aware beamforming.
\begin{figure*}[t]
    \centering
    {\includegraphics[width= 0.9\textwidth]{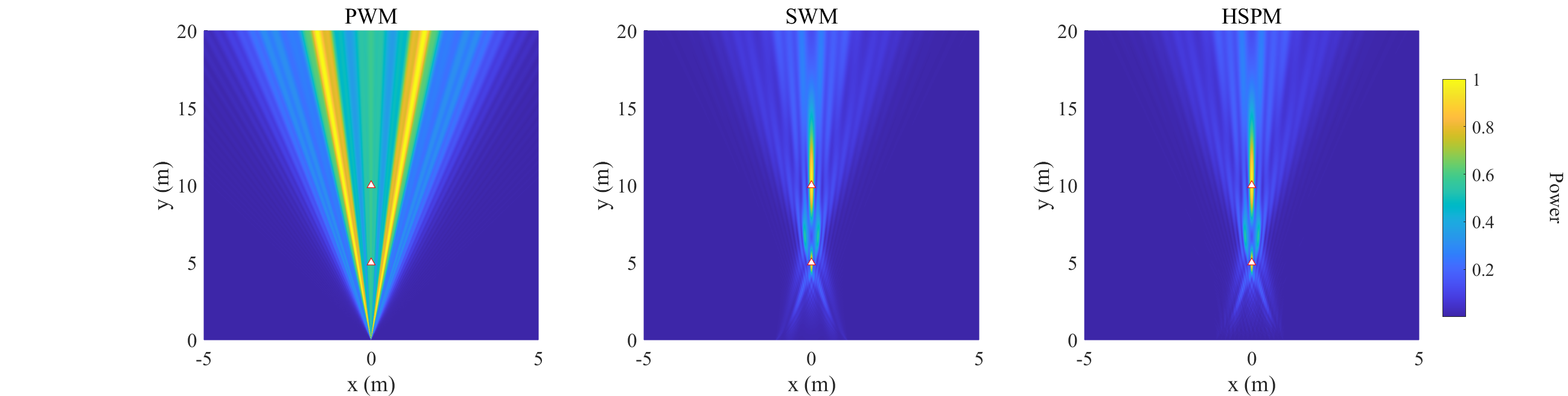}}
    \caption{Normalized power with PWM, SWM, and HSPM-based beamforming for users at the same direction.} 
    \label{fig_MUbeams}
\end{figure*} 
\subsubsection{Field-Independent Beamforming}
In field-independent beamforming, the design of the hybrid precoders/combiners is solely based on the numerical values of the channel matrix, without leveraging the distinct physical characteristics of the electromagnetic wave propagation in specific communication fields. In this case, one commonly adopted approach is the alternating optimization(AO)-based beamforming algorithm, which can effectively address the non-convex beamforming problem by alternately optimizing one of the beamformer variables while fixing the remaining ones in each iteration. 

In~\cite{ref_HDA,ref_HAD_EE}, the authors propose to design the hybrid beamformers by directly solving the SE-maximization problem where the contribution of each element in the beamformer matrix is extracted and updated iteratively. As these methods involve element-wise optimization, the computational complexity tend to be extremely high especially in UM-MIMO systems. Therefore, one alternative is to simplify the original problem into a low-rank matrix approximation (LRMA) problem where the Euclidean distance between the hybrid precoder (combiner) and the optimal precoder (combiner) is minimized~\cite{ref_AO,ref_min_Euclidean2}, i.e., 
\begin{equation}
    \min_{\mathbf{F}_{\rm RF},\mathbf{F}_{\rm BB}} \lVert \mathbf{F}_{\rm opt}-\mathbf{F}_{\rm RF}\mathbf{F}_{\rm BB}\rVert_F^2.
\end{equation}
For P2P communications, the optimal beamformers can be obtained through singular value decomposition (SVD) of the channel matrix, i.e., $\mathbf{H} = \mathbf{U}\mathbf{\Sigma}\mathbf{V}^H$. Then the optimal solution is given by 
\begin{subequations}
\begin{align}
       \mathbf{F}_{\rm opt} & = \mathbf{V}_{N_s}\mathbf{\Gamma},\\ 
       \mathbf{W}_{\rm opt} & = \mathbf{U}_{N_s}, 
\end{align}
\end{subequations}
where $\mathbf{V}_{N_s}$ and $\mathbf{U}_{N_s}$ denotes the first $N_s$ columns of matrix $V$ and $U$, $N_s$ represents the number of data streams, and $\mathbf{\Gamma}$ represents the power allocation matrix. A similar approach is adopted to overcome the wideband beam squint problem in~\cite{ref_beamsquint3}, where an ideal analog precoder is first derived and then approximated by a hybrid precoder implemented with TTD and phase shifters.

For MU communications with inter-user interference, user-specific quality of service (QoS) requirement, and dynamic user distributions, the beamforming optimization problem for fully digital MIMO becomes non-convex, such that the optimal beamformers for fully digital MIMO are generally unclear. To address this issue, authors in~\cite{ref_MMSE1,ref_MMSE2} propose to transform the original sum SE maximization problem into an equivalent weighted MMSE problem. Specifically, this transformation is achieved by introducing auxiliary weight matrix variables, whose design is shown to be a convex optimization problem. Based on this fact, develop different AO algorithms are proposed in~\cite{ref_MMSE_AO1,ref_MMSE_OFDM,ref_MMSE_AO_IRS} to support transmission in various MU scenarios, including narrowband systems, OFDM systems and IRS-aided systems. As AO-based algorithms can achieve near-optimal performance, it often suffers from high computational complexity as they rely on repeated iterations between subproblems. Instead, non-iterative optimization schemes has been widely investigated. In~\cite{ref_non-iterative1}, an equal gain method is applied to find the analog precoder, and the digital precoder is obtained with zero-forcing (ZF) method. In~\cite{ref_HBF}, a low-complexity algorithm is developed for MU scenarios where the block diagonalization (BD) is adopted for the digital beamformers, and a closed-form solution for the analog precoder is derived based on SVD. 
However, in MU communications, applying FF PWM model with only angular domain resolution fails to distinguish users at similar angles, leading to severe user interference. Instead, beamforming with NF SWM brings joint angle-distance domain resolution, which can effectively mitigate the user interference even when users are at the same direction. To illustrate, we consider simultaneous transmission to two users with the same transmission angle, which are located at (0,5)~m and (0,10)~m, respectively. The hybrid beamformer is obtained where the interference cancellation-based BD algorithm is applied at the digital stage, and the analog beam is adopted corresponding to each channel model at the analog stage. As shown in Fig.~\ref{fig_MUbeams}, hybrid beamforming with PWM model results in extremely low power at both users. This is because in order to cancel the interference between such co-directional users with PWM-based beam, the algorithm forces to reduce the signal power to both users. This phenomenon exposes the limitation of the FF PWM-based beam that with only angular domain resolution, it struggles to simultaneously suppress co-directional user interference and preserve desired signal power. In contrast, SWM/HSPM-based beamforming considers the spherical wavefront to eliminate the interference and spatially distinguish users sharing identical angles. This capability enables simultaneous communication for co-directional users, and is particularly advantageous for dense connectivity.

Despite the limitations of the FF PWM-based beams, the above works constructed based on the FF PWM can still be applied directly to NF and CF channel models since they treat the channel as a whole matrix without investigating its field-dependent physical characteristics. In~\cite{ref_BeamFocusing}, the NF SWM model is adopted with field-independent algorithms that are originally proposed for FF communications, where an MMSE-based optimal precoder is first developed, and then the LRMA problem is formulated and solved through AO for fully-digital, hybrid phase shifter-based, and dynamic metasurface antenna (DMA) architectures. In~\cite{ref_dynamic_NF}, the authors propose a dynamic hybrid beamforming architecture in NF to balance the power consumption and the sum rate, where AO-based algorithms is proposed to jointly optimize the switch module and the hybrid precoders. 
For CF wideband communications, a TTD-aided architecture is proposed in~\cite{ref_cross_field1}, where a traditional FF AO-based algorithms is adapted to solve the hybrid TTD-aided precoder.
These works suggest that the incorporation of the spherical wave propagation effect in the NF/CF channel model causes fundamentally different beam patterns where the FF beam aligns on the angle direction, while the NF/CF beam aligns (or focuses) on a specific position. That is to say, the beam patterns generated with field-independent algorithms can automatically match the channel model in use.


\subsubsection{Field-Aware Beamforming}
Although field-independent beamforming can be utilized in general situations regardless of the communication fields, it usually possesses high complexity due to matrix calculation or iterative optimization. 
In contrast, the field-aware beamforming explicitly exploits the structural property of the channel with field-specific FF, NF, and CF beams, to assist analog beamforming design, thus achieving low complexity.

In FF communications, authors in~\cite{ref_Spatially_sparse} demonstrate that the angle domain sparsity in mmWave/THz channels can be utilized to design the beamformer, where the original optimization problem can be simplified into a sparse reconstruction problem. Then, the optimal precoder can be approximated through an OMP-based construction process, where each column of the precoder is selected from a predefined set composed of the FF angle domain array response vectors. Considering wideband systems, authors in~\cite{ref_wideband_sparse} extend the set of the FF array response vectors in~\cite{ref_Spatially_sparse}, and propose a new set of precoding vectors which can effectively alleviate the beam squint effect. 

In~\cite{ref_HBD}, MU communication is considered where the analog beamformer is designed through exhaustive search from a set of FF DFT basis, and in~\cite{ref_sparsity1}, the angular domain orthogonality of different users is analyzed and exploited, based on which the closed form of the asymptotically optimal analog beamformer is derived. 
For wideband beamforming, a two-stage beam steering codebook-based hybrid beamforming scheme is proposed in~\cite{ref_frequencyselective} to jointly maximize the sum of the channel modulus over all subcarriers. Considering the beam split effect, authors in~\cite{ref_splitmultiplexing2} investigate the angular coverage of the wideband beam, based on which a beam split multiplexing scheme is proposed in~\cite{ref_splitmultiplexing1} for dense FF MU communications. For beamforming under partial CSI, authors in~\cite{ref_partial_1,ref_partial_2} design robust beamforming schemes where only the FF path angle information is assumed to be known.

\begin{figure}
\centerline{\includegraphics[width=0.4\textwidth]{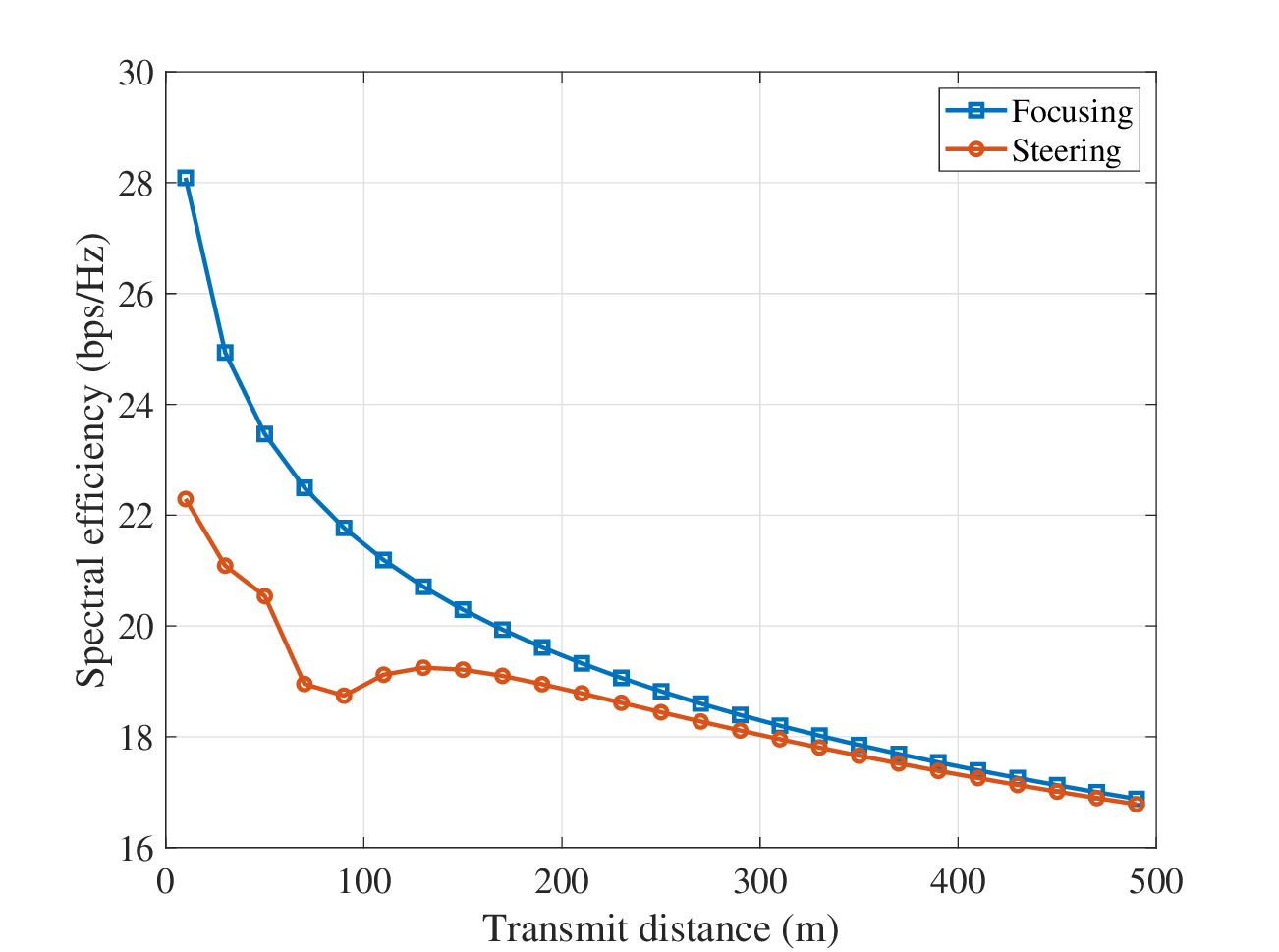}}
    \caption{Spectral efficiency with beam focusing and beam steering. 
}

   \label{fig_SE_focusvssteer}
\end{figure}

In NF scenarios with spherical wave propagation, applying the FF beam can cause significant misalignment and result in high performance loss.
Particularly, Fig.~\ref{fig_SE_focusvssteer} demonstrates that in NF channels, conventional FF steering beams - designed solely for angular alignment - exhibit SE loss over distance relative to NF focusing beams, which simultaneously account for both angular and distance dimensions. 
Inspired by this, some studies propose beam focusing-aware precoding. 
In~\cite{ref_Dai_LDMA}, it is demonstrated that similar to the angular orthogonality of FF beam, NF beam focusing vectors exhibit asymptotic distance-domain orthogonality as the antenna count grows, enabling enhanced spatial multiplexing. Then, a location division multiple access (LDMA) scheme is proposed to serve users at different positions rather than at different angles as in the FF.
Authors in~\cite{ref_NF_WideBF} investigate the NF beam split, which is shown to have a different beam pattern where the beams at different frequencies shift to distinct spatial locations (characterized by unique angle-distance pairs) rather than simply diverging along different directions as observed in FF scenarios. Then, a phase-delay focusing (PDF) method is developed based on a piece-wise far-field model where the channel matrix is constructed with multiple PWM-based submatrices such that the beamforming vector can be constructed for each sub-channel separately. 

Most prior works focus exclusively on either FF or NF beamforming, but recent research has expanded to include CF beamforming for broader applicability. Authors in~\cite{ref_cross_beamforming2} propose to utilize a subarray-based analog beamformer under the CF HSPM channel model, which adapts to both near-field and far-field as the proposed analog vector equals a near-field focusing vector when the number of subarrays is equal to the number of antennas and reduces to a far-field steering vector assuming a single array. In~\cite{ref_FieldawareCF}, the authors propose a two stage cross-field aware beamforming scheme where pre-beamforming is conducted first based on a beam selection with a CF codebook, and then the precoding matrix is designed to further mitigate user interference.

Field-dependent beamforming schemes usually achieve computational efficiency. By exploiting the wave propagation physics, i.e., angular sparsity in FF and joint angle-distance focusing ability of NF/CF beams, the intensive matrix computations and iterative procedures required by field-independent methods can be effectively avoided. However, these methods highly depend on the channel model in use, thus requiring careful identification of the communication field.


\subsection{ML-Based Beamforming}
While traditional optimization-based methods remain foundational in beamforming design, emerging ML-based beamforming introduces complementary strengths to provide robustness against dynamic environments and imperfect CSI while maintaining high computational efficiency~\cite{ref_twenty_five}. In the following, the ML-based beamforming is generally catheterized into the data-driven and the model-drived learning schemes.

In data-driven ML beamforming, the beamformers are constructed directly through a nonlinear mapping from the input channel matrix to the output beamformers. By considering the beamforming procedure as a black box, data-driven methods can learn near-optimal solutions without relying on specific mathematical models.
In~\cite{ref_learning_fast}, a DL-based beamforming prediction network (BPNet) is designed to achieve fast downlink beamforming while maintaining comparable sum rate performance. To mitigate the dependency on accurate CSI, authors in~\cite{ref_learning_imperfect1,ref_learning_imperfect2} consider beamforming under imperfect CSI where an unsupervised deep learning neural network (USDNN) is utilized in~\cite{ref_learning_imperfect1} to reduce the computing resources needed for labeling as in the supervised learning, and a graph neural network (GNN)-based architecture is proposed in~\cite{ref_learning_imperfect2} for joint multicast and unicast transmission. 
These methods treat the beamforming optimization as a direct mapping from the input to the output, thus is applicable in FF, NF and CF fields. However, data-driven ML beamforming typically demands extensive datasets and significant computational resources. 

On the contrary, the model-aided ML beamforming incorporates physical knowledge and mathematical models to the learning schemes to reduce the training time and the requirement of large dataset. For FF communications, authors in~\cite{ref_learning_model3} propose a computationally efficiency GNN-aided deep network architecture based on a 2D DFT codebook. In~\cite{ref_learning_model2}, the authors consider a finite-alphabet precoding problem where a traditional iterative discrete estimation precoder is extended to a neural network, which can achieve low computational complexity. In~\cite{ref_learning_model4}, a support vector machine algorithm (SVM)-based analog beam selection strategy is developed where the angles of each transmission path are taken as the input feature vector to reduce the complexity. For NF communications, authors in~\cite{ref_learning_model1} propose an NF channel latent feature learning module to learn the spherical wave propagation effect in NF channels.
However, as the model-aided ML beamforming heavily depends on prior knowledge of the channel model, it cannot be universally deployed for various communication fields, and its performance may be sensitive to the channel model in use.

\begin{table*}[t]
\centering
\caption{Comparison of Beamforming Methods}
\renewcommand{\arraystretch}{1.2}
\begin{tabular}{|p{3.5cm}|p{2.5cm}|p{2cm}|p{2cm}|p{1.5cm}|p{1.5cm}|}
\hline
\textbf{Method Type} & \multicolumn{3}{c|}{\textbf{Example Studies}} & \textbf{Spectral} \newline\textbf{Efficiency} & \textbf{Complexity} \\
\cline{2-4}
 & \textbf{FF} & \textbf{NF} & \textbf{CF} & & \\
\hline

\textbf{Optimization-based: Field-Independent Beamforming} & 
F. Sohrabi~\cite{ref_HDA}\newline
K. Ardah~\cite{ref_HAD_EE}\newline
X. Yu~\cite{ref_AO}\newline
Z. Luo~\cite{ref_min_Euclidean2}\newline
A. Najjar~\cite{ref_beamsquint3}\newline
Q. Shi~\cite{ref_MMSE1}\newline
S. S. Christense~\cite{ref_MMSE2}\newline
J. Cong~\cite{ref_MMSE_AO1}\newline
J. Du~\cite{ref_MMSE_OFDM}\newline
Q. Wu~\cite{ref_MMSE_AO_IRS}\newline 
L. Liang~\cite{ref_non-iterative1}\newline
X. Wu~\cite{ref_HBF} &
H. Zhang~\cite{ref_BeamFocusing}\newline
M. Liu~\cite{ref_dynamic_NF}&
J. Xie~\cite{ref_cross_field1}
& Near-optimal & Medium-High \\
\hline

\textbf{Optimization-based: Field-Aware Beamforming} & 
O. E. Ayach~\cite{ref_Spatially_sparse}\newline
Q. Wan~\cite{ref_wideband_sparse}\newline
W. Ni~\cite{ref_HBD}\newline
Q. Yue~\cite{ref_sparsity1}\newline
H. Yuan~\cite{ref_frequencyselective}\newline
B. Ning~\cite{ref_splitmultiplexing2}\newline
J. Wang~\cite{ref_splitmultiplexing1}\newline
L. Jiang~\cite{ref_partial_1} \newline
L. Yan~\cite{ref_partial_2}&
Z. Wu~\cite{ref_Dai_LDMA}\newline
M. Cui~\cite{ref_NF_WideBF}&
H. Shen~\cite{ref_cross_beamforming2}\newline
T. Gao~\cite{ref_FieldawareCF}
& Sub-optimal & Low \\
\hline

\textbf{ML-based Beamforming: Data-driven} & 
H. Huang~\cite{ref_learning_fast}\newline
P. Zhang~\cite{ref_learning_imperfect1} \newline
Z. Zhang~\cite{ref_learning_imperfect2}& -- & --
& High & Low-Medium \\
\hline

\textbf{ML-based Beamforming: Model-aided} & 
L. Schynol~\cite{ref_learning_model3} \newline
H. He~\cite{ref_learning_model2} \newline
Y. Long~\cite{ref_learning_model4}&
H. Ting~\cite{ref_learning_model1}& --
& High & Low \\
\hline
\end{tabular}
\label{tab_HBF_comparison}
\end{table*}




\subsection{Potential Research Directions}
Hybrid beamforming technologies typically involve the design of the beamforming architectures and the beamforming algorithms. Design of the architectures usually aims to explore the spatial multiplexing and reduce the hardware cost, while the design of hybrid beamforming target at improving the spectral efficiency while lowering the computational complexity. A comparison of representative hybrid beamforming algorithms is presented in Table~\ref{tab_HBF_comparison}.

The main difference between the FF, NF and CF beamformers is that FF beamformers can only generate beams targeting at a certain direction while NF/CF beamformers can generate beams focus at a specific location. However, some beamformers proposed for the FF can be applied to NF/CF as they rely only on the information of the full channel matrix, and neglect the filed-dependent characteristics of wave propagation. For other methods utilizing the angular sparsity in FF and the joint angle-distance domain focusing ability in NF/CF, the design of the beamformers must identify the communication field.

Although extensive progress has been made towards the hybrid beamforming across different communication distances, new challenges and research directions have emerged.
\begin{itemize}
    \item 
    In high-frequency communications, such as mmWave and THz, the problem of blockage and limited spatial multiplexing remains a major challenge. With Gaussian beams generated by traditional beam steering in FF and beam focusing in NF/CF, these problems are usually addressed through the assistance of the RIS-aided systems, which inevitably increases the design complexity. 
    As an alternative, with the increasing size of NF region, non-Gaussian beam patterns have shown great potential to address these issues, such as the Bessel beams which can reconstruct itself after an obstruction~\cite{ref_Airy}, the airy beam which can propagate along a curved path, thus bypassing an obstacle~\cite{ref_airy1}, and orbital angular momentum (OAM) beam which shows the ability to create multiple orthogonal channels, thus enhancing the system capacity~\cite{ref_OAM}. 
    \item Despite the theoretical advantages, hybrid beamforming integrating these beams have only recently begun to be systematically explored in the context of THz communications systems. To generate these beams in MIMO, high-resolution hardware such as phase shifters and high computational complexity is typically required. Moreover, with THz wideband, different subcarriers will cause phase deviation, leading to distortion of the resulted beam pattern~\cite{ref_Airy}. Therefore, developing efficient beamforming schemes that harness the unique properties of near-field non-Gaussian beams while maintaining manageable system complexity is critical for THz communications, and requires further exploration. 
    \item Rapid development of advanced technologies in modern society, such as high-speed transportation (HSR) systems, has led to highly dynamic wireless channel environments. Existing works have considered IRS-aided high-speed beamforming in FF~\cite{ref_highspeed1} and Doppler compensation beamforming design with NF model~\cite{ref_highspeed2}. While in practical THz communications, transmissions may be rapidly changed between NF and FF propagation regimes. This motivates the development of intelligent unified beamforming schemes to support rapid transition from NF to FF while maintaining low complexity. Furthermore, as CF beamforming shows its potential to adapt to various communication distances, it has rarely been studied in the context of HSR, which remains a promising research direction.
\end{itemize}


\section{BEAM TRACKING}
\label{sec_Beam_tracking}

Beam tracking can be considered a specialized form of beam alignment, focusing on adapting to user movement to ensure intermittent yet reliable transmission. Consequently, despite concerns regarding scanning overhead and complexity, the beam scanning and estimation methods discussed in previous sections can be directly applied to beam tracking.
However, achieving fast and efficient tracking with low overhead requires techniques that exploit the temporal correlation of user movement. 
To address this, various studies have explored training and estimation strategies tailored for beam tracking. 
These approaches can be broadly classified into Bayesian statistics-based methods, ML-based techniques, side-information-assisted strategies, and other alternative solutions.
In the following sections, we provide a detailed discussion of each category.

\subsection{Bayesian Statistics-Based Beam Tracking}
\label{subsec_Bayesian_based}
Bayesian statistics-based beam tracking utilizes Bayesian inference to dynamically estimate and update beam directions based on observations obtained from either the current beams or specially designed beams, thereby reducing scanning overhead. Typically, the approach models the channel state at time instant \( t \), denoted as \( s_t \) (usually contains DoA, DoD, path gain), as a Markov process as 
\begin{equation}
\label{equ_state}
s_t = f(s_{t-1}) + \omega_t,
\end{equation}
where \( f(\cdot) \) represents the state transition function, \( s_{t-1} \) denotes the previous channel state, and \( \omega_t \) is the process noise. The corresponding channel observation at time \( t \), denoted as \( z_t \), is given by
\begin{equation}
\label{equ_observation}
z_t = g(s_{t}) + n_t,
\end{equation}
where \( g(\cdot) \) is the observation function and \( n_t \) represents the observation noise. The tracking process first predicts the next state using the state expression in~\eqref{equ_state} and then updates the state based on new channel observations.

Depending on the modeling of state transitions and observations
Bayesian beam tracking methods can be categorized into the Kalman Filter (KF), Extended Kalman Filter (EKF), Unscented Kalman Filter (UKF), Particle Filter (PF), and partially observable Markov decision process (POMDP)-based approaches.

\subsubsection{KF and EKF}
The KF considers a linear state transition function \( f(\cdot) \) and Gaussian noise. However, since beamformed signals are inherently nonlinear, the KF does not accurately model the real signal behavior in MIMO systems and is thus rarely studied in the literature. 
The EKF extends the KF to nonlinear systems by linearizing the state transition and observation models using a first-order Taylor expansion, making it suitable for mildly nonlinear systems.

For FF communications, the state is usually modeled by the DoA, DoD, and path gain. 
Particularly, an analog beamforming architecture for mmWave communication systems is considered in~\cite{ref_Beam_tracking_EKF}, where the EKF is applied for beam prediction, enabling tracking with a single beam. In~\cite{ref_Beam_tracking_EKF2}, an EKF-based approach updates both the channel state and beamforming weights using a robust MMSE beamformer constrained by the array vector error, which is derived from the EKF-estimated error variance. Furthermore, authors in~\cite{ref_tracking_EKF3} introduce two well-designed beam pairs to eliminate the unknown path gain in the EKF measurement model, ensuring improved tracking performance.

For NF communications, authors in~\cite{ref_Two_Timescale_NF_tracking} propose an EKF-based mmWave beam tracking and localization framework. Unlike FF state representations, which typically include angles and path gains, the NF state \( s_t \) is modeled as a combination of user location and velocity, capturing the curvature of arrival (CoA). This representation accounts for the spherical wavefront, mitigating nonlinearities inherent in NF channels and ensuring stable tracking.
For CF communications, in~\cite{ref_ELAA_TrainingTracking}, the EKF is deployed with CoA state parameters to achieve beam tracking.

\subsubsection{UKF}
Unlike the EKF, which relies on linearization, the UKF approximates nonlinear state transitions by propagating multiple sigma points through the nonlinear function. This approach captures higher-order nonlinearities and improves estimation accuracy, which however, brings higher complexity. 
The UKF for FF mmWave tracking is studied in~\cite{ref_tracking_UKF}, where simulation results show that the UKF outperforms the EKF, providing better tracking accuracy.

\subsubsection{PF}
The PF represents the state using a set of weighted particles, making it suitable for highly nonlinear systems. 
For FF beam tracking, a PF-based approach is investigated in~\cite{ref_beamwidth_control}, where beamwidth adaptation is achieved through partial activation of the antenna array to enhance performance. In~\cite{ref_tracking_nonlinear}, an auxiliary PF is introduced by incorporating an auxiliary variable, improving robustness at the cost of additional computational complexity due to the need for an extra particle update step.
For NF beam tracking, authors in~\cite{ref_NF_Tracking_limit} model the state using CoA and derives the posterior Cramér-Rao lower bound for tracking estimation error. 
After that, various tracking methods, including the EKF, PF, and MLE, are implemented, and the results demonstrate the feasibility and high accuracy of these approaches.

\subsubsection{POMDP}
The POMDP framework enables decision-making in uncertain environments where the full system state is not directly observable. Instead, an agent maintains a belief distribution over possible states and selects actions to maximize long-term rewards. This approach effectively handles state uncertainty, integrates decision-making with tracking, and supports long-term planning under uncertainty.
However, solving POMDP problems is often computationally expensive. Approximate methods, including sampling-based approaches, belief-space planning, and RL (including DL techniques), are commonly used to address this challenge.

For FF communications, in~\cite{ref_POMDP_far}, a dual timescale learning and adaptation framework is proposed for low-overhead beam tracking in mmWave vehicular networks. The short-timescale optimization is formulated as a POMDP and solved using point-based value iteration. In~\cite{ref_POMDP_Tracking}, active sensing and sequential beam tracking for mmWave aerial communications are studied. The problem is modeled as a POMDP to ensure optimal beam alignment despite stochastic mobility. A Bayesian posterior probability-based algorithm is proposed for joint sensing and communication, dynamically selecting beams while balancing mutual information for DoA estimation and spectral efficiency.

As a summary, the Bayesian statistics-based beam tracking methods dynamically estimate and update beam directions, improving tracking accuracy while minimizing scanning overhead. 
Although most studies are conducted for FF communications, it has been shown in~\cite{ref_NF_Tracking_limit} that the primary distinction between FF and NF tracking lies in the state representation, transitioning from direct channel parameters to CoA.

\subsection{ML-Based Beam Tracking}
ML has been widely adopted for beam tracking due to its capability to capture complex temporal and spatial relationships in wireless channels. Existing ML-based beam tracking methods can be broadly categorized into DL-based methods and RL-based methods.

\subsubsection{DL-Assisted Beam Tracking}
In this category, recurrent neural networks (RNNs), particularly long short-term memory (LSTM) networks, are widely used for beam tracking due to their ability to capture the temporal correlation of input signals. Meanwhile, Transformers leverage self-attention mechanisms to model long-range dependencies, making them well-suited for highly dynamic beam tracking.

In FF communications, LSTM-based models have been employed to process sequential channel observations~\cite{ref_milli, ref_Learning_tracking}. While in~\cite{ref_milli}, the beam direction is directly estimated, authors in~\cite{ref_Learning_tracking} predict the beam state, which is then used for adaptive beam training and Bayesian sequential updates. Both methods significantly enhance tracking performance in mobile environments.
For highly dynamic scenarios, such as unmanned aerial vehicle (UAV)-assisted mmWave networks, a hybrid KF and LSTM approach is proposed in \cite{ref_KFLSTM_tracking}, where KF estimates short-term beam variations, and LSTM captures long-term trends, improving tracking accuracy while reducing scanning overhead. To further minimize overhead, authors in~\cite{ref_POMDP_far} introduce a dual-timescale learning and adaptation framework for vehicular networks. This framework utilizes a deep recurrent variational autoencoder—which integrates RNNs with variational inference—to learn a probabilistic model of beam dynamics from noisy feedback, enabling predictive beam tracking.

In NF communications, a digit-aware transformer framework is deployed in \cite{ref_AI_BT_NF} to predict mobile user trajectories, allowing precise beam direction adjustments and dynamic data stream allocation. This method achieves near-perfect CSI estimation, significantly improving throughput and beam focusing gain. Similarly, authors in~\cite{ref_NF_AI_Tracking} integrate a Transformer-based location prediction framework with a DRL-based approach, dynamically optimizing near-field beam focusing and data stream allocation.

\subsubsection{RL-Based Beam Tracking Methods}
RL techniques have been extensively integrated into beam tracking, optimizing long-term beamforming strategies by maximizing cumulative rewards. RL directly interacts with the wireless environment, optimizing beam selection strategies based on reward feedback without requiring labeled data.
RL-based beam tracking methods can be broadly categorized into three major categories, including DRL-based methods, bandit-based methods, as well as Q-learning-based methods.

DRL leverages deep neural networks to approximate value functions or policies, enabling robust beam tracking in dynamic environments.
For FF communications, a dual-path beam tracking framework is proposed in~\cite{ref_Daul_path_tracking}, employing a recurrent constrained DRL (RCDRL) algorithm to maintain two beam clusters, improving robustness against beam blockage. To leverage historical data, LSTM networks are integrated into RCDRL to predict beam cluster centers based on past beam selection and SNR trends, while DRL optimizes beam sweeping and tracking efficiency.

A two-phase RL-based method is introduced in~\cite{ref_beam_pattern_selection}, which jointly optimizes beam tracking and beamforming codebook selection. The inner RL agent updates the transmission beam index based on short-term SNR feedback, while the outer agent selects beamforming codebooks based on long-term SNR performance. This hierarchical approach enhances beam alignment and tracking accuracy.
In addition, authors in~\cite{ref_tracking_Q} introduce a deep Q-network (DQN)-based beam tracking method for mmWave communications, where the optimal receive beam angle is determined using received signals, eliminating the need for prior channel model knowledge and reducing computational complexity.
For NF communications, DQN-based beam tracking is applied in~\cite{ref_Qlearning_THzTracking} for THz communications, eliminating reliance on mobility models. The method dynamically updates beamforming directions to maintain stable performance, particularly under abrupt mobility changes.

Bandit-based methods model beam tracking as a decision-making process where the optimal beam selection is learned through reward feedback.
Particularly, authors in~\cite{ref_MAMBA} formulate beam tracking as a multi-armed bandit (MAB) problem, introducing adaptive Thompson sampling to dynamically select beams and transmission rates based on prior beam-quality data and feedback information.
In~\cite{ref_tracking_bandit}, MAB-based beam tracking is extended with a stochastic bandit-learning framework, where beam index differences are used to guide adaptive beam training, enhancing achievable rate performance.

Q-learning and its extensions optimize beam selection by updating Q-values based on past beam choices and observed SNR.
For FF communications, authors in~\cite{ref_QLearning_tracking} propose a Q-learning-based beam tracking method, where rewards derived from current and past observations optimize beam selection to maximize SINR.

The ML-based beam tracking methods offer significant advantages in addressing dynamic channel variations and user mobility. 
RNN-based models effectively capture temporal correlations, making them suitable for sequential beam tracking.  
RL-assisted methods enable adaptive beam selection and alignment, improving tracking robustness and efficiency.  


\subsection{Side Information-Assisted Beam Tracking}

Side information, such as channel fingerprints and visual data, can enhance beam tracking by reducing the need for frequent beam sweeping and improving tracking accuracy.
For fingerprints-assisted tracking, authors in~\cite{ref_tracking_Fingerprint} map statistical beamforming gains to user locations. The method tracks user movement by estimating the gains of untrained beam configurations, leading to improved beamforming performance.

For vision-assisted tracking, camera data is leveraged to aid beam tracking by providing additional environmental context. In~\cite{ref_hu2021image}, a sequence-to-sequence model is introduced combining CNN and gated recurrent unit to process fused image and beam index data, enabling accurate sequential beam direction prediction.
In~\cite{ref_CV_tracking}, an encoder-decoder machine learning framework is employed that utilizes past visual sensing information to predict future beam directions, significantly reducing beam training overhead. Authors in~\cite{ref_VC_tracking_exp} conduct experiments on vision-aided tracking in IRS systems, where a camera mounted on the IRS captures environmental information to guide beam adjustments. The IRS dynamically updates its reflection coefficients based on pre-designed codebooks, enabling real-time tracking in both NF and FF scenarios.

\begin{table*}[t]
\centering
\caption{Comparison of Beam Tracking Methods}
\renewcommand{\arraystretch}{1.2}
\begin{tabular}{|p{3cm}|p{3cm}|p{2.5cm}|p{1.5cm}|p{1.5cm}|p{1.5cm}|}
\hline
\textbf{Method} & \multicolumn{2}{c|}{\textbf{Example Studies}} & \textbf{Scanning Overhead} & \textbf{Complexity} & \textbf{Alignment Accuracy} \\
\cline{2-3}
 & \textbf{FF} & \textbf{NF/CF} & & & \\
\hline

\textbf{EKF} & 
V. Va~\cite{ref_Beam_tracking_EKF}\newline
S. Jayaprakasam~\cite{ref_Beam_tracking_EKF2}\newline
F. Liu~\cite{ref_tracking_EKF3} &
S. Palmucci~\cite{ref_Two_Timescale_NF_tracking}\newline
K. Chen~\cite{ref_ELAA_TrainingTracking} &
Low & Low & Low \\
\hline

\textbf{UKF} & 
S. G. Larew~\cite{ref_tracking_UKF} &
-- &
Low & Low-Medium & Low-Medium \\
\hline

\textbf{PF} & 
H. Chung~\cite{ref_beamwidth_control}\newline
J. Lim~\cite{ref_tracking_nonlinear} &
A. Guerra~\cite{ref_NF_Tracking_limit} &
Low & High & Medium-High \\
\hline

\textbf{POMDP} & 
M. Hussain~\cite{ref_POMDP_far}\newline
N. Ronquillo~\cite{ref_POMDP_Tracking} &
-- &
Low & High & Medium-High \\
\hline

\textbf{DL-based} & 
Y. Chen~\cite{ref_milli}\newline
M. Hussain~\cite{ref_POMDP_far} \newline
S. H. Lim~\cite{ref_Learning_tracking}\newline
L. Yan~\cite{ref_KFLSTM_tracking} &
M. Zhang~\cite{ref_AI_BT_NF, ref_NF_AI_Tracking} &
Medium & High & Medium-High \\
\hline

\textbf{RL-based} & 
R. Wang~\cite{ref_Daul_path_tracking}\newline
J. Jeong~\cite{ref_beam_pattern_selection}\newline
H. Park~\cite{ref_tracking_Q}\newline
I. Aykin~\cite{ref_MAMBA}\newline 
J. Zhang~\cite{ref_tracking_bandit}\newline
H.-L. Chiang~\cite{ref_QLearning_tracking} &
H. Park~\cite{ref_Qlearning_THzTracking} &
Medium & High & Medium-High \\
\hline

\textbf{Fingerprint-assisted} & 
\multicolumn{2}{c|}{R. Deng~\cite{ref_tracking_Fingerprint}} &
None & Medium & High \\
\hline

\textbf{Vision-assisted} & 
\multicolumn{2}{c|}{Z. Hu~\cite{ref_hu2021image},\newline
S. Jiang~\cite{ref_CV_tracking},\newline
M. Ouyang~\cite{ref_VC_tracking_exp}} &
None & High & Medium \\
\hline

\textbf{Adaptive tracking} & 
Y. Karaçora~\cite{ref_event_tracking}\newline
H. Ding~\cite{ref_Context_Aware_Tracking}\newline
L. Yang~\cite{ref_tracking_UAV}\newline
Z. Xiao~\cite{ref_Tracking_Variable}\newline
D. Zhang~\cite{ref_tracking_codebookopt} &
P. Gavriilidis~\cite{ref_NearTracking} &
Medium & Medium & Medium \\
\hline

\textbf{Hierarchical/predictive tracking} & 
B. Ning~\cite{ref_QUPA}\newline
J. Tan~\cite{ref_wideband_tracking_dai}\newline
G. Stratidakis~\cite{ref_Hierarchical_Beamtracking}
&
M. Cui~\cite{ref_NF_rainbow} &
Medium & Medium & Medium \\
\hline

\textbf{Data-based tracking} & 
J. Palacios~\cite{ref_tracking_Infocom} &
-- &
None & Medium & Low \\
\hline
\end{tabular}
\label{tab_tracking_comparison}
\end{table*}

\subsection{Other Beam Tracking Strategies}  

In addition to the previously discussed methods, several other approaches are explored in the literature to assist beam tracking. These methods can be categorized into adaptive tracking strategies, hierarchical and predictive tracking, and data-based tracking.

\subsubsection{Adaptive Tracking Strategies}  
Adaptive tracking dynamically adjusts the tracking frequency or scanning strategy to reduce scanning overhead while maintaining high tracking accuracy.

For FF communications, the authors in~\cite{ref_event_tracking} consider a THz system and optimize the timing of channel estimation regarding channel variations. This approach minimizes pilot overhead while preventing misalignment and blockage-induced outages. In~\cite{ref_Context_Aware_Tracking}, a context-aware beam update scheme is presented for vehicular communications, which determines when to trigger beam sweeping based on vehicle mobility patterns, avoiding unnecessary overhead and outdated beam selection.
Additionally, an adaptive training-based beam tracking scheme is proposed in~\cite{ref_tracking_UAV} for UAV systems that dynamically adjusts training frequency based on beam coherence time estimated from joint beam training and angular velocity estimation. It also optimizes beamwidth selection to maximize UAV flying range and speed. 
Furthermore, in~\cite{ref_Tracking_Variable}, a variable step beam-based tracking scheme is proposed, where a small portion of beam energy is allocated for tracking while maintaining communication quality.  
During the tracking process, 
the beam main lobe is dynamically adjusted to improve tracking accuracy and communication reliability.
Moreover, in~\cite{ref_tracking_codebookopt}, the tracking error minimization problem is considered. A closed-form upper bound on the average tracking error probability is derived and optimized in terms of beam codebooks, power allocation, and beam pattern design, to improve the beam scanning efficiency. 

For NF communications, the authors in~\cite{ref_NearTracking} propose a threshold-triggered NF beam tracking framework, where beam sweeping is initiated when beamforming gain falls below a threshold. 
In this case, the NF feature, i.e., beamforming gain incorporates both effects from misalignment angles and distance should be taken into account.

\subsubsection{Hierarchical and Predictive Beam Tracking}  
These approaches improve tracking efficiency by integrating hierarchical prediction, beam resolution adaptation, and multi-directional tracking.
For example, for FF communications, the authors in~\cite{ref_Hierarchical_Beamtracking} predict the user's next beam direction and dynamically adjust beam resolution with a multi-resolution codebook, improving tracking efficiency. In~\cite{ref_QUPA}, a two-step tracking process is proposed. First, the beam search is performed in the vicinity of the previously used beam pair. Then, a new beam pair is selected based on the observed variation trend, reducing tracking overhead.
Moreover, authors in~\cite{ref_wideband_tracking_dai} introduce a beam zooming mechanism by deploying split beams in different frequencies to simultaneously track multiple physical directions. The method dynamically adjusts the angular coverage of beams to match potential user movement patterns, improving tracking efficiency.
This method is further extended for NF communications in~\cite{ref_NF_rainbow}.

\subsubsection{Data-Based Beam Tracking}  
This method directly utilizes data packets for beam tracking, eliminating the need for beam scanning.
For example, in~\cite{ref_tracking_Infocom}, the hybrid analog-digital transceivers is deployed to collect channel information from multiple spatial directions simultaneously using data packets. Then, a probabilistic optimization model predicts the temporal evolution of mmWave channels under mobility.

As a conclusion, these design techniques focus on various aspects to reduce the scanning overhead and enhance the tracking performance. 

\subsection{Potential Research Directions}

Several beam tracking methods have been studied in the literature, typically categorized into Bayesian statistics-based, ML-based, side information-assisted, and other approaches. Within each category, key design aspects include the tracking beam pattern, the algorithm for channel parameter estimation, the determination of tracking frequency, and so on. A comparison of representative beam tracking techniques is summarized in Table~\ref{tab_tracking_comparison}.

The main difference between FF and NF/CF beam tracking lies in the underlying channel model. In NF and CF, an additional distance-dependent phase term appears in the array response vector, which impacts the tracking design. For Bayesian statistics-based methods, this leads to a change in state representation—from direct channel parameters in FF to CoA representations in NF. For other methods, while FF tracking typically relies on angular-domain beams, NF/CF designs must consider joint angle–distance domain beams to accommodate the additional spatial resolution.

Despite the progress made in FF beam tracking, studies on NF and CF tracking remain limited. This gap highlights several open research directions for advancing beam tracking under near-field and cross-field propagation conditions.
\begin{itemize}
    \item Regarding the design of tracking beam patterns, NF/CF beam patterns fundamentally differ from those in FF due to the additional distance-domain resolution. Effectively designing tracking beams for NF/CF scenarios is a critical challenge. 
    To address this, a deeper analysis of NF/CF beam patterns under various beamforming structures, as well as adopting beam pattern design to moving scenarios, could help develop effective tracking beams.
    \item In terms of tracking algorithms, many existing FF beam tracking methods can be adapted for NF and CF tracking. For instance, as discussed in Sec.~\ref{sec_Beam_tracking}-\ref{subsec_Bayesian_based}, the primary distinction between FF and NF/CF EKF-based tracking lies in the selection of the state vector. 
    Therefore, identifying a systematic approach to transition from FF to NF/CF tracking to adopt to different algorithms is an important direction. Furthermore, exploring NF/CF-specific beam tracking algorithms remains a promising area for future research.
    \item Although~\cite{ref_NearTracking} has proposed an event-triggered beam tracking approach for NF, frequent interactions between the BS and UE to adjust the tracking frequency may be impractical, due to the requirement of frequent signaling. A promising alternative is the use of channel statistics-assisted beam tracking frequency design for NF/CF, which could enhance efficiency in real-world communication systems.
\end{itemize}

\section{CONCLUSION}
\label{sec_conclusion}
The evolution of 6G and beyond wireless networks introduces new challenges and opportunities driven by the use of higher frequency bands and large-scale MIMO systems, which extend communication regimes beyond the traditional FF into NF and CF regions. These emerging propagation regimes fundamentally alter the characteristics of wireless channels and necessitate a rethinking of BM techniques.
In this survey, we reviewed BM techniques—including beam scanning, CSI estimation, beamforming, and beam tracking—from the perspective of propagation fields. We first established the foundation by modeling FF, NF, and CF channels and analyzing their impact on beam pattern design.
Compared to FF channels, NF and CF channels offer additional resolution in the distance domain, which becomes a key factor influencing BM design differences across regimes.
We categorized existing BM techniques by methodology and examined how their operations vary across FF, NF, and CF regions, emphasizing the implications of angular and distance-domain resolution on performance, complexity, and applicability.

Our analysis shows that while NF-based BM techniques can sometimes be extended to FF and CF scenarios, such extensions often incur high complexity due to the increased parameter space.
Therefore, unified and low-complexity BM strategies—motivated by the concept of CF communication—are essential for practical deployment across diverse field conditions.
These strategies may be implemented through channel approximations or by adapting FF-based methods with appropriate modifications.
Finally, we identified several open challenges and future research directions, including the adaptation of FF techniques to NF/CF regimes, the development of unified BM frameworks, and the need for BM design tailored to emerging communication scenarios.

\section*{REFERENCES}
\def\refname{\vadjust{\vspace*{-1em}}} 
\bibliographystyle{IEEEtran}
\bibliography{main} 

\begin{thebibliography}{100}
\providecommand{\url}[1]{#1}
\csname url@samestyle\endcsname
\providecommand{\newblock}{\relax}
\providecommand{\bibinfo}[2]{#2}
\providecommand{\BIBentrySTDinterwordspacing}{\spaceskip=0pt\relax}
\providecommand{\BIBentryALTinterwordstretchfactor}{4}
\providecommand{\BIBentryALTinterwordspacing}{\spaceskip=\fontdimen2\font plus
\BIBentryALTinterwordstretchfactor\fontdimen3\font minus
  \fontdimen4\font\relax}
\providecommand{\BIBforeignlanguage}[2]{{%
\expandafter\ifx\csname l@#1\endcsname\relax
\typeout{** WARNING: IEEEtran.bst: No hyphenation pattern has been}%
\typeout{** loaded for the language `#1'. Using the pattern for}%
\typeout{** the default language instead.}%
\else
\language=\csname l@#1\endcsname
\fi
#2}}
\providecommand{\BIBdecl}{\relax}
\BIBdecl

\bibitem{ref_6G1}
M.~Z. Chowdhury, M.~Shahjalal, S.~Ahmed, and Y.~M. Jang, ``{6G Wireless
  Communication Systems: Applications, Requirements, Technologies, Challenges,
  and Research Directions},'' \emph{IEEE Open Journal of the Communications
  Society}, vol.~1, pp. 957--975, 2020.

\bibitem{ref_6G2}
P.~Yang, Y.~Xiao, M.~Xiao, and S.~Li, ``{6G Wireless Communications: Vision and
  Potential Techniques},'' \emph{IEEE Network}, vol.~33, no.~4, pp. 70--75,
  2019.

\bibitem{ref_6GKPI}
A.~Mourad, R.~Yang, P.~H. Lehne, and A.~de~la Oliva, ``{Towards 6G: Evolution
  of Key Performance Indicators and Technology Trends},'' in \emph{Proc. of 6G
  Wireless Summit}, 2020, pp. 1--5.

\bibitem{ref_minband}
J.~Zhang, H.~Miao, P.~Tang, L.~Tian, and G.~Liu, ``{New Mid-Band for 6G:
  Several Considerations from the Channel Propagation Characteristics
  Perspective},'' \emph{IEEE Communications Magazine}, vol.~63, no.~1, pp.
  175--180, 2025.

\bibitem{ref_6GmmW_THz}
S.~Tripathi, N.~V. Sabu, A.~K. Gupta, and H.~S. Dhillon, ``{Millimeter-wave and
  terahertz spectrum for 6G wireless},'' in \emph{6G Mobile Wireless
  Networks}.\hskip 1em plus 0.5em minus 0.4em\relax Springer, 2021, pp.
  83--121.

\bibitem{ref_THz}
A.~Shafie, N.~Yang, C.~Han, J.~M. Jornet, M.~Juntti, and T.~Kürner,
  ``Terahertz communications for 6g and beyond wireless networks: Challenges,
  key advancements, and opportunities,'' \emph{IEEE Network}, vol.~37, no.~3,
  pp. 162--169, 2023.

\bibitem{ref_THz_Old_revisit}
I.~F. Akyildiz, C.~Han, Z.~Hu, S.~Nie, and J.~M. Jornet, ``{Terahertz Band
  Communication: An Old Problem Revisited and Research Directions for the Next
  Decade},'' \emph{IEEE Trans. Commun.}, vol.~70, no.~6, pp. 4250--4285, May
  2022.

\bibitem{ref_rol_mmW}
W.~Hong, Z.~H. Jiang, C.~Yu, D.~Hou, H.~Wang, C.~Guo, Y.~Hu, L.~Kuai, Y.~Yu,
  Z.~Jiang, Z.~Chen, J.~Chen, Z.~Yu, J.~Zhai, N.~Zhang, L.~Tian, F.~Wu,
  G.~Yang, Z.-C. Hao, and J.~Y. Zhou, ``{The Role of Millimeter-Wave
  Technologies in 5G/6G Wireless Communications},'' \emph{IEEE Journal of
  Microwaves}, vol.~1, no.~1, pp. 101--122, 2021.

\bibitem{ref_6G_ELAA}
Z.~Wang, J.~Zhang, H.~Du, D.~Niyato, S.~Cui, B.~Ai, M.~Debbah, K.~B. Letaief,
  and H.~V. Poor, ``{A Tutorial on Extremely Large-Scale MIMO for 6G:
  Fundamentals, Signal Processing, and Applications},'' \emph{IEEE
  Communications Surveys \& Tutorials}, vol.~26, no.~3, pp. 1560--1605, 2024.

\bibitem{ref_bjornson2024towards}
E.~Bj{\"o}rnson, C.-B. Chae, R.~W. Heath~Jr, T.~L. Marzetta, A.~Mezghani,
  L.~Sanguinetti, F.~Rusek, M.~R. Castellanos, D.~Jun, and {\"O}.~T. Demir,
  ``{Towards 6G MIMO: Massive Spatial Multiplexing, Dense Arrays, and Interplay
  Between Electromagnetics and Processing},'' \emph{arXiv preprint
  arXiv:2401.02844}, 2024.

\bibitem{ref_6G_network}
M.~Giordani, M.~Polese, M.~Mezzavilla, S.~Rangan, and M.~Zorzi, ``{Toward 6G
  Networks: Use Cases and Technologies},'' \emph{IEEE Communications Magazine},
  vol.~58, no.~3, pp. 55--61, 2020.

\bibitem{ref_6G_frontiers}
C.~D. Alwis, A.~Kalla, Q.-V. Pham, P.~Kumar, K.~Dev, W.-J. Hwang, and
  M.~Liyanage, ``{Survey on 6G Frontiers: Trends, Applications, Requirements,
  Technologies and Future Research},'' \emph{IEEE Open Journal of the
  Communications Society}, vol.~2, pp. 836--886, 2021.

\bibitem{ref_rayleigh_Mag}
S.~Sun, R.~Li, C.~Han, X.~Liu, L.~Xue, and M.~Tao, ``{How to Differentiate
  Between Near Field and Far Field: Revisiting the Rayleigh Distance},''
  \emph{IEEE Communications Magazine}, vol.~63, no.~1, pp. 22--28, 2025.

\bibitem{ref_Near_Cui}
M.~Cui, Z.~Wu, Y.~Lu, X.~Wei, and L.~Dai, ``{Near-Field MIMO Communications for
  6G: Fundamentals, Challenges, Potentials, and Future Directions},''
  \emph{IEEE Communications Magazine}, vol.~61, no.~1, pp. 40--46, 2023.

\bibitem{ref_nearfield_tut}
H.~Lu, Y.~Zeng, C.~You, Y.~Han, J.~Zhang, Z.~Wang, Z.~Dong, S.~Jin, C.-X. Wang,
  T.~Jiang, X.~You, and R.~Zhang, ``{A Tutorial on Near-Field XL-MIMO
  Communications Toward 6G},'' \emph{IEEE Communications Surveys \& Tutorials},
  vol.~26, no.~4, pp. 2213--2257, 2024.

\bibitem{ref_NF_OJCOMS}
Y.~Liu, Z.~Wang, J.~Xu, C.~Ouyang, X.~Mu, and R.~Schober, ``{Near-Field
  Communications: A Tutorial Review},'' \emph{IEEE Open Journal of the
  Communications Society}, vol.~4, pp. 1999--2049, 2023.

\bibitem{ref_Cross}
C.~Han, Y.~Chen, L.~Yan, Z.~Chen, and L.~Dai, ``{Cross Far- and Near-Field
  Wireless Communications in Terahertz Ultra-Large Antenna Array Systems},''
  \emph{IEEE Wireless Communications}, vol.~31, no.~3, pp. 148--154, 2024.

\bibitem{ref_hybrid_field}
Y.~Zhang, Z.~Yang, S.~Yue, L.~Liu, and B.~Di, ``{Hybrid Near-Field and
  Far-Field XL-MIMO: How Many Users Can Be Supported?}'' \emph{IEEE
  Communications Letters}, vol.~28, no.~10, pp. 2402--2406, 2024.

\bibitem{ref_BM_3GPP}
M.~Giordani, M.~Polese, A.~Roy, D.~Castor, and M.~Zorzi, ``{A Tutorial on Beam
  Management for 3GPP NR at mmWave Frequencies},'' \emph{IEEE Communications
  Surveys \& Tutorials}, vol.~21, no.~1, pp. 173--196, 2019.

\bibitem{ref_SW_PW_Modeling}
F.~{Bohagen}, P.~{Orten}, and G.~E. {Oien}, ``{On Spherical vs. Plane Wave
  Modeling of Line-of-sight MIMO Channels},'' \emph{IEEE Trans. Commun.},
  vol.~57, no.~3, pp. 841--849, 2009.

\bibitem{ref_spherical_fronts}
P.~{Zhang}, J.~{Chen}, X.~{Yang}, N.~{Ma}, and Z.~{Zhang}, ``{Recent Research
  on Massive MIMO Propagation Channels: A Survey},'' \emph{IEEE Commun. Mag.},
  vol.~56, no.~12, pp. 22--29, 2018.

\bibitem{ref_HSPM}
Y.~Chen, L.~Yan, and C.~Han, ``{Hybrid Spherical- and Planar-Wave Modeling and
  DCNN-Powered Estimation of Terahertz Ultra-Massive MIMO Channels},''
  \emph{IEEE Transactions on Communications}, vol.~69, no.~10, pp. 7063--7076,
  2021.

\bibitem{ref_two_level}
X.~Song, W.~Rave, N.~Babu, S.~Majhi, and G.~Fettweis, ``{Two-Level Spatial
  Multiplexing Using Hybrid Beamforming for Millimeter-Wave Backhaul},''
  \emph{IEEE Trans. Wireless Commun.}, vol.~17, no.~7, pp. 4830--4844, 2018.

\bibitem{ref_DAoSA}
L.~Yan, C.~Han, and J.~Yuan, ``{A Dynamic Array-of-Subarrays Architecture and
  Hybrid Precoding Algorithms for Terahertz Wireless Communications},''
  \emph{IEEE J. Sel. Areas Commun.}, vol.~38, no.~9, pp. 2041--2056, 2020.

\bibitem{ref_cross_measure}
Y.-Q. Wang, C.~Han, S.~Sun, and J.~Zhang, ``{Cross Far- and Near-Field Channel
  Measurement and Modeling in Extremely Large-scale Movable Antenna Array
  Systems},'' \emph{IEEE Transactions on Antennas and Propagation}, to appear
  2025.

\bibitem{ref_Airy}
A.~Singh, V.~Petrov, H.~Guerboukha, I.~V. Reddy, E.~W. Knightly, D.~M.
  Mittleman, and J.~M. Jornet, ``{Wavefront Engineering: Realizing Efficient
  Terahertz Band Communications in 6G and Beyond},'' \emph{IEEE Wireless
  Communications}, vol.~31, no.~3, pp. 133--139, 2024.

\bibitem{ref_AoSA_training}
C.~Lin, G.~Y. Li, and L.~Wang, ``{Subarray-based Coordinated Beamforming
  Training for mmWave and sub-THz Communications},'' \emph{IEEE J. Sel. Areas
  Commun.}, vol.~35, no.~9, pp. 2115--2126, 2017.

\bibitem{ref_3D_channel}
C.~Han and I.~F. Akyildiz, ``{Three-Dimensional End-to-End Modeling and
  Analysis for Graphene-Enabled Terahertz Band Communications},'' \emph{IEEE
  Trans. Veh. Technol.}, vol.~66, no.~7, pp. 5626--5634, 2017.

\bibitem{ref_Channel_smart_radio}
K.~{Guan}, G.~{Li}, T.~{Kürner}, A.~F. {Molisch}, B.~{Peng}, R.~{He},
  B.~{Hui}, J.~{Kim}, and Z.~{Zhong}, ``{On Millimeter Wave and THz Mobile
  Radio Channel for Smart Rail Mobility},'' \emph{IEEE Trans. Veh. Technol.},
  vol.~66, no.~7, pp. 5658--5674, 2017.

\bibitem{ref_WSMS}
L.~Yan, Y.~Chen, C.~Han, and J.~Yuan, ``{Joint Inter-path and Intra-path
  Multiplexing for Terahertz Widely-spaced Multi-subarray Hybrid Beamforming
  Systems},'' \emph{IEEE Trans. Commun.}, vol.~70, no.~2, pp. 1407--1422, 2022.

\bibitem{ref_hybrid_Channel_Estimation}
X.~Wei and L.~Dai, ``{Channel Estimation for Extremely Large-Scale Massive
  MIMO: Far-Field, Near-Field, or Hybrid-Field?}'' \emph{IEEE Communications
  Letters}, vol.~26, no.~1, pp. 177--181, 2022.

\bibitem{ref_HuHybrid_Field_Channel_Estimation}
Z.~Hu, C.~Chen, Y.~Jin, L.~Zhou, and Q.~Wei, ``{Hybrid-Field Channel Estimation
  for Extremely Large-Scale Massive MIMO System},'' \emph{IEEE Communications
  Letters}, vol.~27, no.~1, pp. 303--307, 2023.

\bibitem{IEEE_Std_802_15_3c}
``{IEEE Standard for Information Technology--Local and Metropolitan Area
  Networks--Specific Requirements--Part 15.3: Amendment 2:
  Millimeter-wave-based Alternative Physical Layer Extension},'' pp. 1--200,
  Oct. 2009, iEEE Std. 802.15.3c-2009.

\bibitem{ref_QUPA}
B.~Ning, Z.~Chen, Z.~Tian, C.~Han, and S.~Li, ``{A Unified 3D Beam Training and
  Tracking Procedure for Terahertz Communication},'' \emph{IEEE Trans. Wireless
  Commun.}, vol.~21, no.~4, pp. 2445--2461, April 2022.

\bibitem{ref_Steering_focusing}
H.~Zhang, N.~Shlezinger, F.~Guidi, D.~Dardari, and Y.~C. Eldar, ``{6G Wireless
  Communications: From Far-Field Beam Steering to Near-Field Beam Focusing},''
  \emph{IEEE Communications Magazine}, vol.~61, no.~4, pp. 72--77, 2023.

\bibitem{ref_codebookdesign_Dai}
X.~Wei, L.~Dai, Y.~Zhao, G.~Yu, and X.~Duan, ``{Codebook design and beam
  training for extremely large-scale RIS: Far-field or near-field?}''
  \emph{China Commun.}, vol.~19, no.~6, pp. 193--204, Jun. 2022.

\bibitem{ref_NearorFar}
M.~Cui and L.~Dai, ``{Channel Estimation for Extremely Large-Scale MIMO:
  Far-Field or Near-Field?}'' \emph{IEEE Transactions on Communications},
  vol.~70, no.~4, pp. 2663--2677, 2022.

\bibitem{ref_near_training_UCA}
Y.~Xie, B.~Ning, L.~Li, and Z.~Chen, ``{Near-Field Beam Training in THz
  Communications: The Merits of Uniform Circular Array},'' \emph{IEEE Wireless
  Commun. Lett.}, vol.~12, no.~4, pp. 575--579, April 2023.

\bibitem{ref_hybrid_beamforming}
C.~Han, L.~Yan, and J.~Yuan, ``{Hybrid Beamforming for Terahertz Wireless
  Communications: Challenges, Architectures, and Open Problems},'' \emph{IEEE
  Wireless Commun.}, vol.~28, no.~4, pp. 198--204, Aug. 2021.

\bibitem{ref_zhang2017codebook}
J.~Zhang, Y.~Huang, Q.~Shi, J.~Wang, and L.~Yang, ``{Codebook design for beam
  alignment in millimeter wave communication systems},'' \emph{IEEE
  Transactions on Communications}, vol.~65, no.~11, pp. 4980--4995, 2017.

\bibitem{ref_Song2017Codebook}
J.~Song, J.~Choi, and D.~J. Love, ``{Common Codebook Millimeter Wave Beam
  Design: Designing Beams for Both Sounding and Communication With Uniform
  Planar Arrays},'' \emph{IEEE Transactions on Communications}, vol.~65, no.~4,
  pp. 1859--1872, 2017.

\bibitem{ref_xiao2016hierarchical}
Z.~Xiao, T.~He, P.~Xia, and X.-G. Xia, ``{Hierarchical codebook design for
  beamforming training in millimeter-wave communication},'' \emph{IEEE
  Transactions on Wireless Communications}, vol.~15, no.~5, pp. 3380--3392,
  2016.

\bibitem{ref_NF_Hie_Training}
Y.~Lu, Z.~Zhang, and L.~Dai, ``{Hierarchical Beam Training for Extremely
  Large-Scale MIMO: From Far-Field to Near-Field},'' \emph{IEEE Transactions on
  Communications}, vol.~72, no.~4, pp. 2247--2259, 2024.

\bibitem{ref_Codebook_NearorFar}
X.~Zhang, H.~Zhang, J.~Zhang, C.~Li, Y.~Huang, and L.~Yang, ``{Codebook Design
  for Extremely Large-Scale MIMO Systems: Near-Field and Far-Field},''
  \emph{IEEE Transactions on Communications}, vol.~72, no.~2, pp. 1191--1206,
  2024.

\bibitem{ref_Near_TWO_Training}
C.~Wu, C.~You, Y.~Liu, L.~Chen, and S.~Shi, ``Two-stage hierarchical beam
  training for near-field communications,'' \emph{IEEE Transactions on
  Vehicular Technology}, vol.~73, no.~2, pp. 2032--2044, 2024.

\bibitem{ref_fast_near_training}
Y.~Zhang, X.~Wu, and C.~You, ``{Fast Near-Field Beam Training for Extremely
  Large-Scale Array},'' \emph{IEEE Wireless Commun. Lett.}, vol.~11, no.~12,
  pp. 2625--2629, Dec. 2022.

\bibitem{ref_NF_Training_Sparse_DFT}
C.~Zhou, C.~Wu, C.~You, J.~Zhou, and S.~Shi, ``{Near-field Beam Training with
  Sparse DFT Codebook},'' \emph{IEEE Transactions on Communications}, pp. 1--1,
  to appear 2024.

\bibitem{ref_NF_Training_2learning}
C.~Weng, X.~Guo, and Y.~Wang, ``{Near-Field Beam Training With Hierarchical
  Codebook: Two-Stage Learning-Based Approach},'' \emph{IEEE Transactions on
  Vehicular Technology}, vol.~73, no.~9, pp. 14\,003--14\,008, 2024.

\bibitem{ref_eltraining_near}
X.~Shi, J.~Wang, Z.~Sun, and J.~Song, ``Spatial-chirp codebook-based
  hierarchical beam training for extremely large-scale massive mimo,''
  \emph{IEEE Transactions on Wireless Communications}, vol.~23, no.~4, pp.
  2824--2838, 2024.

\bibitem{ref_efficient_Hybrid}
J.~Luo, J.~Fan, K.~Xie, and X.~Shi, ``{Efficient Hybrid Near- and Far-Field
  Beam Training for XL-MIMO Communications},'' \emph{IEEE Transactions on
  Vehicular Technology}, vol.~73, no.~12, pp. 19\,785--19\,790, 2024.

\bibitem{ref_triple_training}
K.~Chen, C.~Qi, O.~A. Dobre, and G.~Ye~Li, ``{Triple-Refined Hybrid-Field Beam
  Training for mmWave Extremely Large-Scale MIMO},'' \emph{IEEE Transactions on
  Wireless Communications}, vol.~23, no.~8, pp. 8556--8570, 2024.

\bibitem{ref_wideband_tracking_dai}
J.~Tan and L.~Dai, ``Wideband beam tracking in thz massive mimo systems,''
  \emph{IEEE J. Sel. Areas Commun.}, vol.~39, no.~6, pp. 1693--1710, 2021.

\bibitem{ref_NF_rainbow}
M.~Cui, L.~Dai, Z.~Wang, S.~Zhou, and N.~Ge, ``{Near-Field Rainbow: Wideband
  Beam Training for XL-MIMO},'' \emph{IEEE Transactions on Wireless
  Communications}, vol.~22, no.~6, pp. 3899--3912, 2023.

\bibitem{ref_simu_training_Zhang}
R.~Zhang, H.~Zhang, W.~Xu, and C.~Zhao, ``{A Codebook Based Simultaneous Beam
  Training for mmWave Multi-User MIMO Systems with Split Structures},'' in
  \emph{Proc. of IEEE Global Communications Conference (GLOBECOM)}, 2018, pp.
  1--6.

\bibitem{ref_hie_Qi}
C.~Qi, K.~Chen, O.~A. Dobre, and G.~Y. Li, ``{Hierarchical Codebook-Based
  Multiuser Beam Training for Millimeter Wave Massive MIMO},'' \emph{IEEE
  Transactions on Wireless Communications}, vol.~19, no.~12, pp. 8142--8152,
  2020.

\bibitem{ref_wang2024fast}
H.~Wang, J.~Fang, H.~Duan, and H.~Li, ``{Fast Hybrid Far/Near-Field Beam
  Training For Extremely Large-Scale Millimeter Wave/Terahertz Systems},''
  \emph{IEEE Transactions on Communications}, to appear 2024.

\bibitem{ref_coded_training}
T.~Zheng, J.~Zhu, Q.~Yu, Y.~Yan, and L.~Dai, ``{Coded Beam Training},''
  \emph{IEEE Journal on Selected Areas in Communications}, vol.~43, no.~3, pp.
  928--943, 2025.

\bibitem{ref_Reinforcement_Learning_codebook}
Y.~Zhang, M.~Alrabeiah, and A.~Alkhateeb, ``{Reinforcement Learning of Beam
  Codebooks in Millimeter Wave and Terahertz MIMO Systems},'' \emph{IEEE Trans.
  Commun.}, vol.~70, no.~2, pp. 904--919, Feb. 2022.

\bibitem{ref_neural_codebook}
R.~M. Dreifuerst and R.~W. Heath, ``{Neural Codebook Design for MIMO Network
  Beam Management},'' \emph{IEEE Transactions on Wireless Communications}, pp.
  1--1, to appear 2025.

\bibitem{ref_DRL_sweeping}
J.~Che, Z.~Zhang, Y.~Yang, and Z.~Yang, ``{Efficient Initial Access Based on
  DRL-Empowered Beam Sweeping},'' \emph{IEEE Transactions on Wireless
  Communications}, to appear 2025.

\bibitem{ref_meta_alignment}
Z.~Xu, S.~Wang, and Y.-J.~A. Zhang, ``Scenario-adaptive meta-learning for
  mmwave beam alignment,'' \emph{IEEE Transactions on Wireless Communications},
  to appear 2025.

\bibitem{ref_Near_DL_training}
G.~Jiang and C.~Qi, ``{Near-Field Beam Training Based on Deep Learning for
  Extremely Large-Scale MIMO},'' \emph{IEEE Communications Letters}, vol.~27,
  no.~8, pp. 2063--2067, 2023.

\bibitem{ref_NF_Side_Training}
L.~Liu, C.~You, Y.~Zhang, and T.~Liu, ``{Side Angle Information Assisted
  Near-Field Beam Training for XL-Array Communications},'' \emph{IEEE
  Communications Letters}, vol.~28, no.~9, pp. 2201--2205, 2024.

\bibitem{ref_CV_alignment}
Y.~Ahn, J.~Kim, S.~Kim, K.~Shim, J.~Kim, S.~Kim, and B.~Shim, ``Toward
  intelligent millimeter and terahertz communication for 6g: Computer
  vision-aided beamforming,'' \emph{IEEE Wireless Commun.}, vol.~30, no.~5, pp.
  179--186, Oct. 2023.

\bibitem{ref_fast_training}
P.~Wang, J.~Fang, W.~Zhang, and H.~Li, ``{Fast Beam Training and Alignment for
  IRS-Assisted Millimeter Wave/Terahertz Systems},'' \emph{IEEE Transactions on
  Wireless Communications}, vol.~21, no.~4, pp. 2710--2724, 2022.

\bibitem{ref_Root_MUSIC_HDAPA}
D.~Hu, Y.~Zhang, L.~He, and J.~Wu, ``{Low-Complexity Deep-Learning-Based DOA
  Estimation for Hybrid Massive MIMO Systems With Uniform Circular Arrays},''
  \emph{IEEE Wireless Communications Letters}, vol.~9, no.~1, pp. 83--86, 2020.

\bibitem{ref_milli}
Y.~Chen, L.~Yan, C.~Han, and M.~Tao, ``{Millidegree-Level Direction-of-Arrival
  Estimation and Tracking for Terahertz Ultra-Massive MIMO Systems},''
  \emph{IEEE Transactions on Wireless Communications}, vol.~21, no.~2, pp.
  869--883, 2022.

\bibitem{ref_cross_training}
Y.~Chen, C.~Han, and E.~Bj{\"o}rnson, ``{Can Far-Field Beam Training Be
  Deployed for Cross-Field Beam Alignment in Terahertz UM-MIMO
  Communications?}'' \emph{IEEE Transactions on Wireless Communications},
  vol.~23, no.~10, pp. 14\,972--14\,987, 2024.

\bibitem{ref_DL_NFTraining}
W.~Liu, H.~Ren, C.~Pan, and J.~Wang, ``{Deep Learning Based Beam Training for
  Extremely Large-Scale Massive MIMO in Near-Field Domain},'' \emph{IEEE
  Communications Letters}, vol.~27, no.~1, pp. 170--174, 2023.

\bibitem{ref_MU_NF_BT}
W.~Liu, C.~Pan, H.~Ren, J.~Wang, and R.~Schober, ``{Near-Field Multiuser
  Beam-Training for Extremely Large-Scale MIMO Systems},'' \emph{IEEE
  Transactions on Communications}, pp. 1--1, 2024.

\bibitem{ref_NF_Training_DFT}
X.~Wu, C.~You, J.~Li, and Y.~Zhang, ``{Near-Field Beam Training: Joint Angle
  and Range Estimation With DFT Codebook},'' \emph{IEEE Transactions on
  Wireless Communications}, vol.~23, no.~9, pp. 11\,890--11\,903, 2024.

\bibitem{ref_RIS_NF_lowOverhead_Training}
W.~Liu, C.~Pan, H.~Ren, F.~Shu, S.~Jin, and J.~Wang, ``{Low-Overhead Beam
  Training Scheme for Extremely Large-Scale RIS in Near Field},'' \emph{IEEE
  Transactions on Communications}, vol.~71, no.~8, pp. 4924--4940, 2023.

\bibitem{ref_OMP}
A.~Alkhateeb, O.~El~Ayach, G.~Leus, and R.~W. Heath, ``{Channel Estimation and
  Hybrid Precoding for Millimeter Wave Cellular Systems},'' \emph{IEEE Journal
  of Selected Topics in Signal Processing}, vol.~8, no.~5, pp. 831--846, 2014.

\bibitem{CS_OMP}
R.~Méndez-Rial, C.~Rusu, N.~González-Prelcic, A.~Alkhateeb, and R.~W. Heath,
  ``{Hybrid MIMO Architectures for Millimeter Wave Communications: Phase
  Shifters or Switches?}'' \emph{IEEE Access}, vol.~4, pp. 247--267, 2016.

\bibitem{ref_OMP_THz}
K.~Dovelos, M.~Matthaiou, H.~Q. Ngo, and B.~Bellalta, ``{Channel Estimation and
  Hybrid Combining for Wideband Terahertz Massive MIMO Systems},'' \emph{IEEE
  Journal on Selected Areas in Communications}, vol.~39, no.~6, pp. 1604--1620,
  June 2021.

\bibitem{AMP}
D.~L. Donoho, A.~Javanmard, and A.~Montanari, ``{Information-Theoretically
  Optimal Compressed Sensing via Spatial Coupling and Approximate Message
  Passing},'' \emph{IEEE Transactions on Information Theory}, vol.~59, no.~11,
  pp. 7434--7464, Nov. 2013.

\bibitem{ref_wideband_CE}
K.~Dovelos, M.~Matthaiou, H.~Q. Ngo, and B.~Bellalta, ``{Channel Estimation and
  Hybrid Combining for Wideband Terahertz Massive MIMO Systems},'' \emph{IEEE
  Journal on Selected Areas in Communications}, vol.~39, no.~6, pp. 1604--1620,
  June 2021.

\bibitem{ref_BSA_OMP}
A.~M. Elbir and S.~Chatzinotas, ``Bsa-omp: Beam-split-aware orthogonal matching
  pursuit for thz channel estimation,'' \emph{IEEE Wireless Communications
  Letter}, vol.~12, no.~4, pp. 738--742, April 2023.

\bibitem{ref_IRS_CE_CE}
Z.~Wan, Z.~Gao, F.~Gao, M.~D. Renzo, and M.-S. Alouini, ``{Terahertz Massive
  MIMO With Holographic Reconfigurable Intelligent Surfaces},'' \emph{IEEE
  Transations on Communications}, vol.~69, no.~7, pp. 4732--4750, Jul. 2021.

\bibitem{ref_trice}
K.~Ardah, S.~Gherekhloo, A.~L.~F. de~Almeida, and M.~Haardt, ``{TRICE: A
  Channel Estimation Framework for RIS-Aided Millimeter-Wave MIMO Systems},''
  \emph{IEEE Signal Processing Letter}, vol.~28, pp. 513--517, Feb. 2021.

\bibitem{ref_OMP_IRS}
P.~Wang, J.~Fang, H.~Duan, and H.~Li, ``{Compressed Channel Estimation for
  Intelligent Reflecting Surface-Assisted Millimeter Wave Systems},''
  \emph{IEEE Signal Processing Letter}, vol.~27, pp. 905--909, May 2020.

\bibitem{ref_IRS_CE_CS_THz}
X.~Ma, Z.~Chen, W.~Chen, Z.~Li, Y.~Chi, C.~Han, and S.~Li, ``{Joint Channel
  Estimation and Data Rate Maximization for Intelligent Reflecting Surface
  Assisted Terahertz MIMO Communication Systems},'' \emph{IEEE Access}, vol.~8,
  pp. 99\,565--99\,581, 2020.

\bibitem{ref_IRSCE_sparsity}
T.~Lin, X.~Yu, Y.~Zhu, and R.~Schober, ``{Channel Estimation for IRS-Assisted
  Millimeter-Wave MIMO Systems: Sparsity-Inspired Approaches},'' \emph{IEEE
  Transactions on Communications}, vol.~70, no.~6, pp. 4078--4092, April 2022.

\bibitem{ref_triple}
X.~Shi, J.~Wang, and J.~Song, ``{Triple-Structured Compressive Sensing-Based
  Channel Estimation for RIS-Aided MU-MIMO Systems},'' \emph{IEEE Transactions
  on Wireless Communications}, vol.~21, no.~12, pp. 11\,095--11\,109, Dec.
  2022.

\bibitem{ref_NFCE_JSAC}
S.~Yang, C.~Xie, W.~Lyu, B.~Ning, Z.~Zhang, and C.~Yuen, ``{Near-Field Channel
  Estimation for Extremely Large-Scale Reconfigurable Intelligent Surface
  (XL-RIS)-Aided Wideband mmWave Systems},'' \emph{IEEE Journal on Selected
  Areas in Communications}, vol.~42, no.~6, pp. 1567--1582, 2024.

\bibitem{ref_DSE_SSE}
Y.~Chen, R.~Li, C.~Han, S.~Sun, and M.~Tao, ``{Hybrid Spherical- and
  Planar-Wave Channel Modeling and Estimation for Terahertz Integrated UM-MIMO
  and IRS Systems},'' \emph{IEEE Transactions on Wireless Communications},
  vol.~22, no.~12, pp. 9746--9761, 2023.

\bibitem{ref_Hybrid_CE}
S.~Yue, S.~Zeng, L.~Liu, Y.~C. Eldar, and B.~Di, ``{Hybrid Near-Far Field
  Channel Estimation for Holographic MIMO Communications},'' \emph{IEEE
  Transactions on Wireless Communications}, vol.~23, no.~11, pp.
  15\,798--15\,813, 2024.

\bibitem{ref_Cross_CE}
S.~Tarboush, A.~Ali, and T.~Y. Al-Naffouri, ``{Cross-Field Channel Estimation
  for Ultra Massive-MIMO THz Systems},'' \emph{IEEE Transactions on Wireless
  Communications}, vol.~23, no.~8, pp. 8619--8635, 2024.

\bibitem{ref_Super_ChenHu}
C.~{Hu}, L.~{Dai}, T.~{Mir}, Z.~{Gao}, and J.~{Fang}, ``{Super-Resolution
  Channel Estimation for MmWave Massive MIMO With Hybrid Precoding},''
  \emph{IEEE Transactions on Vehicular Technology}, vol.~67, no.~9, pp.
  8954--8958, Sept. 2018.

\bibitem{ref_EM_Bayesian}
W.~{Shao}, S.~{Zhang}, X.~{Zhang}, J.~{Ma}, N.~{Zhao}, and V.~C.~M. {Leung},
  ``{Massive MIMO Channel Estimation Over the mmWave Systems Through Parameters
  Learning},'' \emph{IEEE Communications Letters}, vol.~23, no.~4, pp.
  672--675, Apr. 2019.

\bibitem{EM_SAGE3}
K.~{Mawatwal}, D.~{Sen}, and R.~{Roy}, ``{A Semi-Blind Channel Estimation
  Algorithm for Massive MIMO Systems},'' \emph{IEEE Wireless Communications
  Letters}, vol.~6, no.~1, pp. 70--73, Feb. 2017.

\bibitem{ref_IRS_CE_ANM}
J.~He, H.~Wymeersch, and M.~Juntti, ``{Channel Estimation for RIS-Aided mmWave
  MIMO Systems via Atomic Norm Minimization},'' \emph{IEEE Transactions on
  Wireless Communications}, vol.~20, no.~9, pp. 5786--5797, Sept. 2021.

\bibitem{2D_B_MUSIC}
Z.~Guo, X.~Wang, and W.~Heng, ``{Millimeter-wave Channel Estimation based on
  2-D Beamspace MUSIC Method},'' \emph{IEEE Transactions on Wireless
  Communications}, vol.~16, no.~8, pp. 5384--5394, Aug. 2017.

\bibitem{2D_Unitary_ESPRIT}
A.~{Liao}, Z.~{Gao}, Y.~{Wu}, H.~{Wang}, and M.~{Alouini}, ``{2D} {Unitary}
  {ESPRIT} {Based} {Super-Resolution} {Channel} {Estimation} for
  {Millimeter-Wave} {Massive} {MIMO} {With Hybrid Precoding},'' \emph{IEEE
  Access}, vol.~5, pp. 24\,747--24\,757, Nov. 2017.

\bibitem{ref_ESPRIT_OFDM}
W.~{Ma}, C.~{Qi}, and G.~Y. {Li}, ``{High-Resolution Channel Estimation for
  Frequency-Selective mmWave Massive MIMO Systems},'' \emph{IEEE Trans.
  Wireless Commun.}, vol.~19, no.~5, pp. 3517--3529, May 2020.

\bibitem{LDAMP}
H.~{He}, C.~{Wen}, S.~{Jin}, and G.~Y. {Li}, ``{Deep Learning-Based Channel
  Estimation for Beamspace mmWave Massive MIMO Systems},'' \emph{IEEE Wireless
  Communications Letters}, vol.~7, no.~5, pp. 852--855, Oct. 2018.

\bibitem{ref_GM_LAMP}
X.~{Wei}, C.~{Hu}, and L.~{Dai}, ``{Deep Learning for Beamspace Channel
  Estimation in Millimeter-Wave Massive MIMO Systems},'' \emph{IEEE
  Transactions on Communications}, vol.~69, no.~1, pp. 182--193, Jan. 2021.

\bibitem{ref_hybrid_THz_CE}
W.~Yu, Y.~Shen, H.~He, X.~Yu, J.~Zhang, and K.~B. Letaief, ``{Hybrid Far- and
  Near-Field Channel Estimation for THz Ultra-Massive MIMO via Fixed Point
  Networks},'' in \emph{Proc. of IEEE Global Commun. Conf.}, Rio de Janeiro,
  Brazil, 2022, pp. 5384--5389.

\bibitem{mmWave_CE_CNN}
P.~{Dong}, H.~{Zhang}, G.~Y. {Li}, I.~S. {Gaspar}, and N.~{NaderiAlizadeh},
  ``{Deep CNN-Based Channel Estimation for mmWave Massive MIMO Systems},''
  \emph{IEEE Journal of Selected Topics in Signal Processing}, vol.~13, no.~5,
  pp. 989--1000, Sept. 2019.

\bibitem{ref_CE_GAN}
E.~Balevi and J.~G. Andrews, ``{Wideband Channel Estimation With a Generative
  Adversarial Network},'' \emph{IEEE Transactions on Wireless Communications},
  vol.~20, no.~5, pp. 3049--3060, Jan. 2021.

\bibitem{ref_sparse_CE_DL}
W.~{Ma}, C.~{Qi}, Z.~{Zhang}, and J.~{Cheng}, ``{Sparse Channel Estimation and
  Hybrid Precoding Using Deep Learning for Millimeter Wave Massive MIMO},''
  \emph{IEEE Transactions on Communications}, vol.~68, no.~5, pp. 2838--2849,
  May 2020.

\bibitem{ref_IRS_CE_DnDL}
S.~Liu, Z.~Gao, J.~Zhang, M.~D. Renzo, and M.-S. Alouini, ``{Deep Denoising
  Neural Network Assisted Compressive Channel Estimation for mmWave Intelligent
  Reflecting Surfaces},'' \emph{IEEE Transactions on Vehicular Technology},
  vol.~69, no.~8, pp. 9223--9228, Aug. 2020.

\bibitem{ref_IRS_CE_active_DL}
A.~Taha, M.~Alrabeiah, and A.~Alkhateeb, ``{Enabling Large Intelligent Surfaces
  With Compressive Sensing and Deep Learning},'' \emph{IEEE Access}, vol.~9,
  pp. 44\,304--44\,321, March 2021.

\bibitem{ref_Multiray}
C.~Han, A.~O. Bicen, and I.~F. Akyildiz, ``{Multi-Ray Channel Modeling and
  Wideband Characterization for Wireless Communications in the Terahertz
  Band},'' \emph{IEEE Trans. Wireless Commun.}, vol.~14, no.~5, pp. 2402--2412,
  2015.

\bibitem{ref_NF_dynamic1}
Z.~Zhang, Y.~Liu, Z.~Wang, J.~Chen, and T.~Q.~S. Quek, ``Dynamic mimo
  architecture design for near-field communications,'' \emph{IEEE Transactions
  on Wireless Communications}, vol.~23, no.~10, pp. 14\,669--14\,684, 2024.

\bibitem{ref_THzPrism}
B.~Zhai, Y.~Zhu, A.~Tang, and X.~Wang, ``{THzPrism: Frequency-Based Beam
  Spreading for Terahertz Communication Systems},'' \emph{IEEE Wireless
  Communications Letters}, vol.~9, no.~6, pp. 897--900, 2020.

\bibitem{ref_DSFTTD}
L.~Yan, C.~Han, and J.~Yuan, ``{Energy-Efficient Dynamic-Subarray With Fixed
  True-Time-Delay Design for Terahertz Wideband Hybrid Beamforming},''
  \emph{IEEE Journal on Selected Areas in Communications}, vol.~40, no.~10, pp.
  2840--2854, 2022.

\bibitem{ref_cross_field1}
J.~Xie, C.~Luo, and Y.~Luo, ``Cross near- and far-field beamforming for
  wideband multi-user terahertz communications,'' \emph{IEEE Communications
  Letters}, vol.~28, no.~10, pp. 2397--2401, 2024.

\bibitem{ref_HDA}
F.~Sohrabi and W.~Yu, ``Hybrid digital and analog beamforming design for
  large-scale antenna arrays,'' \emph{IEEE Journal of Selected Topics in Signal
  Processing}, vol.~10, no.~3, pp. 501--513, 2016.

\bibitem{ref_HAD_EE}
K.~Ardah, G.~Fodor, Y.~C.~B. Silva, W.~C. Freitas, and A.~L.~F. de~Almeida,
  ``Hybrid analog-digital beamforming design for se and ee maximization in
  massive mimo networks,'' \emph{IEEE Transactions on Vehicular Technology},
  vol.~69, no.~1, pp. 377--389, 2020.

\bibitem{ref_AO}
X.~Yu, J.-C. Shen, J.~Zhang, and K.~B. Letaief, ``Alternating minimization
  algorithms for hybrid precoding in millimeter wave mimo systems,'' \emph{IEEE
  Journal of Selected Topics in Signal Processing}, vol.~10, no.~3, pp.
  485--500, 2016.

\bibitem{ref_min_Euclidean2}
Z.~Luo, L.~Zhao, H.~Liu, and Y.~Li, ``Robust hybrid beamforming in millimeter
  wave systems with closed-form least-square solutions,'' \emph{IEEE Wireless
  Communications Letters}, vol.~10, no.~1, pp. 156--160, 2021.

\bibitem{ref_beamsquint3}
A.~Najjar, M.~El-Absi, and T.~Kaiser, ``Hybrid delay-phase precoding in
  wideband um-mimo systems under true time delay and phase shifter hardware
  limitations,'' \emph{IEEE Transactions on Wireless Communications}, vol.~23,
  no.~7, pp. 7246--7262, 2024.

\bibitem{ref_MMSE1}
Q.~Shi, M.~Razaviyayn, Z.-Q. Luo, and C.~He, ``An iteratively weighted mmse
  approach to distributed sum-utility maximization for a mimo interfering
  broadcast channel,'' \emph{IEEE Transactions on Signal Processing}, vol.~59,
  no.~9, pp. 4331--4340, 2011.

\bibitem{ref_MMSE2}
S.~S. Christensen, R.~Agarwal, E.~De~Carvalho, and J.~M. Cioffi, ``Weighted
  sum-rate maximization using weighted mmse for mimo-bc beamforming design,''
  \emph{IEEE Transactions on Wireless Communications}, vol.~7, no.~12, pp.
  4792--4799, 2008.

\bibitem{ref_MMSE_AO1}
J.~Cong, T.~Lin, and Y.~Zhu, ``Hybrid mmse beamforming for multiuser
  millimeter-wave communication systems,'' \emph{IEEE Communications Letters},
  vol.~22, no.~11, pp. 2390--2393, 2018.

\bibitem{ref_MMSE_OFDM}
J.~Du, W.~Xu, C.~Zhao, and L.~Vandendorpe, ``Weighted spectral efficiency
  optimization for hybrid beamforming in multiuser massive mimo-ofdm systems,''
  \emph{IEEE Transactions on Vehicular Technology}, vol.~68, no.~10, pp.
  9698--9712, 2019.

\bibitem{ref_MMSE_AO_IRS}
Q.~Wu and R.~Zhang, ``Intelligent reflecting surface enhanced wireless network
  via joint active and passive beamforming,'' \emph{IEEE Transactions on
  Wireless Communications}, vol.~18, no.~11, pp. 5394--5409, 2019.

\bibitem{ref_non-iterative1}
L.~Liang, W.~Xu, and X.~Dong, ``Low-complexity hybrid precoding in massive
  multiuser mimo systems,'' \emph{IEEE Wireless Communications Letters},
  vol.~3, no.~6, pp. 653--656, 2014.

\bibitem{ref_HBF}
X.~Wu, D.~Liu, and F.~Yin, ``Hybrid beamforming for multi-user massive mimo
  systems,'' \emph{IEEE Transactions on Communications}, vol.~66, no.~9, pp.
  3879--3891, 2018.

\bibitem{ref_BeamFocusing}
H.~Zhang, N.~Shlezinger, F.~Guidi, D.~Dardari, M.~F. Imani, and Y.~C. Eldar,
  ``{Beam Focusing for Near-Field Multiuser MIMO Communications},'' \emph{IEEE
  Transactions on Wireless Communications}, vol.~21, no.~9, pp. 7476--7490,
  2022.

\bibitem{ref_dynamic_NF}
M.~Liu, M.~Li, R.~Liu, and Q.~Liu, ``Dynamic hybrid beamforming designs for
  elaa near-field communications,'' \emph{IEEE Journal on Selected Areas in
  Communications}, vol.~43, no.~3, pp. 644--658, 2025.

\bibitem{ref_Spatially_sparse}
O.~E. Ayach, S.~Rajagopal, S.~Abu-Surra, Z.~Pi, and R.~W. Heath, ``Spatially
  sparse precoding in millimeter wave mimo systems,'' \emph{IEEE Transactions
  on Wireless Communications}, vol.~13, no.~3, pp. 1499--1513, 2014.

\bibitem{ref_wideband_sparse}
Q.~Wan, J.~Fang, Z.~Chen, and H.~Li, ``Hybrid precoding and combining for
  millimeter wave/sub-thz mimo-ofdm systems with beam squint effects,''
  \emph{IEEE Transactions on Vehicular Technology}, vol.~70, no.~8, pp.
  8314--8319, 2021.

\bibitem{ref_HBD}
W.~Ni and X.~Dong, ``Hybrid block diagonalization for massive multiuser mimo
  systems,'' \emph{IEEE Transactions on Communications}, vol.~64, no.~1, pp.
  201--211, 2016.

\bibitem{ref_sparsity1}
Q.~Yue, J.~Hu, T.~Shui, Q.~Huang, K.~Yang, and L.~Hanzo, ``Hybrid terahertz
  beamforming relying on channel sparsity and angular orthogonality,''
  \emph{IEEE Transactions on Vehicular Technology}, vol.~73, no.~4, pp.
  4759--4773, 2024.

\bibitem{ref_frequencyselective}
H.~Yuan, N.~Yang, K.~Yang, C.~Han, and J.~An, ``Hybrid beamforming for
  terahertz multi-carrier systems over frequency selective fading,'' \emph{IEEE
  Transactions on Communications}, vol.~68, no.~10, pp. 6186--6199, 2020.

\bibitem{ref_splitmultiplexing2}
B.~Ning, L.~Li, W.~Chen, and Z.~Chen, ``Wideband terahertz communications with
  aosa: Beam split aggregation and multiplexing,'' in \emph{GLOBECOM 2022 -
  2022 IEEE Global Communications Conference}, 2022, pp. 1709--1714.

\bibitem{ref_splitmultiplexing1}
J.~Wang and M.~Kaneko, ``Exploiting beam split-based multi-user diversity in
  terahertz mimo-ofdm systems,'' \emph{IEEE Wireless Communications Letters},
  vol.~14, no.~1, pp. 28--32, 2025.

\bibitem{ref_partial_1}
L.~Jiang and H.~Jafarkhani, ``Multi-user analog beamforming in millimeter wave
  mimo systems based on path angle information,'' \emph{IEEE Transactions on
  Wireless Communications}, vol.~18, no.~1, pp. 608--619, 2019.

\bibitem{ref_partial_2}
L.~Yan, C.~Han, N.~Yang, and J.~Yuan, ``Dynamic-subarray with fixed phase
  shifters for energy-efficient terahertz hybrid beamforming under partial
  csi,'' \emph{IEEE Transactions on Wireless Communications}, vol.~22, no.~5,
  pp. 3231--3245, 2023.

\bibitem{ref_Dai_LDMA}
Z.~Wu and L.~Dai, ``{Multiple Access for Near-Field Communications: SDMA or
  LDMA?}'' \emph{IEEE Journal on Selected Areas in Communications}, vol.~41,
  no.~6, pp. 1918--1935, 2023.

\bibitem{ref_NF_WideBF}
M.~Cui and L.~Dai, ``{Near-Field Wideband Beamforming for Extremely Large
  Antenna Arrays},'' \emph{IEEE Transactions on Wireless Communications},
  vol.~23, no.~10, pp. 13\,110--13\,124, 2024.

\bibitem{ref_cross_beamforming2}
H.~Shen, Y.~Chen, C.~Han, and J.~Yuan, ``{Hybrid Beamforming with
  Widely-spaced-array for Multi-user Cross-Near-and-Far-Field
  Communications},'' \emph{IEEE Transactions on Communications}, pp. 1--1, to
  appear 2025.

\bibitem{ref_FieldawareCF}
T.~Gao, Y.~Song, C.~Liu, Z.~Yin, N.~Cheng, and D.~Ge, ``Hybrid-field-aware
  two-stage beamforming scheme for xl-mimo systems,'' \emph{IEEE Internet of
  Things Journal}, pp. 1--1, 2025.

\bibitem{ref_twenty_five}
A.~M. Elbir, K.~V. Mishra, S.~A. Vorobyov, and R.~W. Heath, ``Twenty-five years
  of advances in beamforming: From convex and nonconvex optimization to
  learning techniques,'' \emph{IEEE Signal Processing Magazine}, vol.~40,
  no.~4, pp. 118--131, 2023.

\bibitem{ref_learning_fast}
H.~Huang, Y.~Peng, J.~Yang, W.~Xia, and G.~Gui, ``Fast beamforming design via
  deep learning,'' \emph{IEEE Transactions on Vehicular Technology}, vol.~69,
  no.~1, pp. 1065--1069, 2020.

\bibitem{ref_learning_imperfect1}
P.~Zhang, L.~Pan, T.~Laohapensaeng, and M.~Chongcheawchamnan, ``Hybrid
  beamforming based on an unsupervised deep learning network for downlink
  channels with imperfect csi,'' \emph{IEEE Wireless Communications Letters},
  vol.~11, no.~7, pp. 1543--1547, 2022.

\bibitem{ref_learning_imperfect2}
Z.~Zhang, M.~Tao, and Y.-F. Liu, ``Learning to beamform in joint multicast and
  unicast transmission with imperfect csi,'' \emph{IEEE Transactions on
  Communications}, vol.~71, no.~5, pp. 2711--2723, 2023.

\bibitem{ref_learning_model3}
L.~Schynol, M.~Hemsing, and M.~Pesavento, ``Codebook-based downlink beamforming
  with imperfect csi using model-driven deep learning,'' in \emph{2024 58th
  Asilomar Conference on Signals, Systems, and Computers}, 2024, pp. 583--590.

\bibitem{ref_learning_model2}
H.~He, M.~Zhang, S.~Jin, C.-K. Wen, and G.~Y. Li, ``Model-driven deep learning
  for massive mu-mimo with finite-alphabet precoding,'' \emph{IEEE
  Communications Letters}, vol.~24, no.~10, pp. 2216--2220, 2020.

\bibitem{ref_learning_model4}
Y.~Long, Z.~Chen, J.~Fang, and C.~Tellambura, ``Data-driven-based analog beam
  selection for hybrid beamforming under mm-wave channels,'' \emph{IEEE Journal
  of Selected Topics in Signal Processing}, vol.~12, no.~2, pp. 340--352, 2018.

\bibitem{ref_learning_model1}
H.~Ting, Z.~Wang, and Y.~Liu, ``Adaptive ttd configurations for near-field
  communications: An unsupervised transformer approach,'' \emph{IEEE
  Transactions on Wireless Communications}, vol.~24, no.~1, pp. 277--292, 2025.

\bibitem{ref_airy1}
V.~Petrov, H.~Guerboukha, A.~Singh, and J.~M. Jornet, ``Wavefront hopping for
  physical layer security in 6g and beyond near-field thz communications,''
  \emph{IEEE Transactions on Communications}, pp. 1--1, 2024.

\bibitem{ref_OAM}
A.~Mehra and S.~Arnon, ``Optimal receiver design for orbital angular momentum
  (oam) communication via sampled beam analysis,'' \emph{Journal of Lightwave
  Technology}, pp. 1--10, 2025.

\bibitem{ref_highspeed1}
J.~Feng, B.~Zheng, C.~You, X.~Xiong, J.~Tang, F.~Chen, and R.~Zhang,
  ``Irs-aided wireless relaying for high-speed train communication: Beamforming
  design and channel estimation,'' \emph{IEEE Transactions on Wireless
  Communications}, vol.~23, no.~12, pp. 18\,380--18\,393, 2024.

\bibitem{ref_highspeed2}
H.~Tong, T.~Cheng, X.~Wang, T.~Li, X.~Su, and Y.~Xu, ``Near-field beamforming
  and doppler compensation for mmwave internet of vehicles,'' \emph{IEEE
  Communications Letters}, vol.~28, no.~3, pp. 702--706, 2024.

\bibitem{ref_Beam_tracking_EKF}
V.~Va, H.~Vikalo, and R.~W. Heath, ``{Beam tracking for mobile millimeter wave
  communication systems},'' in \emph{Proc. of IEEE Global Conference on Signal
  and Information Processing}, 2016, pp. 743--747.

\bibitem{ref_Beam_tracking_EKF2}
S.~Jayaprakasam, X.~Ma, J.~W. Choi, and S.~Kim, ``{Robust Beam-Tracking for
  mmWave Mobile Communications},'' \emph{IEEE Communications Letters}, vol.~21,
  no.~12, pp. 2654--2657, 2017.

\bibitem{ref_tracking_EKF3}
F.~Liu, P.~Zhao, and Z.~Wang, ``{EKF-Based Beam Tracking for mmWave MIMO
  Systems},'' \emph{IEEE Communications Letters}, vol.~23, no.~12, pp.
  2390--2393, 2019.

\bibitem{ref_Two_Timescale_NF_tracking}
S.~Palmucci, A.~Guerra, A.~Abrardo, and D.~Dardari, ``{Two-Timescale Joint
  Precoding Design and RIS Optimization for User Tracking in Near-Field MIMO
  Systems},'' \emph{IEEE Transactions on Signal Processing}, vol.~71, pp.
  3067--3082, 2023.

\bibitem{ref_ELAA_TrainingTracking}
K.~Chen, C.~Qi, C.-X. Wang, and G.~Y. Li, ``{Beam Training and Tracking for
  Extremely Large-Scale MIMO Communications},'' \emph{IEEE Transactions on
  Wireless Communications}, vol.~23, no.~5, pp. 5048--5062, 2024.

\bibitem{ref_tracking_UKF}
S.~G. Larew and D.~J. Love, ``{Adaptive Beam Tracking With the Unscented Kalman
  Filter for Millimeter Wave Communication},'' \emph{IEEE Signal Processing
  Letters}, vol.~26, no.~11, pp. 1658--1662, 2019.

\bibitem{ref_beamwidth_control}
H.~Chung, J.~Kang, H.~Kim, Y.~M. Park, and S.~Kim, ``{Adaptive Beamwidth
  Control for mmWave Beam Tracking},'' \emph{IEEE Communications Letters},
  vol.~25, no.~1, pp. 137--141, 2021.

\bibitem{ref_tracking_nonlinear}
J.~Lim, H.-M. Park, and D.~Hong, ``{Beam Tracking Under Highly Nonlinear Mobile
  Millimeter-Wave Channel},'' \emph{IEEE Communications Letters}, vol.~23,
  no.~3, pp. 450--453, 2019.

\bibitem{ref_NF_Tracking_limit}
A.~Guerra, F.~Guidi, D.~Dardari, and P.~M. Djurić, ``{Near-Field Tracking With
  Large Antenna Arrays: Fundamental Limits and Practical Algorithms},''
  \emph{IEEE Transactions on Signal Processing}, vol.~69, pp. 5723--5738, 2021.

\bibitem{ref_POMDP_far}
M.~Hussain and N.~Michelusi, ``{Learning and Adaptation for Millimeter-Wave
  Beam Tracking and Training: A Dual Timescale Variational Framework},''
  \emph{IEEE Journal on Selected Areas in Communications}, vol.~40, no.~1, pp.
  37--53, 2022.

\bibitem{ref_POMDP_Tracking}
N.~Ronquillo, C.-S. Gau, and T.~Javidi, ``{Integrated Beam Tracking and
  Communication for (Sub-)mmWave Links With Stochastic Mobility},'' \emph{IEEE
  Journal on Selected Areas in Information Theory}, vol.~4, pp. 94--111, 2023.

\bibitem{ref_Learning_tracking}
S.~H. Lim, S.~Kim, B.~Shim, and J.~W. Choi, ``{Deep Learning-Based Beam
  Tracking for Millimeter-Wave Communications Under Mobility},'' \emph{IEEE
  Transactions on Communications}, vol.~69, no.~11, pp. 7458--7469, 2021.

\bibitem{ref_KFLSTM_tracking}
L.~Yan, X.~Fang, Y.~Fang, L.~Hao, Q.~Xue, and C.~Xu, ``{KF-LSTM Based Beam
  Tracking for UAV-Assisted mmWave HSR Wireless Networks},'' \emph{IEEE
  Transactions on Vehicular Technology}, vol.~71, no.~10, pp. 10\,796--10\,807,
  2022.

\bibitem{ref_AI_BT_NF}
M.~Zhang, R.~Zhong, X.~Mu, and Y.~Liu, ``{AI-Empowered Beam Tracking for
  Near-Field Communications},'' in \emph{Proc. of IEEE International Conference
  on Communications}, 2024, pp. 1643--1648.

\bibitem{ref_NF_AI_Tracking}
------, ``{AI-Empowered Beam Tracking for Near-Field Communications},'' in
  \emph{Proc. of IEEE International Conference on Communications}, 2024, pp.
  1643--1648.

\bibitem{ref_Daul_path_tracking}
R.~Wang, C.~She, C.~Li, Y.~Li, and B.~Vucetic, ``{Dual-Path Beam Tracking for
  Service Continuity of Ultra-Reliable and Low-Latency Communications},''
  \emph{IEEE Transactions on Communications}, vol.~73, no.~1, pp. 524--539,
  2025.

\bibitem{ref_beam_pattern_selection}
J.~Jeong, S.~H. Lim, Y.~Song, and S.-W. Jeon, ``{Online Learning for Joint Beam
  Tracking and Pattern Optimization in Massive MIMO Systems},'' in \emph{Proc.
  of IEEE Conference on Computer Communications}, 2020, pp. 764--773.

\bibitem{ref_tracking_Q}
H.~Park, J.~Kang, S.~Lee, J.~W. Choi, and S.~Kim, ``{Deep Q-Network Based Beam
  Tracking for Mobile Millimeter-Wave Communications},'' \emph{IEEE
  Transactions on Wireless Communications}, vol.~22, no.~2, pp. 961--971, 2023.

\bibitem{ref_Qlearning_THzTracking}
H.~Park, H.~Chung, A.~Conti, M.~Z. Win, and S.~Kim, ``{Robust Near-field Beam
  Tracking via Deep Q-network for THz Communications},'' in \emph{Proc. of
  International Conference on Information Fusion}, 2024, pp. 1--5.

\bibitem{ref_MAMBA}
I.~Aykin, B.~Akgun, M.~Feng, and M.~Krunz, ``{MAMBA: A Multi-armed Bandit
  Framework for Beam Tracking in Millimeter-wave Systems},'' in \emph{IEEE
  INFOCOM 2020 - IEEE Conference on Computer Communications}, 2020, pp.
  1469--1478.

\bibitem{ref_tracking_bandit}
J.~Zhang, Y.~Huang, Y.~Zhou, and X.~You, ``{Beam Alignment and Tracking for
  Millimeter Wave Communications via Bandit Learning},'' \emph{IEEE
  Transactions on Communications}, vol.~68, no.~9, pp. 5519--5533, 2020.

\bibitem{ref_QLearning_tracking}
H.-L. Chiang, K.-C. Chen, W.~Rave, M.~Khalili~Marandi, and G.~Fettweis,
  ``{Machine-Learning Beam Tracking and Weight Optimization for mmWave
  Multi-UAV Links},'' \emph{IEEE Transactions on Wireless Communications},
  vol.~20, no.~8, pp. 5481--5494, 2021.

\bibitem{ref_tracking_Fingerprint}
R.~Deng, S.~Chen, S.~Zhou, Z.~Niu, and W.~Zhang, ``{Channel Fingerprint Based
  Beam Tracking for Millimeter Wave Communications},'' \emph{IEEE
  Communications Letters}, vol.~24, no.~3, pp. 639--643, 2020.

\bibitem{ref_hu2021image}
Z.~Hu and C.~Han, ``{Image and index fused sequence-to-sequence algorithm for
  vision-aided millimeter-wave beam tracking},'' in \emph{Proc. of ACM Workshop
  on Millimeter-Wave and Terahertz Networks and Sensing Systems}, 2021, pp.
  7--12.

\bibitem{ref_CV_tracking}
S.~Jiang and A.~Alkhateeb, ``{Computer Vision Aided Beam Tracking in A
  Real-World Millimeter Wave Deployment},'' in \emph{Proc. of IEEE Globecom
  Workshops}, 2022, pp. 142--147.

\bibitem{ref_VC_tracking_exp}
M.~Ouyang, F.~Gao, Y.~Wang, S.~Zhang, P.~Li, and J.~Ren, ``{Computer
  Vision-Aided Reconfigurable Intelligent Surface-Based Beam Tracking:
  Prototyping and Experimental Results},'' \emph{IEEE Transactions on Wireless
  Communications}, vol.~22, no.~12, pp. 8681--8693, 2023.

\bibitem{ref_event_tracking}
Y.~Karaçora, C.~Chaccour, A.~Sezgin, and W.~Saad, ``{Event-Based Beam Tracking
  With Dynamic Beamwidth Adaptation in Terahertz (THz) Communications},''
  \emph{IEEE Transactions on Communications}, vol.~71, no.~10, pp. 6195--6210,
  2023.

\bibitem{ref_Context_Aware_Tracking}
H.~Ding and K.~G. Shin, ``{Context-Aware Beam Tracking for 5G mmWave V2I
  Communications},'' \emph{IEEE Transactions on Mobile Computing}, vol.~22,
  no.~6, pp. 3257--3269, 2023.

\bibitem{ref_tracking_UAV}
L.~Yang and W.~Zhang, ``{Beam Tracking and Optimization for UAV
  Communications},'' \emph{IEEE Transactions on Wireless Communications},
  vol.~18, no.~11, pp. 5367--5379, 2019.

\bibitem{ref_Tracking_Variable}
Z.~Xiao, C.~Qi, and J.~Nie, ``Beam tracking based on variable step beam for
  millimeter wave massive mimo,'' \emph{IEEE Communications Letters}, vol.~27,
  no.~9, pp. 2417--2421, 2023.

\bibitem{ref_tracking_codebookopt}
D.~Zhang, A.~Li, M.~Shirvanimoghaddam, P.~Cheng, Y.~Li, and B.~Vucetic,
  ``{Codebook-Based Training Beam Sequence Design for Millimeter-Wave Tracking
  Systems},'' \emph{IEEE Transactions on Wireless Communications}, vol.~18,
  no.~11, pp. 5333--5349, 2019.

\bibitem{ref_NearTracking}
P.~Gavriilidis and G.~C. Alexandropoulos, ``{Near-Field Beam Tracking with
  Extremely Large Dynamic Metasurface Antennas},'' \emph{arXiv preprint
  arXiv:2406.01488}, 2024.

\bibitem{ref_Hierarchical_Beamtracking}
G.~Stratidakis, G.~D. Ntouni, A.~A. Boulogeorgos, D.~Kritharidis, and
  A.~Alexiou, ``{A Low-Overhead Hierarchical Beam-tracking Algorithm for THz
  Wireless Systems},'' in \emph{Proc. of European Conf. Netw. Commun.},
  Dubrovnik, Croatia, 2020, pp. 74--78.

\bibitem{ref_tracking_Infocom}
J.~Palacios, D.~De~Donno, and J.~Widmer, ``{Tracking mm-Wave channel dynamics:
  Fast beam training strategies under mobility},'' in \emph{Proc. of IEEE
  International Conference on Computer Communications}, 2017, pp. 1--9.

\end{thebibliography}

\end{document}